\documentclass{article}
\pdfoutput=1
\usepackage[a4paper, total={16cm, 21cm}]{geometry}
\usepackage[utf8]{inputenc}
\usepackage[title]{appendix}
\usepackage{subfiles, hyperref, stackengine, enumitem, diagbox}
\usepackage{amsmath, amsthm, amsfonts, bm}
\usepackage[dvipsnames]{xcolor}
\usepackage{graphicx, svg, tikz}
\usepackage[position=bottom]{subfig}
\usepackage{array, longtable, multirow, rotating, subcaption}
\usepackage{epsfig, packages/symbols}

\theoremstyle{plain}
\newtheorem{theorem}{Theorem}
\newtheorem{lemma}[theorem]{Lemma}
\newtheorem{proposition}[theorem]{Proposition}
\newtheorem{corollary}[theorem]{Corollary}
\theoremstyle{definition}
\newtheorem{definition}[theorem]{Definition}
\newtheorem{example}[theorem]{Example}

\usepackage[style=phys, biblabel=brackets, sorting=none]{biblatex}
\bibliography{refs.bib}

\usepackage[blocks]{authblk}

\title{Undecidability of the spectral gap in \\ rotationally symmetric Hamiltonians}
\usepackage{orcidlink}
\author[1]{Laura Castilla-Castellano\,\orcidlink{0009-0007-6934-0758} \thanks{{\tt laurca04@ucm.es}}}
\author[1,2,3]{Angelo Lucia\,\orcidlink{0000-0003-1709-1220}\thanks{{\tt angelo.lucia@polimi.it}}}
   
\affil[1]{
    Departamento de Análisis Matemático y Matemática Aplicada, 
    Universidad Complutense de Madrid,\newline 28040 Madrid, Spain
}

\affil[2]{
    Instituto de Ciencias Matemáticas,
    28049 Madrid, Spain
}

\affil[3]{
    Dipartimento di Matematica, Politecnico di Milano, 20131 Milano, Italy
}
    
\date{\today}

\begin{document}

\maketitle

\begin{abstract}
    The problem of determining the existence of a spectral gap in a lattice quantum spin system was previously shown to be undecidable for one \cite{Bausch_2020} or more dimensions \cite{Cubitt2015,CPW22}. In these works, families of nearest-neighbor interactions are constructed whose spectral gap depends on the outcome of a Turing machine Halting problem, therefore making it impossible for an algorithm to predict its existence. While these models are translationally invariant, they are not invariant under the other symmetries of the lattice, a property which is commonly found in physically relevant cases. This poses the question of whether the spectral gap problem could be decidable for Hamiltonians with stronger symmetry constraints. We give a negative answer to this question, in the case of models with 4-body (plaquette) interactions on the square lattice satisfying rotation, but not reflection, symmetry: rotational symmetry is not enough to make the problem decidable.
\end{abstract}

\clearpage

\tableofcontents

\clearpage

\section{Introduction}

Many-body quantum systems can present highly complex behaviors, despite the fact that their defining interactions are often quite simple. Understanding how to predict properties of such systems, given a description of their constituent elements and their interactions, is a crucial task in physics. In recent years, an increasing number of results have appeared, showing that there is an intrinsic limitation to what algorithms can predict about physical models: many relevant properties of physical systems are \emph{undecidable}, in the sense that there is no general algorithm capable of predicting, from a finite description of the model, whether the desired property is present or not. 
Early results regarding undecidability in physics include macroscopical distinguishability of states in Quantum Field Theories \cite{Komar64}, solutions to the wave equation given the initial conditions \cite{Pourel81},
evolution of three-dimensional dynamical systems \cite{Moore90, Bennett90} or of elastic collisions between hard balls \cite{Fredkin82}.
Further results exploited connections with the undecidability of cellular automata \cite{Domany84, Omohundro84, Gu09}, while others leveraged reductions to the tiling problem, a notoriously undecidable classical problem \cite{Wang, Berger1966, Kanter90}. 

With the appearance of quantum information science, the connection between computability and quantum physics has become even stronger, and this renowned interest has brought to light many undecidable properties in numerous areas of quantum physics: unitary operators \cite{Lloyd, Lloyd94}, quantum measurements \cite{Eisert12}, tensor network states \cite{Kliesch14, Morton12, Delascuevas16, Scarpa2020}, measurement-based quantum computation \cite{VandenNest08}, channel capacities \cite{Elkouss2018}, quantum correlations \cite{Bendersky16, 2101.11087}, quantum control \cite{Bondar2020}, quantum field theory \cite{Tachikawa2023}, quantum randomness \cite{AgeroTrejo2024}. See \cite{CPW22}, \cite{Perales_Eceiza_2025} for a review of undecidable properties in classical and quantum physics and their history.

In the setting of quantum spin systems, the main breakthrough has been the proof that the spectral gap (the difference between the two lowest energy levels of a quantum Hamiltonian) is an undecidable quantity \cite{CPW22, Cubitt2015, Bausch_2020}. Since the spectral gap is a crucial quantity in much of the low-energy behavior of these models, from the decay of correlation to the presence of a phase transition \cite{Young14}, it is not surprising that many other results have been obtained in this setting \cite{Bausch_2017, Bausch2021, 2105.13350, 2105.09854, Watson2022, lipics.stacs.2023.54}. 

The general strategy employed in most of these results could be summarized as follows: if we can encode a specific computational problem, already known to be undecidable (usually the Halting problem for Turing machines), into a family of physical models, and if we can do this in such a way that a given property of interest is controlled by the solution of the undecidable problem, then the task of predicting this property will also be undecidable. This procedure results in models which are often convoluted, and as a consequence they do not resemble naturally occurring or physically-inspired models. Notwithstanding the tremendous importance of these results in delimiting the possibility of computational exploration of the physical world, it is an important open question to understand where exactly the limits of computability lie: do some of these undecidable properties become computable if we restrict our task to sufficiently simple or realistic physical models?

As a way to make the constructions arising from undecidability proofs closer to  realistic cases, we consider one of the core concepts of physical theories, namely the presence of certain symmetries of the interactions. We ask the question of whether the undecidability results hold if we restrict our problem only to models which respect a certain symmetry. Previous results show that incorporating a symmetry to a problem can make it easier to solve. Some examples include (i) the classical tiling problem, where the 2-dimensional case with rotation symmetry is known to be of polynomial complexity, as opposed to the exponential non-deterministic complexity of the equivalent non-symmetric problem \cite{GI10}, (ii) the Heisenberg XXX/XXZ models, where the presence of symmetries are the key ingredient in order to proof its integrability \cite{faddeev1996algebraicbetheansatzworks}, (iii) MPS constructions, which are greatly simplified by the presence of reflection symmetry \cite{Pollmann_2010}, \cite{perezgarcia2007matrixproductstaterepresentations}, (iv) classification tasks in 1-dimensional systems that become tractable under symmetries \cite{PhysRevB.83.035107}, and (v) poly-time algorithms for ground state energy of commuting 1-dimensional Hamiltonians \cite{bravyi2003commutative}, even when the general problem is QMA-complete \cite{10.1007/978-3-540-30538-5_31}.

Specifically, as we will consider spin models on the 2D square lattice, we will focus on the discrete spatial symmetries of the lattice. We will consider the problem of determining whether a given quantum spin Hamiltonian on a 2D square lattice has a spectral gap or not, in the sense of whether the difference between its two smallest energy levels is lower bounded by a strictly positive constant uniformly in the size of the lattice. 

Without taking into account symmetries, this problem is undecidable, as it was shown in \cite{Cubitt2015}. In that work, the authors construct a family, indexed by a parameter $n\in \mathbb{N}$, of one-body $h^{(i)}(n)$ and two-body interactions, $h^{col}(n)$ for the vertical direction and $h^{row}(n)$ for the horizontal direction, in such a way that the spectral gap problem for the corresponding family of local translation invariant Hamiltonians $H(n)$ is undecidable. While this construction is, by definition, translation invariant, it does not respect any of the other discrete symmetries of the lattice: specifically, it is not \emph{rotation invariant}, i.e., $h^{col}(n) \neq h^{row}(n)$.

As it is common in lattice models for the local interactions to respect the discrete symmetries of the underlying lattice, we ask the question of whether the undecidability result holds if we impose such symmetries on the local interactions. Specifically, in this work we are interested in the spectral gap problem for \emph{rotationally symmetric} Hamiltonians, i.e., for Hamiltonians whose local interactions are invariant under $\frac{\pi}{2}$, $\pi$, and $\frac{3}{2}\pi$ rotations of the square lattice (but specifically not necessarily invariant under reflection symmetry across any of the coordinates).

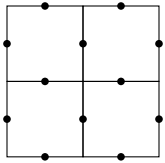
\begin{figure}[htbp]

\centering

\begin{tikzpicture}

    \draw[thin] (0,0) rectangle (1,1);
    \draw[thin] (0,1) rectangle (1,2);
    \draw[thin] (1,0) rectangle (2,1);
    \draw[thin] (1,1) rectangle (2,2);
    
    \fill[black] (0,0.5) circle (0.05);
    \fill[black] (0,1.5) circle (0.05);
    \fill[black] (1,0.5) circle (0.05);
    \fill[black] (1,1.5) circle (0.05);
    \fill[black] (2,0.5) circle (0.05);
    \fill[black] (2,1.5) circle (0.05);
    
    \fill[black] (0.5,0) circle (0.05);
    \fill[black] (1.5,0) circle (0.05);
    \fill[black] (0.5,1) circle (0.05);
    \fill[black] (1.5,1) circle (0.05);
    \fill[black] (0.5,2) circle (0.05);
    \fill[black] (1.5,2) circle (0.05);

\end{tikzpicture}

\caption{Example $2\times 2$ lattice system with sites at the edges.}
\label{fig:lattice_system}

\end{figure}

In contrast to the previous example of classical tiling problems, we show that the spectral gap problem is undecidable with a rotational symmetry, at least when we consider 4-body, plaquette Hamiltonians. Informally, our result could be summarized as follows:

\textbf{Main result (Theorem \ref{theorem:undecidability_spectral_gap}).} Consider a quantum spin system where the spins are sitting on the midpoints of the edges of the square lattice (see Figure~\ref{fig:lattice_system}). For this lattice, there exists a family, indexed by a parameter $n\in\mathbb{N}$, of one-body $h^{(i)}(n)$ and 4-body plaquette interactions $h^{(i,j,k,l)}(n)$, with the property that every $h^{(i,j,k,l)}(n)$ is invariant under cyclic permutation of the indices. Then, the spectral gap problem for the family of local translation invariant Hamiltonians $H(n)$ constructed from such interactions is an undecidable problem.

The relaxation of the problem from 2-body to 4-body interactions is crucial for our result, and we leave as an open question whether the undecidability result still holds for rotationally symmetric 2-body interactions.

Our construction relies heavily on the one from \cite{Cubitt2015}, but we modify it in important and crucial ways in order to obtain a symmetric model at the end. A summary of the key points and main differences from the previous construction is the following:

\begin{enumerate}
    \item An encoding of a Quantum Turing Machine (QTM) into a 1D spin Hamiltonian, whose ground state energy on a finite chain of size $L$ with open boundary conditions depends on the behavior of the QTM. In particular, if the QTM halts on space less than $O(L)$ and time less than $O(c^L)$, the ground state energy is $0$. Otherwise, it has energy larger than $c^{-L}$.

    We modify the construction from \cite{Cubitt2015} by applying an idea from \cite{GI10} (Section 6), in order to make this 1D Hamiltonian reflection symmetric. As a consequence of this modification, the ground state subspace is not unique anymore, but it is degenerate and allows for both a ``forward'' and a ``backwards'' representation of the QTM computation.

    \item An encoding of the classical Tiling problem into a 2D Hamiltonian on the square lattice, in such a way that the ground state of the model gives a description of a classical tiling.

    The Hamiltonian from the previous step has zero energy in the thermodynamic limit in both cases, but using a construction based on 2-dimensional tilings, one can amplify the energy of the halting instances, in order to stop it from vanishing in the limit. The construction from \cite{Cubitt2015} uses the aperiodic Robinson tiling \cite{robinson1971undecidability}: while the tile set is itself invariant under rotations, the 2D Hamiltonian defined in \cite{Cubitt2015} is crucially not rotationally invariant.

    Instead, we define a different encoding of the tiling problem into a 2D Hamiltonian, which is rotationally invariant as long as the underlying tiling set is, but with the downside of having to consider 4-body plaquette interactions instead of nearest neighbor ones.

    \item Finally, in order to merge the Robinson tiling with the QTM Hamiltonian, we initialize a 1D computation on each side of each ``red square'' of size $4^n$, where red squares are features appearing in the Robinson tiling with non-zero density for each value of the parameter $n$. We do so by detecting the corners of such squares, initializing a 1D computation on each side, and fusing them with two specific Hamiltonians that guarantee that the final 2D model has a unique ground state in the gapped case, which is also invariant under rotations (but not under reflections).
\end{enumerate}

As a consequence of the last point, the spectral gap problem we consider is the strong version defined in \cite{Cubitt2015}: to distinguish between the case in which $H(n)$ has a unique ground state with a constant energy gap separating it from any excited state, or has a dense spectrum in an interval containing the ground state energy in the limit of large system size. By construction the family of models defined only falls in one of the two cases.

The paper is organized as follows. In Section~\ref{sec:onedencoding}, we discuss the encoding of a QTM into a 1D local Hamiltonian with reflection symmetry. In Section~\ref{sec:tiling}, we present a 4-body Tiling Hamiltonian which encodes a classical tiling problem, and that is rotationally invariant as long as the underlying tiling set is as well. In Section~\ref{sec:fusion}, we discuss how to connect these two pieces together, initializing a 1D computation on each side of every ``red square'' of the Robinson tiling. Finally, in Section~\ref{sec:undecidability} we put all the pieces together to obtain our undecidability theorem.
\section{Preliminaries}\label{sec:preliminaries}
\subsection{Notation and definitions}
We will be using the standard framework for describing finite spin chains and lattice models. In a chain of $L$ particles, we associate to each site $i$ a Hilbert space $\mathcal{H}^{(i)} \simeq \mathbb{C}^d$, and we will have the following local interactions:
\begin{itemize}
    \item On-site interactions $h^{(i)} \in \mathcal{B}(\mathcal{H}^{(i)})$ for site $i$ in the chain.
    \item Interactions between nearest-neighbor pairs $(i,i+1)$, which are given by $h^{(i,i+1)} \in \mathcal{B}(\mathcal{H}^{(i)} \otimes \mathcal{H}^{(i+1)})$, for all $i \in [1, L-1]$.
\end{itemize}
We will consider the case in which local interactions are translations of each other. The resulting Hamiltonian over the chain is then called \textit{translationally invariant}. We also recall the definitions of frustration-free and classical Hamiltonians, that will be needed when constructing the 1-dimensional Hamiltonian in Section \ref{sec:onedencoding}.

\begin{definition}\label{def:1d_hamiltonian_definitions}
Given on-site interactions $h^{(i)}$ and nearest-neighbor interactions $h^{(i,i+1)}$, we say that the translationally invariant Hamiltonian
    \begin{equation}
        H(L) =  \sum_{i=1}^L h^{(i)} + \sum_{i=1}^{L-1} h^{(i,i+1)}
    \end{equation}
over a $1$-dimensional chain of size $L$ is: 
    \begin{enumerate}
        \item \textit{Frustration-free} if its ground state energy is zero while all $h^{(i)},h^{(i,i+1)}$ are positive semi-definite. That is, a ground state of a frustration-free Hamiltonian minimizes the energy of each interaction term individually.
        \item \textit{Classical} if its defining interactions $h^{(i)},h^{(i,i+1)}$ are diagonal in a given product basis (e.g., the canonical one).
    \end{enumerate}
\end{definition}

Additionally, this Hamiltonian will be reflection invariant (Definition \ref{def:reflection_invariance}). This will be later used to enforce the rotational invariance in the final 2-dimensional lattice Hamiltonian.

\begin{definition}\label{def:reflection_invariance}
For a nearest-neighbor interaction  $h^{(i,i+1)} \in \mathcal{B}(\mathcal{H}^{(i)} \otimes \mathcal{H}^{(i+1)})$, we define its reflection $h^{(i+1,i)} \in \mathcal{B}(\mathcal{H}^{(i)} \otimes \mathcal{H}^{(i+i)})$ as $h^{(i+1,i)} = U_R h^{(i,i+1)} U_R^\dag$, where $U_R\ket{x}\ket{y} = \ket{y}\ket{x}$. 
If $h^{(i,i+1)}=h^{(i+1,i)}$ for all $i \in [1, L - 1]$, the Hamiltonian $H(L)$ is said to be \textit{invariant under reflection}.
\end{definition}

In the 2D lattice setting, our sites will be at the edges of the squares of the $\mathbb{Z}^2$ lattice, instead of at the vertices. Denote as $\Lambda(L\times H)$ the set of sites that rest in the middle point of the edges of a square of length $L \in \mathbb{N}$ unit cells and width $H \in \mathbb{N}$ unit cells, with $L, H \ge 1$. The number of sites in this lattice description is $|\Lambda (L\times H)| = L(H+1)+H(L+1)$. We will denote the square of size $L$ as $\Lambda(L) = \Lambda(L\times L)$. Four sites $i,j,k,l$ form a \emph{plaquette} if they are at the four edges of the same unit cell in the lattice. We use the convention that the four sites are ordered in a clockwise fashion, starting from the top edge, and that the plaquette is denoted by $(i,j,k,l)$.

To each site $i \in \Lambda(L\times H)$, we associate a Hilbert space $\mathcal{H}^{(i)} \simeq \mathbb{C}^d$, and to any subset $S \subseteq \Lambda(L\times H)$ the tensor product $\bigotimes_{i \in S} \mathcal{H}^{(i)}$. We will have the following local interactions:
\begin{itemize}
    \item On-site interactions $h^{(i)} \in \mathcal{B}(\mathcal{H}^{(i)})$ for a site $i \in \Lambda(L\times H)$.
    \item Plaquette interactions $h^{(i,j,k,l)} \in \mathcal{B}(\mathcal{H}^{(i)} \otimes \dots \otimes \mathcal{H}^{(l)})$, for each plaquette $(i,j,k,l)$ (where we are ordering the sites with the convention explained above).
\end{itemize}
Once again, we will consider the translation invariant case, in which each type of local interaction is invariant under translations of the lattice. The corresponding translation invariant Hamiltonian is then given by
\begin{equation}\label{eq:2D_Hamiltonian}
        H^{\Lambda(L)}=\sum_{i \in \Lambda(L)}h^{(i)}+\sum_{(i,j,k,l)\in \Lambda(L)}h^{(i,j,k,l)},
\end{equation}
where with a slight abuse of notation we denoted by $\sum_{(i,j,k,l)\in \Lambda(L)}$ the sum over all plaquettes $(i,j,k,l)$ in $\Lambda(L)$.

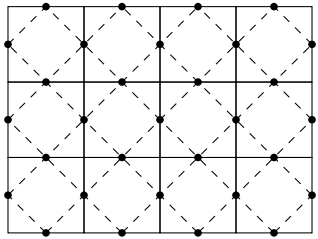
\begin{figure}[htbp]

\centering

\begin{tikzpicture}

    \draw[thin] (0,0) rectangle (1,1);
    \draw[thin] (0,1) rectangle (1,2);
    \draw[thin] (0,2) rectangle (1,3);
    \draw[thin] (1,0) rectangle (2,1);
    \draw[thin] (1,1) rectangle (2,2);
    \draw[thin] (1,2) rectangle (2,3);
    \draw[thin] (2,0) rectangle (3,1);
    \draw[thin] (2,1) rectangle (3,2);
    \draw[thin] (2,2) rectangle (3,3);
    \draw[thin] (3,0) rectangle (4,1);
    \draw[thin] (3,1) rectangle (4,2);
    \draw[thin] (3,2) rectangle (4,3);

    \draw[dashed] (0,2.5) -- (0.5,3);
    \draw[dashed] (0,1.5) -- (1.5,3);
    \draw[dashed] (0,0.5) -- (2.5,3);
    \draw[dashed] (0.5,0) -- (3.5,3);
    \draw[dashed] (1.5,0) -- (4,2.5);
    \draw[dashed] (2.5,0) -- (4,1.5);
    \draw[dashed] (3.5,0) -- (4,0.5);
    \draw[dashed] (0.5,0) -- (0,0.5);
    \draw[dashed] (1.5,0) -- (0,1.5);
    \draw[dashed] (2.5,0) -- (0,2.5);
    \draw[dashed] (3.5,0) -- (0.5,3);
    \draw[dashed] (4,0.5) -- (1.5,3);
    \draw[dashed] (4,1.5) -- (2.5,3);
    \draw[dashed] (4,2.5) -- (3.5,3);
    
    \fill[black] (0,0.5) circle (0.05);
    \fill[black] (0,1.5) circle (0.05);
    \fill[black] (0,2.5) circle (0.05);
    \fill[black] (1,0.5) circle (0.05);
    \fill[black] (1,1.5) circle (0.05);
    \fill[black] (1,2.5) circle (0.05);
    \fill[black] (2,0.5) circle (0.05);
    \fill[black] (2,1.5) circle (0.05);
    \fill[black] (2,2.5) circle (0.05);
    \fill[black] (3,0.5) circle (0.05);
    \fill[black] (3,1.5) circle (0.05);
    \fill[black] (3,2.5) circle (0.05);
    \fill[black] (4,0.5) circle (0.05);
    \fill[black] (4,1.5) circle (0.05);
    \fill[black] (4,2.5) circle (0.05);
    
    \fill[black] (0.5,0) circle (0.05);
    \fill[black] (1.5,0) circle (0.05);
    \fill[black] (2.5,0) circle (0.05);
    \fill[black] (3.5,0) circle (0.05);
    \fill[black] (0.5,1) circle (0.05);
    \fill[black] (1.5,1) circle (0.05);
    \fill[black] (2.5,1) circle (0.05);
    \fill[black] (3.5,1) circle (0.05);
    \fill[black] (0.5,2) circle (0.05);
    \fill[black] (1.5,2) circle (0.05);
    \fill[black] (2.5,2) circle (0.05);
    \fill[black] (3.5,2) circle (0.05);
    \fill[black] (0.5,3) circle (0.05);
    \fill[black] (1.5,3) circle (0.05);
    \fill[black] (2.5,3) circle (0.05);
    \fill[black] (3.5,3) circle (0.05);

\end{tikzpicture}

\caption{Example lattice system, where the sites rest at the middle point of the squares' edges that form the lattice. Here, $L=4$ and $H=3$, with a total of $L(H+1)+H(L+1)=31$ sites. The dashed lines correspond to the edges of a standard lattice description (with sites at the vertices), so this system can also be seen as a $\frac{\pi}{4}$ rotation of the usual $\mathbb{Z}^2$.}
\label{fig:plaquette_system}

\end{figure}

In this work, we first construct a 1-dimensional Hamiltonian with on-site and 2-body terms with the previously defined properties. This will be latter embedded in plaquette form (see Section \ref{sec:fusion}), and will be part of a 2-dimensional Hamiltonian on a square lattice $\Lambda(L)$ with on-site and plaquette (4-body) interactions. Moreover, the constructed Hamiltonian $H^{\Lambda(L)}$ will be rotationally symmetric, as described in Definition \ref{def:rotational_invariance}.

\begin{definition}\label{def:rotational_invariance}
For a plaquette interaction $h^{(i,j,k,l)} \in \mathcal{B}(\mathcal{H}^{(i)} \otimes \dots \otimes \mathcal{H}^{(l)})$, we define its $\frac{\pi}{2}$ rotation as $U_{\pi/2} h^{(i,j,k,l)} U_{\pi/2}^\dag$, where
\[
U_{\pi/2} \ket{x}\ket{y}\ket{w}\ket{z} = \ket{z}\ket{x}\ket{y}\ket{w},
\]
We say that the Hamiltonian $H^{\Lambda(L)}$ is \textit{invariant under rotations} if $h^{(i,j,k,l)} = U_{\pi/2} h^{(i,j,k,l)} U_{\pi/2}^\dag$.
\end{definition}
Note that $U_{\pi/2}$ also generates the $U_{\pi} = U_{\pi/2}^2$ and $U_{3\pi/2}=U_{\pi/2}^3$ rotations, so the Hamiltonian is invariant under the $U_{\pi/2}$ rotation if and only if is invariant under $U_{\theta}$ rotations for $\theta \in \{\pi/2,\pi,3\pi/2\}$.

The set of eigenvalues of the Hamiltonian $H^{\Lambda(L)}$ will be denoted by $\spec H^{\Lambda(L)}:=\{\lambda_0 (H^{\Lambda(L)}), \lambda_1(H^{\Lambda(L)}),\dots\}$ or simply $\{\lambda_0,\lambda_1,\dots\}$ if the Hamiltonian is clear from context. They are always assumed to be listed in non-decreasing order. The smallest eigenvalue $\lambda_0(H^{\Lambda(L)})$ is called the \textit{ground state energy} and the corresponding eigenvectors, \textit{ground states}. One can then define the following concepts related to the ground state energy as in Definition \ref{def:gs_energy_definitions}, where the notions of gap and gapless are the same as the ones considered in \cite{CPW22}.

\begin{definition}\label{def:gs_energy_definitions}
Consider a family of $2$-dimensional plaquette Hamiltonians on a square lattice $\Lambda(L)$, for sizes $L \in \mathbb{N}$. We have the following energy definitions:
    \begin{enumerate}
        \item The \textit{ground state energy density} of Hamiltonian $H^{\Lambda(L)}$ is
            \begin{equation*}
                E_{\rho}:=\lim_{L\rightarrow \infty} E_\rho(L), \quad \text{where } E_\rho(L):=\frac{\lambda_0(H^{\Lambda(L)})}{2L(L+1)}.
            \end{equation*}
        \item The \textit{spectral gap} of $H^{\Lambda(L)}$ is $\Delta(H^{\Lambda(L)}):= \lambda_1(H^{\Lambda(L)})-\lambda_0(H^{\Lambda(L)})$.
        \item The family $\{H^{\Lambda(L)}: L \in \mathbb{N}\}$ is \textit{gapped}, if there is a constant $\gamma>0$ and a system size $L_0$ such that for all $L>L_0$, $\lambda_0(H^{\Lambda(L)})$ is non-degenerate and $\Delta(H^{\Lambda(L)}) \geq \gamma$. In this case, we say that the spectral gap is at least $\gamma$.
        \item The family $\{H^{\Lambda(L)}: L \in \mathbb{N}\}$ is \textit{gapless}, if there is a constant $c>0$ such that for all $\varepsilon>0$ there is an $L_0\in\mathbb{N}$ so that for all $L>L_0$, any point in $[\lambda_0(H^{\Lambda(L)}),\lambda_0(H^{\Lambda(L)})+c]$ is within distance $\varepsilon$ from $\spec H^{\Lambda(L)}$.
    \end{enumerate}
\end{definition}

We finish this section by giving some examples of gapped and gapless Hamiltonians having rotational symmetry, which will also be useful later when construction the final result in Section \ref{sec:undecidability}.

\begin{example}\label{ex:gapped_hamiltonian}
Let the on-site Hilbert space $\mathcal{H}^{(i)} \simeq \mathbb{C}^2$ for every site $i \in \mathbb{Z}^2$.
Consider a single-site interaction $h_0^{(i)}=(\mathbbm{1}_i - \sigma_z^{(i)})/2$, where $\sigma_z$ is the Pauli-Z matrix. We can define a plaquette interaction by:
    \begin{equation}
        h_0^{(i,j,k,l)} = h_0^{(i)} \otimes \mathbbm{1}_{jkl} + h_0^{(j)} \otimes \mathbbm{1}_{ikl} + h_0^{(k)} \otimes \mathbbm{1}_{ijl} + h_0^{(l)} \otimes \mathbbm{1}_{ijk},
    \end{equation}
where $\mathbbm{1}_{jkl}$ is the identity over the subspace $\mathcal{H}^{(j)} \otimes \mathcal{H}^{(k)}  \otimes \mathcal{H}^{(l)} $ (and analogously defined in the other cases). The Hamiltonian $H_0^{\Lambda(L)}$ obtained in this way has unique ground state for every $L$: $\ket{0}^{\otimes \abs{\Lambda(L)}}$, with a ground state energy of $0$. Any site in state $\ket{1}$ will contribute with energy $2$, so it is a gapped Hamiltonian, with a spectral gap of $2$.
\end{example}

\begin{example}\label{ex:gapless_hamiltonian}
Let the on-site Hilbert space $\mathcal{H}^{(i)} \simeq \mathbb{C}^2$ for every site $i \in \mathbb{Z}^2$.
Consider the spin-1/2 ferromagnetic Heisenberg model on the square lattice. Its two-body interaction is given by
    \begin{equation}
        h_d^{(i,i+1)} = -\frac{1}{4}(\sigma_x^{(i)} \otimes \sigma_x^{(i+1)} + \sigma_y^{(i)} \otimes \sigma_y^{(i+1)} + \sigma_z^{(i)} \otimes \sigma_z^{(i+1)}),
    \end{equation}
where $\{\sigma_x, \sigma_y, \sigma_z\}$ are the Pauli matrices. This model presents spontaneous symmetry breaking \cite{Beekman2019} of the continuous $SU(2)$ symmetry, and is therefore gapless due to the Goldstone's theorem \cite{Landau1981, Koma1994}. Its ground state energy per particle is $-1/2$, so we could set its ground state energy to zero by a simple energy shift. Based on this 2-body interaction, we define the following plaquette interaction:
    \begin{equation}
        h_d^{(i,j,k,l)} = \mathbbm{1}_{ijkl} + h_d^{(i,j)} \otimes \mathbbm{1}_{kl} + h_d^{(j,k)} \otimes \mathbbm{1}_{il} + h_d^{(k,l)} \otimes \mathbbm{1}_{ij} + h_d^{(l,i)} \otimes \mathbbm{1}_{jk},
    \end{equation}
where $\mathbbm{1}_{kl}$ is the identity over the subspace $\mathcal{H}_k \otimes \mathcal{H}_l$ (and analogously defined in the other cases). This plaquette term consists of a $h_d^{(i,i+1)}$ interaction between each pair of sites at distance $1/\sqrt{2}$. Following the depiction of Figure \ref{fig:plaquette_system}, this corresponds to nearest-neighbor sites as connected by the dashed lines. With this description, the Hamiltonian $H_d^{\Lambda(L)}$ defined by this local plaquette interaction $h_d^{(i,j,k,l)}$ is rotationally invariant and gapless according to Definition~\ref{def:gs_energy_definitions}.
\end{example}

\subsection{Turing Machines}
In 1936, Alan Turing showed that the Halting problem is an undecidable problem (\cite{turing1936computable}). One common strategy to prove that other problems are also undecidable is to encode the Halting problem in them, and this is precisely what it is done in \cite{CPW22}, and also here. First, we recall the definitions of both undecidability and the Halting problem.

\begin{definition}\label{def:undecidable_problem}
    An \textit{undecidable problem} is a decision problem (one with a yes/no output) that has been proven to have no general algorithm for its resolution.
\end{definition}

\begin{definition}\label{def:halting_problem}
    Given an arbitrary pair $(P, I)$ describing a program and an input, the \textit{Halting problem} is the problem of determining if the program $P$ on input $I$ will finish in a finite amount of steps or will continue to run forever (i.e., $P(I)=0$ or $P(I)=1$).
\end{definition}

The standard way of representing computational problems $P$ is the general model of Turing Machines (TM). This is not restricted to the classical setting, and therefore one can find their equivalent for quantum problems: Quantum Turing Machines (QTM). For completeness, we recall below the definitions\footnote{All the definitions for this section are taken from \cite{CPW22} (Section 3.1) and \cite{BV93} (Section 3).}, as well as some of its properties that are needed to guarantee the final result in Section \ref{sec:onedencoding}. Even so, these properties are not restricting, as they can be found in the general class of universal QTMs (\cite{BV93}).

\begin{definition}\label{def:turing_machine}
  A \textit{(deterministic) Turing Machine} (TM) is defined by a triplet $(\Sigma, Q, \delta)$ where $\Sigma$ is a finite alphabet with an identified blank symbol $\#$, $Q$ is a finite set of states with an identified initial state $q_0$ and final state $q_f\not = q_0$, and $\delta$ is a transition function
      \begin{equation}\label{eq:classical_transition_function}
        \delta: Q\times \Sigma \rightarrow \Sigma\times Q\times \{L,R\}.
      \end{equation}
  The TM has a two-way infinite tape of cells indexed by $\mathbb{Z}$ and a single read/write tape head that moves along the tape. A configuration of the TM is a complete description of the contents of the tape, the location of the tape head and the state $q\in Q$ of the finite control. At any time, only a finite number of the tape cells may contain non-blank symbols.
  
  For any configuration $c$ of the TM, the successor configuration $c'$ is defined by applying the transition function to the current state and the symbol scanned by the head, replacing them by those specified in the transition function and moving the head left (L) or right (R) according to $\delta$.

  By convention, the initial configuration satisfies the following conditions: the head is in cell $0$, called the \textit{starting cell}, and the machine is in state $q_0$. We say that an initial configuration has input $x\in (\Sigma\setminus\#)^*$ if $x$ is written on the tape in positions $0,1,2,\dots$ and all other tape cells are blank. The TM halts on input $x$ if it eventually enters the final state $q_f$. The number of steps a TM takes to halt on input $x$ is its running time on input $x$. If a TM halts, then its output is the string in $\Sigma^*$ consisting of those tape contents from the leftmost non-blank symbol to the rightmost non-blank symbol, or the empty string if the entire tape is blank. A TM is called \textit{reversible} if each configuration has at most one predecessor.
\end{definition}

\begin{definition}
    An \textit{Universal Turing Machine} (UTM) is a Turing Machine capable of simulating any other Turing Machine. That is, $UTM(TM, n) = TM(n)$ for every Turing Machine TM and input $n$.
\end{definition}

\begin{definition}
  Call $\tilde{\mathbb{C}}$ to the set of $\alpha \in \mathbb{C}$ such that there is a deterministic algorithm that computes the real and imaginary parts of $\alpha$ to within $2^{-n}$ in time polynomial in $n$. A \textit{Quantum Turing Machine} (QTM) is defined by a triplet $(\Sigma, Q, \delta)$ where $\Sigma$ is a finite alphabet with an identified blank symbol $\#$, $Q$ is a finite set of states with an identified initial state $q_0$ and final state $q_f\neq q_0$, and a quantum transition function
  \begin{equation}\label{eq:quantum_transition_function}
    \delta: Q\times \Sigma \rightarrow \tilde{\mathbb{C}}^{\Sigma\times Q\times \{L,R\}}.
  \end{equation}
  The QTM has a two-way infinite tape of cells indexed by $\mathbb{Z}$ and a single read/write tape head that moves along the tape. We define configurations, initial configurations and final configurations exactly as for deterministic TMs.

  Let $\mathcal{S}$ be the inner-product space of finite complex linear combinations of configurations of the QTM, which we call $M$, with the Euclidean norm. We call each element $\phi\in \mathcal{S}$ a superposition of $M$. The QTM $M$ defines a linear operator $U_M: \mathcal{S}\rightarrow\mathcal{S}$, called the time evolution operator of $M$, as follows: if $M$ starts in configuration $c$ with current state $p$ and scanned symbol $\sigma$, then after one step $M$ will be in superposition of configurations $\psi=\sum_i\alpha_ic_i$, where each nonzero $\alpha_i$ corresponds to the amplitude $\delta(p,\sigma,\tau,q,d)$ of $\ket{\tau}\ket{q}\ket{d}$ in the transition $\delta(p,\sigma)$ and $c_i$ is the new configuration obtained by writing $\tau$, changing the internal state to $q$ and moving the head in the direction of $d$. Extending this map to the entire $\mathcal{S}$ through linearity gives the linear time evolution operator $U_M$.
\end{definition}

\begin{definition}
    Equivalently to the classical case, we say that a \textit{Universal Quantum Turing Machine} (UQTM) is a Quantum Turing Machine capable of simulating any other Quantum Turing Machine. That is, $UQTM(QMT, n) = QTM(n)$ for every Quantum Turing Machine QTM and input $n$.
\end{definition}

\begin{definition}
We say that a QTM $M=(\Sigma,Q,\delta)$ is:
    \begin{enumerate}
        \item \textit{Well-formed}\footnote{Any reversible TM is also a well-formed QTM where the quantum transition function $\delta(p,\sigma, q,\tau, d)=1$ if $\delta(p,\sigma)=(q,\tau,d)$ for the reversible TM and $0$ otherwise (\cite{BV93}, Theorem 4.2).}, if its time evolution operator is an isometry, that is, it preserves the Euclidean norm.
        \item In \textit{normal form}, if it is a well-formed QTM (or reversible TM) and $\forall\sigma\in\Sigma$, $\delta(q_f,\sigma) = \ket{\sigma}\ket{q_0}\ket{N}$.
        \item \textit{Unidirectional}, if each state can be entered from only one direction. In other words, if $\delta(p_1, \sigma_1, \tau_1, q, d_1)$ and $\delta(p_2, \sigma_2, \tau_2, q, d_2)$ are both non-zero, then $d_1$ = $d_2$.
    \end{enumerate}
\end{definition}

\begin{definition}
    A \textit{generalised TM} or \textit{generalised QTM} is defined exactly as a standard TM or QTM, except that the head can also stay still, as well as move left or right. That is, the set of head movement directions is $\{L,R,N\}$ instead of just $\{L,R\}$.
\end{definition}
\section{Encoding the Halting problem in a 1D reflection symmetric Hamiltonian}\label{sec:onedencoding}
If one can construct Hamiltonian interactions that result in different energy behaviors (gapped/gapless) depending on the output of the Halting problem (halts/does not halt), one concludes that the problem of determining the spectral gap of such class of Hamiltonians is undecidable, as it depends on the answer to an undecidable problem. But for this, a way of linking computational problems and Hamiltonians is needed.

This was firstly done in \cite{GI10}, albeit for a different question (the complexity of the ground state problem in 1D). In that work, the authors define local interactions according to the states and transitions of a Universal Turing Machine (UTM), by penalizing the appearance of certain configurations over others. If the UTM runs for $K$ steps, this results in a single zero energy ground state: the superposition of all the UTM states over steps $[0,K]$.

This idea of penalizing configurations was adapted in \cite{CPW22} for the different goal of spectral gap undecidability, resulting in a $1$-dimensional translationally-invariant and nearest-neighbor Hamiltonian, constructed in such a way that the local Hilbert space dimension is a function only of the alphabet size and number of internal states (i.e., finite) of the UTM. This allows them to encode any Universal Turing Machine in a Hamiltonian with fixed local dimension.

Nevertheless, the issue is that the Halting problem (Definition \ref{def:halting_problem}) relies on the arbitrariness of the program-input pair $(P,I)$. The Turing Machine will act in accordance to $P$, but it needs to have the input $I$ written on tape (see Definition \ref{def:turing_machine}). So, before diving into the link between computational evolution and ground states, one needs to face a previous issue: to be able to deterministically write any possible (binary) input. This question was solved in Section 3 in \cite{CPW22}, and we briefly discuss it in Section \ref{subsec:qpe_machine}. Once this preliminary step is done, one can start to build the desired 1-dimensional computational Hamiltonian.

However, the computational Hamiltonian, as defined in Section 4 of \cite{CPW22} does not present any symmetric properties. This will be the goal of this section: modify the computational Hamiltonian in \cite{CPW22}, and make it invariant under reflection. Coincidentally, we can do so by fusing with an idea that traces back again to \cite{GI10}, Section 6, where the authors present (again, for their purpose of complexity of the ground state) a key idea involving 1-dimensional chains and reflection symmetry: for any QTM (as described in their work, and consequently as in the later \cite{CPW22}), one can construct an associated reflection invariant $1$-dimensional Hamiltonian whose ground state still encodes the same evolution of said QTM.

We start by addressing the issue of the input problem, in Section \ref{subsec:qpe_machine}. After that, in Section \ref{subsec:encoding_qtms}, we present the main ideas behind the construction that links Hamiltonian ground states and Turing machines. Then, in Section \ref{subsec:system_building}, we see how to modify these ideas in order to incorporate the notion of symmetry. Finally, in Section \ref{subsec:system_behavior}, we study the ground state energy behavior of the constructed Hamiltonian.

It is important to remark that due to this symmetry, our 1-dimensional Hamiltonian does not have a unique ground state anymore, but two different ground states with the same energy. However, this duplication will not be a further problem: we still need to add more ingredients to the construction, and in Section \ref{sec:undecidability} we will see how the desired final Hamiltonian, which presents undecidable behavior, has a unique ground state if gapped.

\subsection{QPE-based machine for writing inputs}\label{subsec:qpe_machine}
The problem of generating any desired input to feed a Universal Turing Machine (UTM) is solved in Section 3 in \cite{CPW22}, where the authors design a QTM based on the Quantum Phase Estimation (QPE) algorithm. Their purpose is to guarantee that for every $n \in \mathbb{N}$, its binary representation $|n|$ can be written to tape, and after that, the machine halts deterministically.

This is detailed in Theorem 10 of \cite{CPW22}, and as the authors state, it is the only quantum ingredient in the result. Some special requirements (well-formed, normal form, unidirectional QTMs) for the machine are needed, but they explicitly construct this particular family of QTMs that fulfills all of them, denoted as $P_n$. We will use this same $P_n$ in our result, without any modifications. As in their work, this will also be our only quantum gadget.

Afterwards, by concatenating\footnote{Turing Machines can be concatenated by using the Dovetailing Lemma (Lemma 4.9) from \cite{BV93}.} $P_n$ and the UTM, one can obtain a family of (universal) Quantum Turing Machines, QTM($n$), of fixed alphabet size and number of internal states. This QTM($n$) first writes deterministically the binary expansion of $n$, and then uses it as input for the UTM.

As every input $I$ in Definition \ref{def:halting_problem} can be represented as a binary string, and the UTM can simulate any possible program $P$, the family of machines QTM($n$) is indeed a representation of any possible program-input pair $(P, I)$. And, as in \cite{CPW22}, we want to encode in the ground state the Halting problem associated to $(P,I)$.

\subsection{Encoding QTMs in local Hamiltonians}\label{subsec:encoding_qtms}

The key object of encoding quantum computations into ground states is the computational history state, which encodes the entire history of the computation in superposition. This idea goes back to Feynman \cite{feynman1986quantum}, and was developed into its modern form by Kitaev \cite{kitaev2002classical}. We recall its definition and its associated Hamiltonian, as stated in \cite{CPW22}.

\begin{definition}\label{def:history_state}
A \textit{computational history state} $\ket{\psi}_{CQ} \in \mathcal{H}_C \otimes \mathcal{H}_Q$ is a state of the form
\begin{equation}\label{eq:computational_history_state}
\ket{\psi}_{CQ} = \frac{1}{\sqrt{T}} \sum_{t=0}^{T}\ket{t}\ket{\psi_t},
\end{equation}
where $\{\ket{t}\}$ is an orthonormal basis for $\mathcal{H}_C$ and $\ket{\psi_t} = \prod_{i=1}^t U_i\ket{\psi_0}$ for some initial state $\ket{\psi_0} \in \mathcal{H}_Q$ and set of unitaries $U_i\in\cB(\HS_Q)$.
$\mathcal{H}_C$ is called the clock register and $\mathcal{H}_Q$ is called the computational register. If $U_t$ is the unitary transformation corresponding the $t$'th step of a quantum computation, then $\ket{\psi_t}$ is the state of the computation after $t$ steps. We say that the history state $\ket{\psi}$ encodes the evolution of this quantum computation of $T$ steps.
\end{definition}

In order to obtain a Hamiltonian whose ground space is spanned by the set of computational history states, one first focuses on the clock register $\mathcal{H}_C$, and looks for a Hamiltonian that has as ground state the superposition of all clock time steps:
\begin{equation}\label{eq:clock_state}
\ket[C]{\psi} = \frac{1}{\sqrt{T}} \sum_{t=1}^{T}\ket{t},
\end{equation}
which can be enforced by a standard hopping Hamiltonian:
\begin{equation}\label{eq:hopping_Hamiltonian}
    H_C=\sum_{t=1}^T \left(\ket{t}-\ket{t-1}\right) \left(\bra{t}-\bra{t-1}\right).
\end{equation}
To obtain the history state, one applies the controlled unitary $U_{CQ}=\sum_{t} \ketbra{t}{t} \otimes U_t U_{t-1}\cdots U_1$ on $\ket[C]{\psi}\ket[Q]{\psi_0}$, and that results in the final Hamiltonian:
\begin{equation}\label{eq:history_state_Hamiltonian}
    H_{CQ}= \sum_{t=1}^T \left(\ketbra{t}{t} \otimes \1 + \ketbra{t-1}{t-1} \otimes \1 - \ketbra{t}{t-1}\otimes U_t - \ketbra{t-1}{t}\otimes U_t^\dagger \right).
\end{equation}
The difficulty arises when one wants to implement this construction with a Hamiltonian that is local, one-dimensional and translational invariant. These issues were addressed consecutively in \cite{kitaev2002classical}, \cite{AGIK09} and \cite{GI10}. In fact, the purpose of the construction in \cite{GI10} was to show the hardness of finding the ground state energy, but the same ideas were later used as a base to build up the undecidability result in \cite{CPW22} (Section 6.2).

The hopping terms in the Hamiltonian are called \textit{evolution} or \textit{transition rule terms}, and they enforce the evolution of the clock. Transition rule terms have the form $\tfrac{1}{2}(\ket{\psi}-\ket{\varphi})(\bra{\psi}-\bra{\varphi})$, where $\ket{\psi},\ket{\varphi}$ are states on the same pair of adjacent sites. This forces any zero-energy eigenstate with amplitude on a configuration containing neighboring states $ab$ to also have equal amplitude on the configuration in which those $ab$ is replaced by $cd$ (and therefore also being a zero-energy state). Following the notation in \cite{GI10}, we will denote transition rule Hamiltonian terms by their associated transitions $ab\rightarrow cd$ or, more generally, $\ket{\psi} \rightarrow \ket{\varphi}$.

But apart from the transition terms enforcing the evolution of the clock, one also needs to also include penalty terms that restrict the type of configurations that can appear in the ground state, by giving a positive energy to configurations that do not appear in the clock oscillation cycle.

The type of constraints that can be enforced by penalty terms is characterised by the notion of regular expressions\footnote{See, for example, Definition 33 in \cite{CPW22} for a rigorous definition of the notion of regular expressions.}. A regular expression denotes a (possibly infinite) subset of finite-length strings over a finite alphabet. Equivalently, it can be thought of as a pattern that matches all the strings in the subset and no others. In our case, our alphabets will be different sets of states, that will be called \textit{standard basis states}.

Penalty terms have the form $\proj{ab}$ where $a,b$ are standard basis states. This adds a positive energy contribution to any configuration containing $a$ to the left of $b$. We call $ab$ an illegal pair, and denote a penalty term $\proj{ab}$ in the Hamiltonian by its corresponding illegal pair. We call a configuration of states over the chain \textit{legal} if it does not contain any illegal pairs, and \textit{illegal} otherwise.

As the construction is divided over different subspaces (called ``Tracks'', defined later in in Section \ref{sec:onedencoding}), we will sometimes also make use of single-site illegal states $a$, but note that single-site illegal states are easily implemented in terms of illegal pairs, by adding penalty terms $\proj{ax}$ and $\proj{xa}$ for all pairs $ax$ and $xa$ in which the single-site state appears. 

For example, by using single-site illegal states one can enforce that, on all legal states, the marker $\ket{\symbar}$ can only ever appear simultaneously on all tracks. The single-site illegal states enforcing this are shown in Table \ref{tab:end_markers_all_tracks}. Illegal pairs are also used to initialize the tracks in a desired configuration, as per Lemma 5.2 in \cite{GI10}, which ensures that penalty terms can be used in order to restrict to configurations that match a regular expression.

\setlength{\tabcolsep}{2pt}
\renewcommand{\arraystretch}{0.75}
\newcolumntype{P}[1]{>{\centering\arraybackslash}p{#1}}
\begin{longtable}{P{10cm}}\label{tab:end_markers_all_tracks}
    $\twocellsvert{\symbar}{\neg\symbar}$,\,
    $\twocellsvert{\neg\symbar}{\symbar}$ \\
    \caption{Track i-j single-site illegal states, for all pairs $i \not = j$}
\end{longtable}

We will later (in Section \ref{sec:fusion}) precisely enforce this: the sites at the two ends will always be in state $\symbar$ in all tracks. Therefore, we can restrict our analysis just to the subspace $S_{br}$ spanned by these configurations. These configurations are called ``bracketed'' in \cite{CPW22}, as they use two different boundary terms represented by the brackets $\leftend$ and $\rightend$. However, for our purposes, we go back to Section 6 of \cite{GI10}, where the authors introduce a ``A/B labeling'' symmetry idea. We use their description of a system with reflection symmetry, so our single boundary state is indeed $\symbar$. We turn now to explain this ``A/B labeling'', and how to merge it with the system already depicted in Section 4 of \cite{CPW22}.

\subsection{Building the system}\label{subsec:system_building}

We will use the construction in Section 4 of \cite{CPW22} as a starting point. The chain is divided into 6 different tracks (or subspaces). However, we add an additional Track 0, which is the same from Section 6 in \cite{GI10}, and will have the same purpose: making the construction invariant under reflections. For that reason, we also use a single boundary term: $\symbar$, as in \cite{GI10}, and opposed to \cite{CPW22}.

Tracks 1-6 behave as in \cite{CPW22}, with Tracks 1-3 corresponding to the clock register $C$ and Tracks 4-6 to the computational register $Q$ in the computational history state. Track 1 acts as a ``second hand'', and Track 3 as the ``minute hand'', getting incremented by one for every cycle of Track 1. Track 2 stores the counter TM head and internal state needed to implement this incrementing operation. This clock drives the given QTM $M$ as described later in Tracks 4-6. For a detailed review of Tracks $1$-$6$, see Appendix \ref{appendix:tables}.

The new Track 0 will determine the direction of the computation. In the construction of \cite{CPW22}, what guides the computation is the ``arrow state'' in Track 1, which initializes to the left end of the chain, and starts sweeping left to right (and then back). We will call this the \textit{canonical orientation}. With Track 0, we will allow the system to initialize in the right end, and then start running in a right to left direction (and back). We will call it the \textit{reverse orientation}.

In the construction, the different tracks evolve according to some transition rules\footnote{For a detailed explanation of the reasoning behind the construction of Tracks $1$ to $6$, see section $4$ of \cite{CPW22}.}. We will use the same $ab \rightarrow cd$ notation as in the previous works. This represents the transition rules that applies to neighboring sites ($i$, $i+1$). If they are in states $a$ and $b$ respectively, they transform to states $c$ and $d$ on the next time step. This can be interpreted as reading the evolution from left to right (canonical orientation). In the reverse orientation, $a$ also transforms to $c$ and $b$, to $d$. However, sites $i$ and $i+1$ are now interchanged, so the notation is now $ba \rightarrow dc$, or, equivalently, $dc \leftarrow ba$, as if we were reading from right to left.

Arrow states in Track 0 (called ``control particles'' in \cite{GI10}) will only appear in the positions where an arrow state is simultaneously present in Track 1. The basic idea is to force these arrow states to have a $\pa$ on one side and $\pb$ on the other, creating an asymmetry between the two directions that can be distinguished by looking at the label pointed by the arrow. We will require that our chain has an even number of particles $L$, and then construct Track 0 to behave as follows:

\begin{itemize}
    \item There is only one control particle, located in the same position of the arrow in Track 1.
    \item The control particle initializes as $\ar$, with $\symbar$ at one side and $\pa$ to the other.
    \item The rest of the chain alternates between $\pa$ and $\pb$ states.
    \item The control particle is flanked by an $\pa$ state on one side and by an $\pb$ state or $\symbar$ on the other.
    \item When moving, the control particle will change labels, and the arrow orientation will follow the same principles as in Track 1 (only switching when reaching boundaries).
    \item The evolution will consist of the control particle moving through the chain. Non-control states will not change labels, but will be displaced in relation to the control particle.
\end{itemize}

\begin{figure}[htbp]
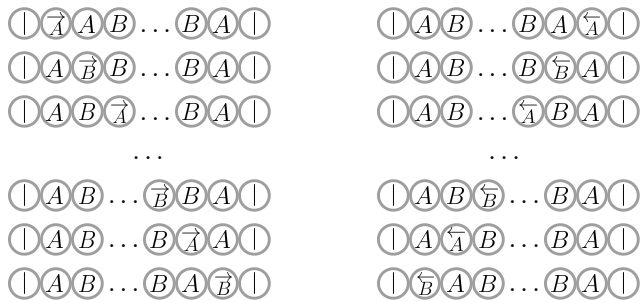

      \centering
        \begin{equation*}
            \begin{split}
            \symbar \ar \pa \pb \dots \pb \pa \symbar \qquad & \qquad \symbar \pa \pb \dots \pb \pa \al \symbar \\
            \symbar \pa \br \pb \dots \pb \pa \symbar \qquad & \qquad \symbar \pa \pb \dots \pb \bl \pa \symbar \\
            \symbar \pa \pb \ar \dots \pb \pa \symbar \qquad & \qquad \symbar \pa \pb \dots \al \pb \pa \symbar \\
            \dots \qquad \qquad \qquad & \qquad \qquad \qquad \dots \\
            \symbar \pa \pb \dots \br \pb \pa \symbar \qquad & \qquad \symbar \pa \pb \bl \dots \pb \pa \symbar \\
            \symbar \pa \pb \dots \pb \ar \pa \symbar \qquad & \qquad \symbar \pa \al \pb \dots \pb \pa \symbar \\
            \symbar \pa \pb \dots \pb \pa \br \symbar \qquad & \qquad \symbar \bl \pa \pb \dots \pb \pa \symbar \\
            \end{split}
        \end{equation*}
      \caption{\centering Behavior of Track 0 on canonical orientation. Adapted from Section 6.1 in \cite{GI10}.}
      \label{fig:T0_example}
\end{figure}

For example, in a $6$-chain, the initialization of Track 0 could look like $\symbar\ar\pa\pb\pa\symbar$ (canonical orientation) or $\symbar\pa\pb\pa\ar\symbar$ (reverse orientation). An example of an iteration of Track 0 in canonical orientation can be found in Figure \ref{fig:T0_example}. On the canonical orientation, arrow states always point to states with the same label as them (or the boundary). In the reverse orientation, they point to opposite labeled states (or the boundary).

Track 0 will also allow us to unify the two different end markers ($\leftend$, $\rightend$) used in \cite{CPW22} as $\symbar$. Additionally, as the size of the chain is even, we can locally distinguish the direction of the computation on the corners, as explained in Table \ref{tab:ends}.

\begin{table}[!htbp]
    \centering
   \begin{tabular}{|c|c|c|}
        \hline
        \rule{0pt}{15pt} & Points to $\symbar$ & Does not point to $\symbar$ \\
        \hline 
        \rule{0pt}{15pt} $\ar$ & reverse start & start \\
        \hline
        \rule{0pt}{15pt} $\br$ & return & reverse return \\
        \hline
        \rule{0pt}{15pt} $\al$ & reverse return & return \\
        \hline
        \rule{0pt}{15pt} $\bl$ & start & reverse start \\
        \hline
    \end{tabular}
    \caption{Depending on the arrow and the label on the state encountered next to the boundary, we can locally determine if we are in the starting or the returning end of the chain, as this depends on the direction of the computation. ``Start'' means the starting point of the $ab \rightarrow cd$ direction, which is the leftmost particle. ``Reverse return'' also refers to the leftmost particle, but seen as the turning back point from the $dc \leftarrow ba$ direction.}
    \label{tab:ends}
\end{table}

In summary, this new track unifies the boundary term, and is able to determine the direction of the computation in Track 1, via the arrowed states, only by nearest-neighbor interactions. Therefore, two totally reflection-symmetric orientations of computation should be permitted, so we need to add the corresponding rules: two new versions of the original ones for Tracks 1-6, for each of the orientations, that include Track 0. The behavior of Tracks 1-6 is not modified, but they are now attached to this new ``orientation track''. The full set of modified rules and illegal pairs can be found in Appendix \ref{appendix:tables}: for the canonical orientation, see Tables \ref{tab:can_rules}-\ref{tab:can_quantum_illegal}. For the reverse orientation, see Tables \ref{tab:rev_rules}-\ref{tab:rev_quantum_illegal}. Each rule and term has also two versions, one for each label, A and B, except the ones involving a single site or the boundary of the chain: as the number of particles on the chain, $L$, is even, when the arrow particle encounters the end of the chain, it will always be in a B state (which later will switch to A, and change its arrow orientation too).

Due to the nature of Track 0, we then have the following result about the evolution of the system as described in this section, which can be seen as the analogous of Lemma 39 in \cite{CPW22}, but in the reflection-symmetric setting. The importance of this result becomes apparent in Theorem \ref{th:qtm_hamiltonian}: each defined configuration uniquely belongs to one orientation, and if the two subsets of rules are closed and independent of each other, the evolution of the system will be fully restricted to one of the two subspaces.

\begin{lemma}\label{lemma:rule_behavior}
The two sets of rules (and illegal pairs) form two closed subspaces that do not overlap with each other: given a legal configuration of the chain, only one set of rules will apply to it and all its transitions. No canonically oriented pair has a transition to a reversed pair and vice versa. Additionally, each state has at most one possible transition.
\end{lemma}
\begin{proof}
By Lemma 39 in \cite{CPW22}, canonically oriented rules for Tracks 1-6 has at most one possible transition. All these rules are reversed, and added to the existent ones. This new set of rules could break this property, but Track 0 is added to make the orientation clear in every neighboring pair of states that have a possible transition. As a legal configuration must have, particularly, a legal configuration on Track 0, the orientation is then fixed. Therefore, this splits the new rules in two different sets, and each one is analogous to the canonically oriented rules. 
\end{proof}

\subsection{Behavior of the system}\label{subsec:system_behavior}

If a state does not follow the correct evolution, it picks up an energy penalty from the transition rule terms. With penalty terms, we detect some undesired configurations, and also give them an energy penalty. However, there are other undesirable configurations that are not detected by any of those, as we are only using nearest-neighbor terms. However, these configurations all evolve under the transition rule terms into configurations that do pick up an energy penalty. This idea was introduced in \cite{AGIK09}, and was called the \textit{Clairvoyance Lemma}. 

As the result is dependent on the specific construction, a version of the Clairvoyance Lemma was proven in \cite{CPW22} (in particular, their Lemma 43), which works for any Hamiltonian with a specific set properties (they call them \textit{standard form Hamiltonian}, see Definition \ref{def:standard_form_Hamiltonian} below). Our Hamiltonian is built in the same way as theirs, so we can apply Lemma 43 from \cite{CPW22} to check that our construction also evolves the remaining undesirable configurations to ones with an energy penalty.

\begin{definition}\label{def:standard_form_Hamiltonian}
We say that a Hamiltonian $H(L) = H_{trans}(L) + H_{pen}(L)$ acting on a Hilbert space $\mathcal{H} = (\mathbb{C}^C\otimes\mathbb{C}^Q)^{\otimes L} = (\mathbb{C}^C)^{\otimes L}\otimes(\mathbb{C}^Q)^{\otimes L} := \mathcal{H}_C\otimes\mathcal{H}_Q$ is of \textit{standard form} if $H_{trans,pen}(L) = \sum_{i=1}^{L-1} h_{trans,pen}^{(i,i+1)}$, and $h_{trans,pen}^{(i,i+1)}$ satisfy the following conditions:
\begin{enumerate}
    \item $h_{trans}^{(i,i+1)} \in \mathcal{B}\left((\mathbb{C}^C\otimes\mathbb{C}^Q)^{\otimes 2}\right)$ is a sum of transition rule terms, where all the transition rules act diagonally on $\mathbb{C}^C\otimes\mathbb{C}^C$ in the following sense. Given standard basis states $a,b,c,d\in\mathbb{C}^C$, exactly one of the following holds:
    \begin{itemize}
        \item There is no transition from $ab$ to $cd$ at all.
        \item $a,b,c,d\in\mathbb{C}^C$ and there exists a unitary $U_{abcd}$ acting on $\mathbb{C}^Q\otimes\mathbb{C}^Q$ together with an orthonormal basis $\{\ket{\psi_{abcd}^i}\}_i$ for $\mathbb{C}^Q\otimes\mathbb{C}^Q$, both depending only on $a,b,c,d$, such that the transition rules from $ab$ to $cd$ appearing in $h_{trans}$ are exactly $\ket{ab}\ket{\psi^i_{abcd}}\rightarrow \ket{cd}U_{abcd}\ket{\psi^i_{abcd}}$ for all $i$.
        \begin{equation}\label{eq:transition_local_interaction}
            h_{trans}^{(i,i+1)} = \sum_{ab\rightarrow cd} (\ket{ab}-\ket{cd})(\bra{ab}-\bra{cd}) \ox (2\1-U_{abcd}-U_{abcd}^\dagger)\;.
        \end{equation}
    \end{itemize}
  \item $h_{pen}^{(i,i+1)} \in \mathcal{B}\left((\mathbb{C}^C\otimes\mathbb{C}^Q)^{\otimes 2}\right)$ is a sum of penalty terms which act non-trivially only on $(\mathbb{C}^C)^{\otimes 2}$ and are diagonal in the standard basis.
\end{enumerate}
\end{definition}

$\mathbb{C}^C$ corresponds to Tracks 0-3 and $\mathbb{C}^Q$ to Tracks 4-6. Transitions from $ab$ to $cd$ will correspond exactly to the transitions shown for Tracks 0-3 in Tables \ref{tab:can_rules} and \ref{tab:rev_rules}. While adding new transitions involving Tracks 4-6, we will need to remove some of the Tracks 0-3 transitions (marked with an asterisk in the Tables), but they will be recovered as restrictions of the new rules to those tracks.

Furthermore, in order to apply this desired version of the Clairvoyance Lemma, one also needs to guarantee that the $U_{abcd}$ appearing in the computational Hamiltonian description are unitaries, and not only partial isometries. This is however guaranteed by restricting to a special class of QTMs: well-formed, normal-form and unidirectional, the same properties needed to construct the family $P_n$ of Section \ref{subsec:qpe_machine}. However, as stated in \cite{BV93} (Section 7), this is not restricting, as is general enough to admit universal QTMs.

As the same properties as the original construction still hold, we summarize the behavior of the encoded QTM in an analogous result to Theorem 32 of \cite{CPW22}, with the addition of reflection invariance and two different ground states with the same energy, which encode the very same computation, only differing on their orientation.

\begin{theorem}\label{th:qtm_hamiltonian}
Let $\mathbb{C}^d=\mathbb{C}^C\otimes\mathbb{C}^Q$ be the local Hilbert space of a $1$-dimensional chain of length $L=2m$ for some $m \in \mathbb{N}$, with special marker state $\ket{\symbar}$. For any well-formed, unidirectional Quantum Turing Machine $M = (\Sigma, Q, \delta)$ and any constant\footnote{$K$ is a constant needed for the QPE machine $P_n$ mentioned in \ref{subsec:qpe_machine}} $K > 0$, we can construct a two-body interaction $h_{q_0}\in\mathcal{B}(\mathbb{C}^d\otimes\mathbb{C}^d)$ such that on a chain of length $L \geq K + 3$, the Hamiltonian $H_{q_0}(L)=\sum_{i=1}^{L-1}h^{(i,i+1)} \in \mathcal{B}(\mathcal{H}(L))$ has the following properties:
    \begin{enumerate}
        \item The Hamiltonian is $1$-dimensional, translationally invariant, nearest-neighbor and reflection invariant as in Definition \ref{def:reflection_invariance}.
        \item $d$ depends only on the alphabet size and number of internal states $M$.
        \item $h_{q_0} \geq 0$ and the overall Hamiltonian $H(L)$ is frustration-free for all $L$.
        \item Denote $\mathcal{H}(L-2)=(\mathbb{C}^{d-2})^{\otimes L-2}$ and define $S_{br}=\linspan(\ket{\symbar}) \otimes \mathcal{H}(L-2) \otimes \linspan(\ket{\symbar}) \subset \mathcal{H}$. Then, the two ground states of $H_{q_0}(L)|_{S_{br}}$ are computational history states encoding the evolution of $M$ on input corresponding to the unary representation of the number $L-K-3$, running on a finite tape segment of length $L-3$.

        If $M$ is proper on input given by $L-K-3$ in unary representation, then:
        
        \item The computational history states always encode $\Omega(\zeta^L)$ time-steps, where $\zeta=\abs{\Sigma\times Q}$ \footnote{This choice of $\zeta$ guarantees that the QTM $M$ has enough time to halt (if it is going to halt) within this number of time-steps in the finite tape segment available.}. If $M$ halts in fewer than the number of encoded time steps, two $\ket{\psi_t}$ states have support on a state $\ket{\top}$ that encodes a halting state of the QTM. The remaining time steps of the evolution encoded in the history state leave $M$'s tape unaltered, and have zero overlap with $\ket{\top}$.
        \item If $M$ runs out of tape within a time $T$ less than the number of encoded time steps (i.e., in time-step $T+1$ it would move its head before the starting cell or beyond cell $L-3$), the computational history states only encode the evolution of $M$ up to time $T$. The remaining steps of the evolution encoded in the computational history states leave $M$'s tape unaltered.
    \end{enumerate}
\end{theorem}
\begin{proof}
The addition of Track 0 makes the $1$-dimensional construction reflection invariant: every transition rules has a reflected version, locally distinguishable. No energy bonus are added, only penalties, so $h_{q_0} \geq 0$.

By Lemma \ref{lemma:rule_behavior}, any well-formed state (i.e., one matching the regular expression defined by the illegal terms from \ref{tab:can_illegal}) evolves to another well-formed state under the only set of transition rules it belongs to: canonical or reverse. As Track 0 of a well-formed state uniquely determines the set of rules used, our construction restricted to that subspace is analogous to the one in Section 4 of \cite{CPW22}. Thus, we can use Lemma 40 in \cite{CPW22} to guarantee that evolving any clock state $\ket{\phi_t}$ using the transition rules will never reach an illegal configuration, but all other well-formed states that do not correspond to valid clock configurations (a correct evolution from an initial clock state) will evolve to illegal configurations.

With these two properties, the fact that we also have a standard form Hamiltonian, and the fact that unitarity of the $U_t$ are guaranteed, we can use the Clairvoyance Lemma 43 in \cite{CPW22} to see that there is only a unique ground state per set of rules (the computational history state of that orientation) with ground state energy 0. The rest follows from Theorem 32 in \cite{CPW22}.
\end{proof}

We denote the Hamiltonian constructed in Theorem \ref{th:qtm_hamiltonian} by $H_{q_0}$. The subscript $q$ is commonly used in the literature just to indicate that the Hamiltonian encodes a computational behavior: this does not mean that $q$ is any specific state. We use $H_{q_0}$ to further imply that this is still not the final step in the 1-dimensional construction. The reason is simple: $H_{q_0}$ has a ground state energy of zero, in both Halting and non Halting instances. In order to achieve a different behavior between instances, we will introduce an energy penalty in the Halting case. In the following sections, the computational Hamiltonian considered is $H_q$, with local interaction
\begin{equation}\label{eq:computational_local_interaction}
    h_q^{(i,i+1)} = h_{q_0}^{(i,i+1)} + \ketbra{\top}{\top}_i \otimes \mathbbm{1}_{i+1} + \mathbbm{1}_i \otimes \ketbra{\top}{\top}_{i+1}.
\end{equation}
Let $\ket{\psi} = \frac{1}{\sqrt{T}}\sum_{t=1}^T\ket{\phi_t}\ket{\psi_t}$ be the normalized computational history state for the QTM, where, by Theorem \ref{th:qtm_hamiltonian}, is a 0-energy state of $H_{q_0}(L)$, and encodes $T=\Omega(\abs{\Sigma\times Q}^L)$ steps. At most two states $\ket{\psi_t}$ can have support on $\ket{\top}$, so due to the exponential dependency on $L$ of the number of time steps encoded, the energy of $H_q(L)$ is positive but bounded by a constant exponentially small on $L$, as 
\begin{equation}\label{eq:computational_Hamiltonian_energy_bound}
\begin{split}
\bra{\psi}H_q(L)\ket{\psi} &= \bra{\psi}\left(\sum_i h_{q_0}^{(i,i+1)}(L) + \ketbra{\top}{\top}_i \otimes \mathbbm{1}_{i+1} + \mathbbm{1}_i \otimes \ketbra{\top}{\top}_{i+1}\right) \ket{\psi}\\
&=\sum_{t=1}^T \frac{1}{T}
         \bra{\psi_t}\left(\ketbra{\top}{\top}_i \otimes \mathbbm{1}_{i+1} + \mathbbm{1}_i \otimes \ketbra{\top}{\top}_{i+1}\right) \ket{\psi_t}
      \leq O\left(\frac{1}{\abs{\Sigma\times Q}^L}\right).
\end{split}
\end{equation}
\section{Encoding a Tiling problem in a 2D rotationally symmetric Hamiltonian}\label{sec:tiling}
We have seen in section \ref{sec:onedencoding} how one can construct a Hamiltonian $H_q$ over a chain of size $L$, in order to make its ground state energy dependent of the outcome of the Halting problem:
\begin{itemize}
    \item If the encoded problem does not halt, ground state energy is $0$.
    \item If the encoded problem halts, ground state energy is $c^{-L}$, as seen by Equation \ref{eq:computational_Hamiltonian_energy_bound}.
\end{itemize}

This behavior has a very obvious limitation: in the thermodynamic limit, the energy of both instances converge to zero. To be able to observe different behaviors in this limit, we use an additional construction, based on 2-dimensional tilings. This will allow us to amplify the energy of the halting instances, so it stops vanishing in the limit.

This idea is also used in \cite{CPW22} (Section 5), where the authors use the rigid geometrical structure of Robinson's quasi-periodic tiling (\cite{robinson1971undecidability}). This tiling proves useful for the following reason: from a finite set of tiles, the arising structure is a set of interlocking squares of increasing size. By using all of these sides as chains for our 1-dimensional Hamiltonian $H_q$, the authors in \cite{CPW22} construct a final 2-dimensional Hamiltonian that has a clear distinction between Halting and non-Halting instances in the thermodynamic limit: positive in the first case, negative in the second.

We would like to exploit the same idea. However, in order to do so while ensuring rotation invariance, we need to modify the construction. To see why, we first turn to a very simple example (Section \ref{subsec:simple_example}), while reviewing the basics of encoding tiling problems in Hamiltonians. Later, in Section \ref{subsec:the_tiles} we explain the structure of the Robinson tiling, and how the quasi-periodic structure appears. Finally, in Section \ref{subsec:hamiltonian_description}, we describe how to transform our tiling problem into a Hamiltonian ground state problem.

\subsection{Simple example}\label{subsec:simple_example}

The tiling problem asks the following: given a finite set of tiles and matching rules between them, can we arrange them properly to cover the plane, or a portion of it? Imagine we have a set composed of three tiles, labeled as $\ket{0}$, $\ket{1}$ and $\ket{2}$ as in Figure \ref{fig:simple_usual}. The matching rule is also simple: matching edges must have the same color.

Its extension to a Hamiltonian description is also very natural: a valid tiling is a zero-energy ground state. This is achieved by just adding an energy penalty to invalid matches. In this example, those would be tile $\ket{1}$ next to $\ket{0}$ or $\ket{2}$ in rows, but $\ket{2}$ next to $\ket{0}$ or $\ket{1}$ in columns. Consequently, row and column terms are not the same, as seen in Figure \ref{fig:simple_usual}, meaning that the tiling Hamiltonian is not rotationally invariant. It is also worth mentioning that this is happening even when the tile set itself is closed under rotations, as tile $\ket{2}$ is the rotation of tile $\ket{1}$ (and vice versa): a rotationally closed tile set does not immediately yield a rotationally invariant Hamiltonian.

\begin{figure}[htbp]
    \centering
    \begin{minipage}{0.3\textwidth}
        \centering
        \captionsetup[subfloat]{labelformat=empty}
\centering
\begin{subfloat}[][$\ket{0}$] {
\begin{tikzpicture}
    \draw[thick] (0, 0) rectangle (1, 1);
    \draw[thick] (0, 0) -- (1, 1);
    \draw[thick] (0, 1) -- (1, 0);
\end{tikzpicture}}
\end{subfloat}\hspace{1em}%
\begin{subfloat}[][$\ket{1}$]{
\begin{tikzpicture}
    \fill[SkyBlue] (0, 0) -- (0, 1) -- (0.5, 0.5);
    \fill[SkyBlue] (1, 0) -- (1, 1) -- (0.5, 0.5);
    \draw[thick] (0, 0) rectangle (1, 1);
    \draw[thick] (0, 0) -- (1, 1);
    \draw[thick] (0, 1) -- (1, 0);
\end{tikzpicture}}
\end{subfloat}\hspace{1em}%
\begin{subfloat}[][$\ket{2}$]{
\begin{tikzpicture}
    \fill[SkyBlue] (0,1) -- (1, 1) -- (0.5, 0.5);
    \fill[SkyBlue] (0,0) -- (1, 0) -- (0.5, 0.5);
    \draw[thick] (0, 0) rectangle (1, 1);
    \draw[thick] (0, 0) -- (1, 1);
    \draw[thick] (0, 1) -- (1, 0);
\end{tikzpicture}}
\end{subfloat}
        \caption*{(a) The tiles}
    \end{minipage}
    \begin{minipage}{0.3\textwidth}
        \centering
        \begin{tikzpicture}
    \draw[thick] (0, 0) rectangle (1, 1);
    \draw[thick] (0, 0) -- (1, 1);
    \draw[thick] (0, 1) -- (1, 0);
    \fill[orange] (0.5, 0.5) circle (0.07cm);
    \draw[thick] (0.5, 0.5) circle (0.07cm);

    \draw[thick] (0, 1) rectangle (1, 2);
    \draw[thick] (0, 1) -- (1, 2);
    \draw[thick] (0, 2) -- (1, 1);
    \fill[orange] (0.5, 1.5) circle (0.07cm);
    \draw[thick] (0.5, 1.5) circle (0.07cm);

    \fill[SkyBlue] (1, 1) -- (1.5, 0.5) -- (2, 1);
    \fill[SkyBlue] (1, 0) -- (1.5, 0.5) -- (2, 0);
    \draw[thick] (1, 0) rectangle (2, 1);
    \draw[thick] (1, 0) -- (2, 1);
    \draw[thick] (1, 1) -- (2, 0);
    \fill[orange] (1.5, 0.5) circle (0.07cm);
    \draw[thick] (1.5, 0.5) circle (0.07cm);

    \fill[SkyBlue] (1, 2) -- (1.5, 1.5) -- (2, 2);
    \fill[SkyBlue] (1, 1) -- (1.5, 1.5) -- (2, 1);
    \draw[thick] (1, 1) rectangle (2, 2);
    \draw[thick] (1, 1) -- (2, 2);
    \draw[thick] (1, 2) -- (2, 1);
    \fill[orange] (1.5, 1.5) circle (0.07cm);
    \draw[thick] (1.5, 1.5) circle (0.07cm);
\end{tikzpicture}
        \caption*{(b) Example tiling}
    \end{minipage}
    \begin{minipage}{0.3\textwidth}
        \centering
        \begin{align*}
        h_{row} = & \ket{01}\bra{01} + \ket{10}\bra{10} \\ &+ \ket{12}\bra{12} + \ket{21}\bra{21}\text{,} \\
        h_{col} = &\ket{02}\bra{02} + \ket{20}\bra{20} \\ &+ \ket{12}\bra{12} + \ket{21}\bra{21}
        \end{align*}
        \caption*{(c) The interactions: matching edges must be of the same color.}
    \end{minipage}
    \caption{If the tiles are described as a unit, sites are at the center of the tile, and then $h_{row} \not = h_{col}$.}
    \label{fig:simple_usual}
\end{figure}

However, by describing the problem in a different fashion, we can solve this problem. As seen in Figure \ref{fig:simple_sides}, we now label colors, and say that a tile is a 4-tuple of colors. To describe a tile, we use the convention of doing it clockwise, starting from the top side. Now, we can think of the tiling problem as starting by assigning a color to each edge, and then checking if the arising $4$-interaction (usually called a \textit{plaquette}) forms a valid tile: that is, one belonging to the tile set. If it is not, it then adds an energy penalty.

\begin{figure}[htbp]
    \centering
    \begin{minipage}{0.3\textwidth}
        \centering
        \captionsetup[subfloat]{labelformat=empty}
\centering
\begin{subfloat}[][$\ket{0000}$] {
\begin{tikzpicture}
    \draw[thick] (0, 0) rectangle (1, 1);
    \draw[thick] (0, 0) -- (1, 1);
    \draw[thick] (0, 1) -- (1, 0);
\end{tikzpicture}}
\end{subfloat}\hspace{1em}%
\begin{subfloat}[][$\ket{0101}$]{
\begin{tikzpicture}
    \fill[SkyBlue] (0, 0) -- (0, 1) -- (0.5, 0.5);
    \fill[SkyBlue] (1, 0) -- (1, 1) -- (0.5, 0.5);
    \draw[thick] (0, 0) rectangle (1, 1);
    \draw[thick] (0, 0) -- (1, 1);
    \draw[thick] (0, 1) -- (1, 0);
\end{tikzpicture}}
\end{subfloat}\hspace{1em}%
\begin{subfloat}[][$\ket{1010}$]{
\begin{tikzpicture}
    \fill[SkyBlue] (0,1) -- (1, 1) -- (0.5, 0.5);
    \fill[SkyBlue] (0,0) -- (1, 0) -- (0.5, 0.5);
    \draw[thick] (0, 0) rectangle (1, 1);
    \draw[thick] (0, 0) -- (1, 1);
    \draw[thick] (0, 1) -- (1, 0);
\end{tikzpicture}}
\end{subfloat}
        \caption*{(a) A tile is formed by 4 colors}
    \end{minipage}
    \begin{minipage}{0.3\textwidth}
        \centering
        \begin{tikzpicture}
    \draw[thick] (0, 0) rectangle (1, 1);
    \draw[thick] (0, 0) -- (1, 1);
    \draw[thick] (0, 1) -- (1, 0);
    
    \draw[thick] (0, 1) rectangle (1, 2);
    \draw[thick] (0, 1) -- (1, 2);
    \draw[thick] (0, 2) -- (1, 1);

    \fill[SkyBlue] (1, 1) -- (1.5, 0.5) -- (2, 1);
    \fill[SkyBlue] (1, 0) -- (1.5, 0.5) -- (2, 0);
    \draw[thick] (1, 0) rectangle (2, 1);
    \draw[thick] (1, 0) -- (2, 1);
    \draw[thick] (1, 1) -- (2, 0);

    \fill[SkyBlue] (1, 2) -- (1.5, 1.5) -- (2, 2);
    \fill[SkyBlue] (1, 1) -- (1.5, 1.5) -- (2, 1);
    \draw[thick] (1, 1) rectangle (2, 2);
    \draw[thick] (1, 1) -- (2, 2);
    \draw[thick] (1, 2) -- (2, 1);
    
    \fill[orange] (0.5, 0) circle (0.07cm);
    \fill[orange] (1.5, 0) circle (0.07cm);
    
    \fill[orange] (0, 0.5) circle (0.07cm);
    \fill[orange] (1, 0.5) circle (0.07cm);
    \fill[orange] (2, 0.5) circle (0.07cm);

    \fill[orange] (0.5, 1) circle (0.07cm);
    \fill[orange] (1.5, 1) circle (0.07cm);

    \fill[orange] (0, 1.5) circle (0.07cm);
    \fill[orange] (1, 1.5) circle (0.07cm);
    \fill[orange] (2, 1.5) circle (0.07cm);

    \fill[orange] (0.5, 2) circle (0.07cm);
    \fill[orange] (1.5, 2) circle (0.07cm);

    \draw[thick] (0.5, 0) circle (0.07cm);
    \draw[thick] (1.5, 0) circle (0.07cm);
    
    \draw[thick] (0, 0.5) circle (0.07cm);
    \draw[thick] (1, 0.5) circle (0.07cm);
    \draw[thick] (2, 0.5) circle (0.07cm);

    \draw[thick] (0.5, 1) circle (0.07cm);
    \draw[thick] (1.5, 1) circle (0.07cm);

    \draw[thick] (0, 1.5) circle (0.07cm);
    \draw[thick] (1, 1.5) circle (0.07cm);
    \draw[thick] (2, 1.5) circle (0.07cm);

    \draw[thick] (0.5, 2) circle (0.07cm);
    \draw[thick] (1.5, 2) circle (0.07cm);
\end{tikzpicture}
        \caption*{(b) Example tiling}
    \end{minipage}
    \begin{minipage}{0.3\textwidth}
        \centering
        \[h = \identity - (\ket{0000}\bra{0000} \]\[+ \ket{0101}\bra{0101} + \ket{1010}\bra{1010})\]
        \caption*{(c) The interaction: arising tiles must belong to our tile set.}
    \end{minipage}
    \caption{If we describe white as $\ket{0}$ and blue as $\ket{1}$, a tile is a $4$-interaction of sides (with a convention, for example, of starting on the top and going clockwise). Now sites are at the edges, but $h = h_{row} = h_{col}$.}
    \label{fig:simple_sides}
\end{figure}

With this description, the interaction represents the problem of existence of the tile. Therefore, for any rotationally invariant set, one can use this description to obtain a rotationally invariant tiling Hamiltonian. However, this comes with a downside, which is having to use 4-interactions instead of 2-interactions.

\subsection{The tiles}\label{subsec:the_tiles}
In this section, we describe the structure of Robinson's quasi-periodic tiling, first presented by Robinson in 1971 (\cite{robinson1971undecidability}). In particular, it is a tiling that constructs a family of red squares of sides $4^n$ for all $n \in \mathbb{N}$. This material can be found in \cite{robinson1971undecidability} and \cite{CPW22}, but, for a better understanding of the final picture, we include it to recall how the construction works, and how the described squares arise.

\begin{figure}[htbp]
  \centering
  \includegraphics[width=0.5\textwidth]{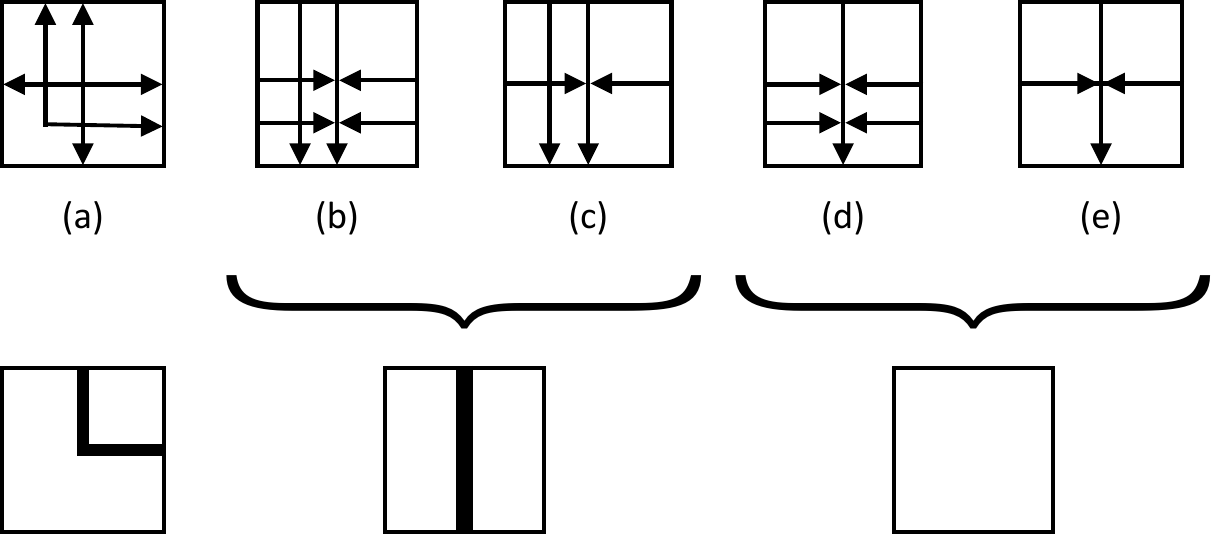}
  \caption{The five basic tiles of Robinson's tiling and their schematic representation. Image taken from Section 5 in \cite{CPW22}}
  \label{fig:robinson_basic_tiles}
\end{figure}

The set of tiles used is based on the five basic Robinson tiles shown in Figure \ref{fig:robinson_basic_tiles}, as well as all of their rotations and reflections. In a valid tiling, arrow heads must meet arrow tails. Tile (a) is called a \textit{cross}, and tiles (b)-(e), an \textit{arm}. In particular, tiles (b) and (c) will represent segments, as we can see in the schematics, whereas (d) and (e) represent blank tiles. Arrows that do not start or end in the mid-point of a square's side are called \textit{side arrows}, in contrast to the \textit{central arrows}. An arm is said to point in the direction of its unique complete central arrow. Lastly, the particular cross depicted in (a) is said to \textit{face} up/right.

We also introduce two possible colors for the side arrows, red or green, but fulfilling the following restrictions:

\begin{itemize}
    \item Colors must match between adjacent arrows of different tiles.
    \item In crosses, the same color must be used in both directions.
    \item In (b) tiles, one color must be used horizontally and the other color vertically.
    \item Green crosses must appear in alternate positions in alternate rows. That is, if there are green crosses at row $i$ in positions $j$, $j+2$, $j+4$... (alternate positions) there must be green crosses in positions $j$, $j+2$, $j+4$... in row $i+2$ (alternate rows).
\end{itemize}

\begin{figure}[htbp]
      \centering
      \includegraphics[width=0.5\textwidth]{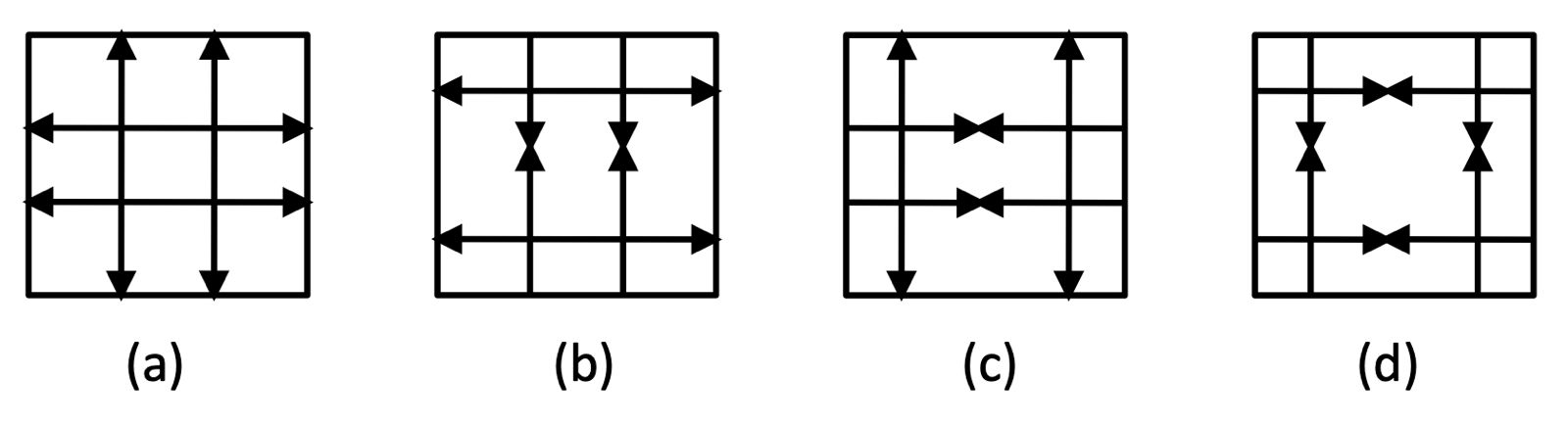}
      \caption{Parity tiles. (a) is associated with green crosses, (b) with any tile presenting a vertical green arrow and (c) with any tile presenting a horizontal green arrow. (d) is associated with all tiles. Image taken from Section 5 of \cite{CPW22}.}
      \label{fig:parity_tiles}
\end{figure}

The last restriction can be achieved by adding extra marking to the tiles, called the parity marking, shown in Figure \ref{fig:parity_tiles}. This four basic tiles already form a closed set under rotation and reflection, and live on a separate ``layer'': parity marking arrows must match only those from parity markings, and arrows from the basic tiles must match only those from the basic tiles.

\begin{figure}[hbtp]
\centering
  \begin{tabular}{cc}
    \includegraphics[width=0.4\columnwidth]{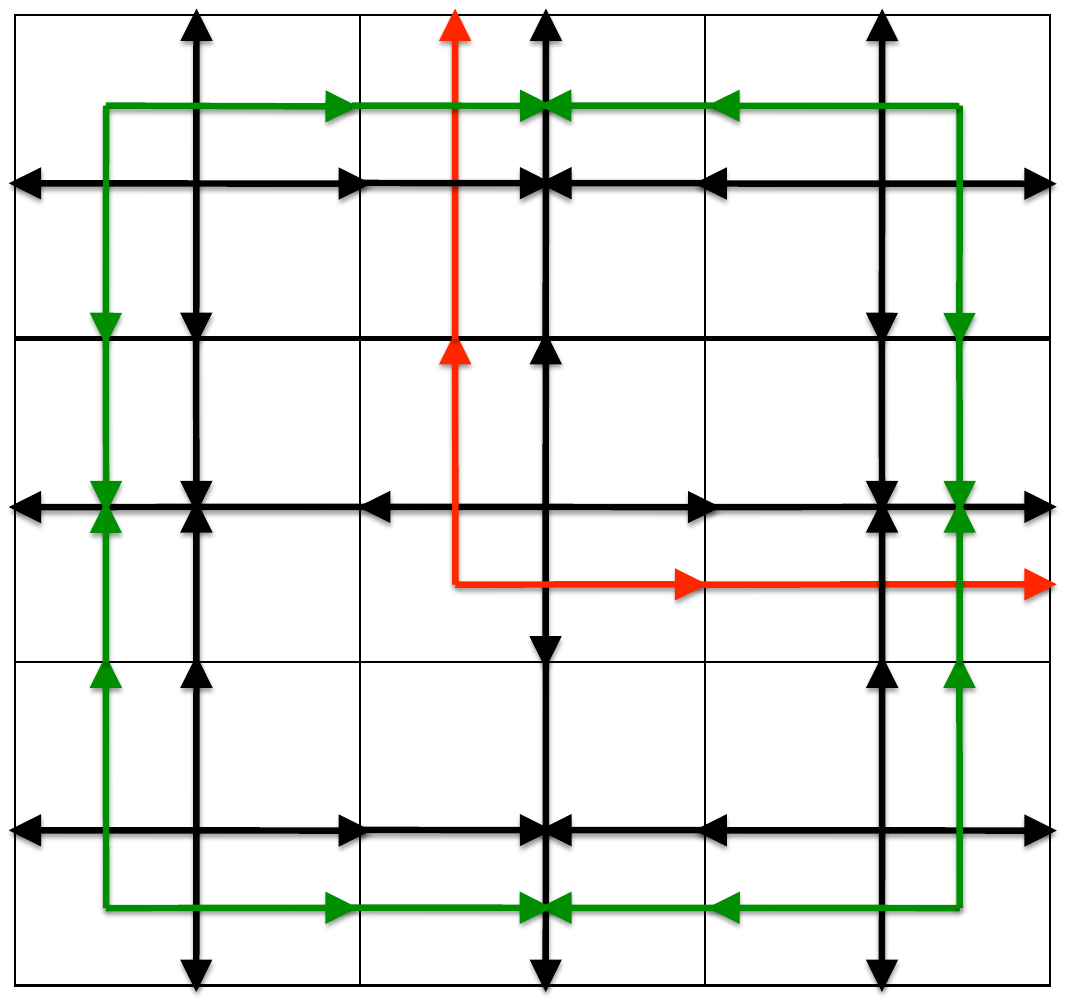} & \includegraphics[width=0.4\columnwidth]{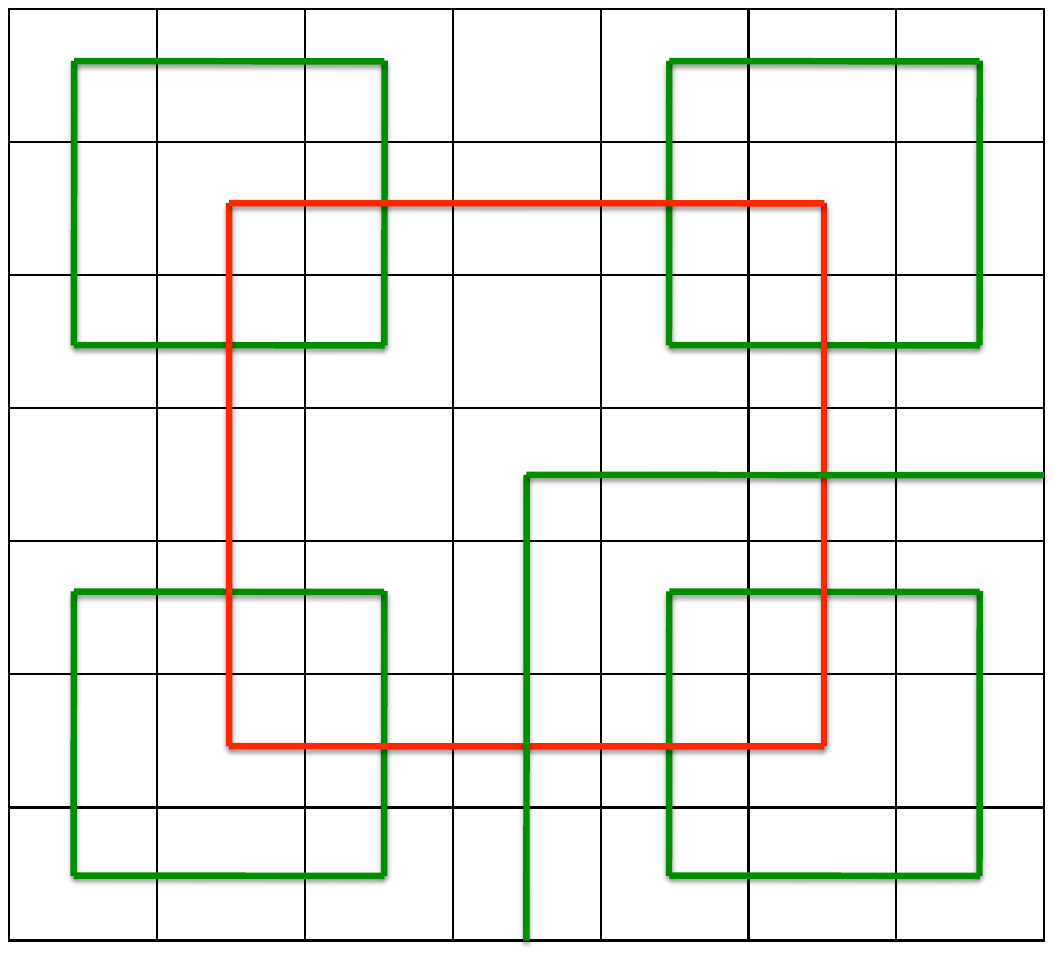}
  \end{tabular}
  \caption{On the left, a $3\times 3$ square of the Robinson tiling, with the choice of a right/up facing cross in the middle. On the right, a $7\times 7$ square of the Robinson tiling, with the choice of a right/down facing cross in the middle. To avoid confusion, only the coloured lines are drawn in this second figure.}
  \label{fig:small_tiling_examples}
\end{figure}

\begin{figure}[htbp]
      \centering
      \includegraphics[width=0.4\textwidth]{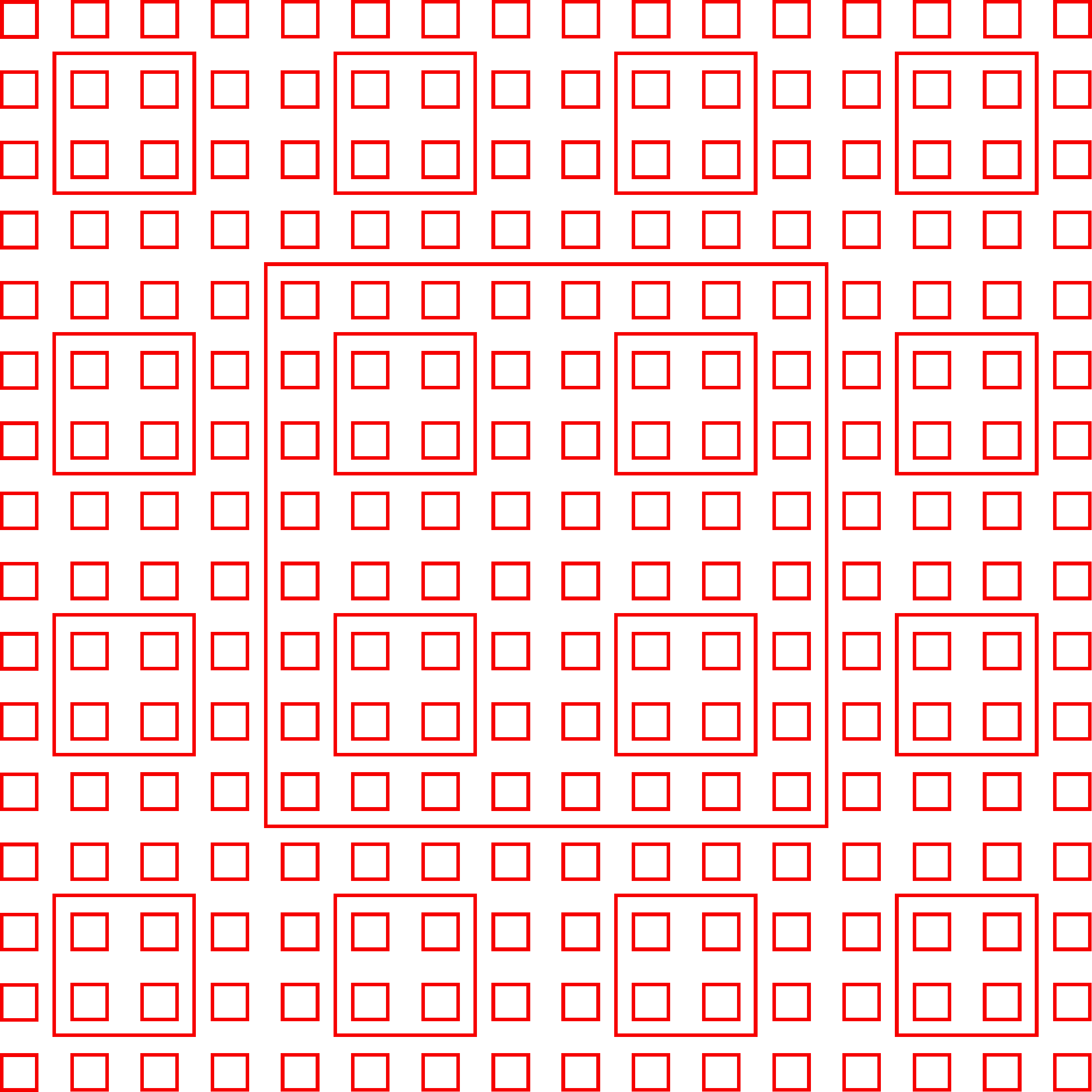}
      \caption{A possible Robinson tiling of the plane. Image taken from Section 5 of \cite{CPW22}.}
      \label{fig:big_tiling_example}
\end{figure}

Fusing the five basic colored tiles with this parity layer gives a set of tiles that covers the plane with a structure of interlocking squares of increasing size. The fact that green crosses must appear in alternate rows in alternate positions means that any given cross completely determines the structure of the $3\times 3$ square constructed in the direction it faces. By the form of the tiling, the central tile must be a red cross, with the only freedom being choosing its facing. Once fixed, the $7\times 7$ red square it forms is also determined, with a green cross in the middle, having again a choice for the direction it faces. An example of this can be seen in Figure \ref{fig:small_tiling_examples}, and continuing this procedure gives a tiling similar to the one in Figure \ref{fig:big_tiling_example}: a quasi-periodic structure of red squares with sides of size $4^n$ for all $n\in \mathbb{N}$.

\begin{figure}[hbtp]
\centering
    \includegraphics[width=0.4\columnwidth]{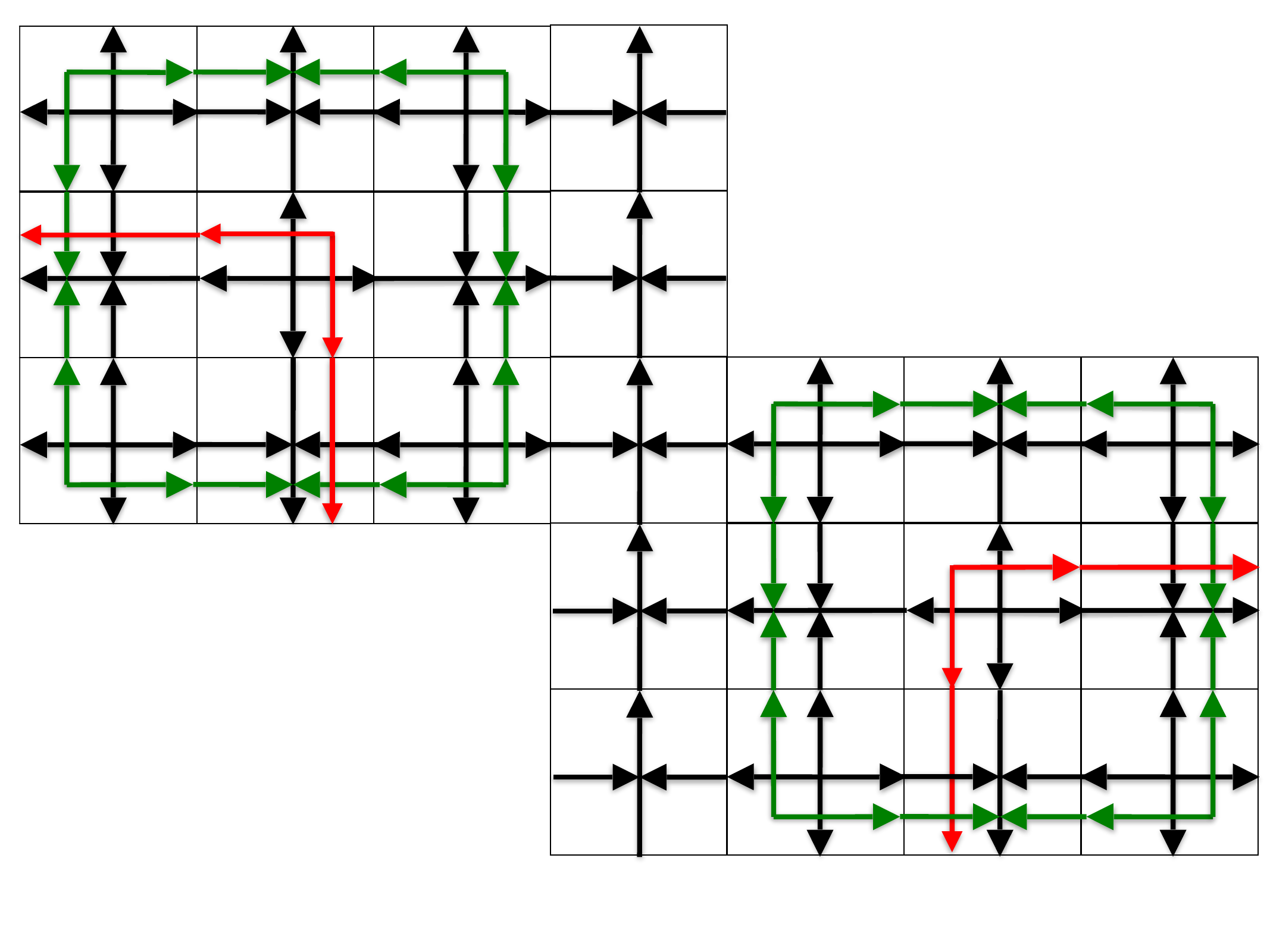} \includegraphics[width=0.4\columnwidth]{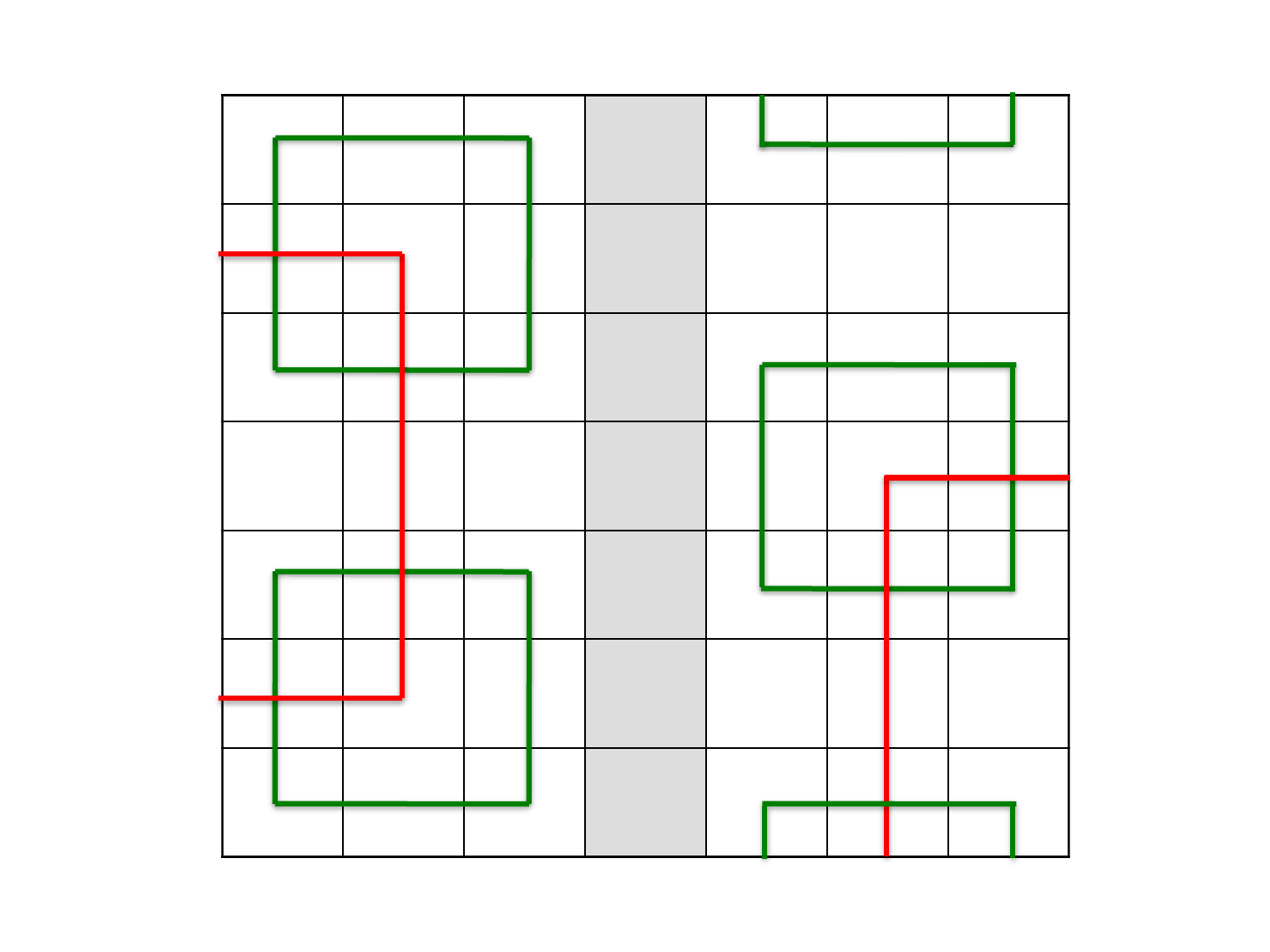}
  \caption{An example of a fault line, in which the patterns are shifted (schematic on the right). Image taken from Section 5 of \cite{CPW22}.}
  \label{fig:fault_lines}
\end{figure}

\begin{figure}[htbp]
  \centering
  \includegraphics[width=0.7\textwidth]{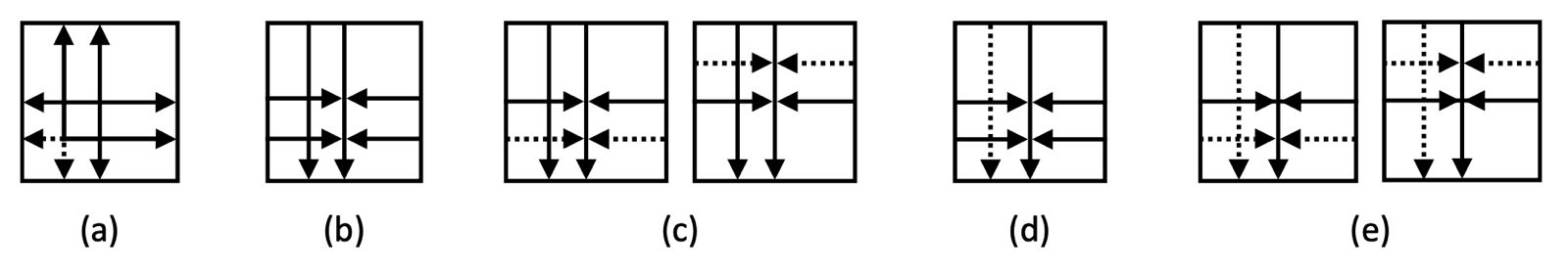}
  \caption{Basic modified Robinson's tiles. Image taken from Section 5 of \cite{CPW22}.}
  \label{fig:modified_robinson_tiles}
\end{figure}

However, with the original Robinson tiles, the characteristic quasi-periodic pattern of squares does not necessarily extend throughout the entire plane. As we are sequentially choosing the orientation of the central crosses, this can cause a fault line between half-planes or quarter-planes (see Figure \ref{fig:fault_lines}). In Section 5 of \cite{CPW22}, the original tiles were modified in order to prevent this, achieving an enforced tiling of the whole plane. These modified tiles (Figure \ref{fig:modified_robinson_tiles}) still present the original markings, in addition to some dashed side arrows, the ones that force adjacent squares to be aligned, and thus, avoid the appearance of fault lines.

The tile set used is then based on this new dashed tiles:

\begin{itemize}
    \item We extend them to the full set by rotation and reflection.
    \item We add the red and green coloring to the solid side arrows.
    \item We add the additional parity layer.
\end{itemize}

\subsection{Hamiltonian description}\label{subsec:hamiltonian_description}
Working with this tile set, the undecidability result from \cite{CPW22} (Section 6) is achieved by ``overlapping'' the 1-dimensional computational Hamiltonian on the top edge of every red square. We will see what this ``overlapping'' means in Section \ref{sec:fusion}, but for the time being, it is sufficient to imagine that they mark the top segment in a particular way. However, the tiling Hamiltonian the authors use to describe the Robinson tiling is not itself invariant under rotations: the basis of the local Hilbert space are the tiles, and therefore, the matching rules are spatial dependent, as we have seen in Section \ref{subsec:simple_example} (Figure \ref{fig:simple_usual}).

In order to solve this problem, we turn to the plaquette interaction setting (as shown in Figure \ref{fig:simple_sides}) and use them for describing the Hamiltonian in this alternative way, where the basis states are the correct matches between sides of the tiles instead of the tiles themselves. However, in the Robinson tiling, two sides can legally be put into contact if their adjacency rules are compatible according to the description in \ref{subsec:the_tiles}, which is different from the simpler rule of matching colors and markings. To understand how to adapt it, we start by studying the tiles' edges, and notice that if facing a tile edge from outside, we can only distinguish several possible cases, depending on:

\begin{itemize}
    \item Arrow orientation. Each edge has one central arrow and one side arrow. Both have the same orientation: arrow tails or arrow heads (2 choices).
    \item Location of the side arrow. If facing the edge from outside, we can see the side arrow either right of the central one or left of it (2 choices).
    \item Style of the side arrow. Side arrows can be solid or dashed (2 choices).
    \item Color, if side arrow is solid. As defined, it could be red or green (2 choices per solid arrow configuration).
    \item Parity layer. We can find central or lateral pairs of arrows with the same heads or tails configuration (4 choices).
\end{itemize}

So we have $16$ possible configurations for dashed arrows (heads/tails - right/left - 4 choices of parity) and $32$ for the solid ones (heads/tails - right/left - red/green - 4 choices of parity). So, a tile edge could in principle have one of this $48$ configurations. And for the tiling to work, each edge only fits with exactly another one. However, as not all parities are associated with all tiles, only $k<48$ valid edges occur. We denote by $\mathcal{S}$ this set of possible edges. However, only $12$ elements from $\mathcal{S}$ have their valid match in $\mathcal{S}$. We give labels (numbers from 1 to 12) to each of these edges. Their description and matching can be found in Table \ref{tab:edge_matching}.

\begin{table}[!htbp]
    \centering
    \begin{tabular}{|c|c|}
    \hline & \\
         1. Dashed right head, lateral tails & 12. Dashed left tail, lateral heads \\ \hline & \\
         2. Dashed left head, lateral tails & 11. Dashed right tail, lateral heads \\ \hline & \\
         3. Solid left green head, central heads & 10. Solid right green tail, central tails \\ \hline & \\
         4. Solid right green head, central heads & 9. Solid left green tail, central tails \\ \hline & \\
         5. Solid left red head, lateral tails & 8. Solid right red tail, lateral heads \\ \hline & \\
         6. Solid right red head, lateral tails & 7. Solid left red tail, lateral heads \\ \hline
    \end{tabular}
    \caption{Edge matching of the modified Robinson tiles. Each edge is comprised of a standard layer configuration (e.g., dashed right read) and a parity layer configuration (e.g., lateral tails). Each row is a match and we have given each edge configuration a distinctive number tag. The function $match$ can then be defined as $match(s_1,s_2) = 1 \iff s_1+s_2=13$. All other existent edges in $\mathcal{S}$ can be assigned an arbitrary label outside $[1,12]$, as no matching is available for them.}
    \label{tab:edge_matching}
\end{table}

Therefore, following a a clockwise orientation from topmost edge to leftmost edge, we can uniquely describe a tile by an ordered 4-tuple $(s_1, s_2, s_3, s_4)$, where all $s_i \in \mathcal{S}$. On the other hand, each edge is associated with another by a unique pairing. However, we are not matching edges of the same type ($1$ with $1$, $2$ with $2$, etc.), but with a ``complementary'' matching instead: $1$ to $12$, $2$ to $11$...

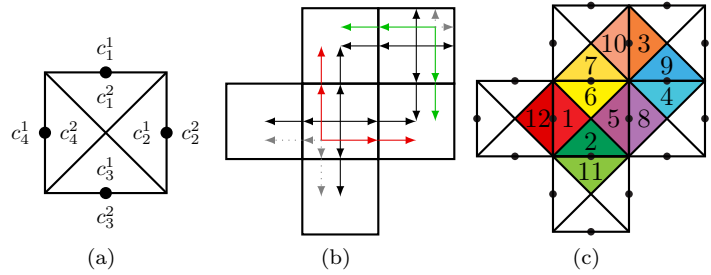
\begin{figure}[htbp]
\centering
\begin{subfloat}[] {
\begin{tikzpicture}[scale=0.8, every node/.style={scale=0.8}]
    \draw[thick] (0,0) rectangle (2,2);
    \fill[black] (1,2) circle (0.1);
    \fill[black] (2,1) circle (0.1);
    \fill[black] (1,0) circle (0.1);
    \fill[black] (0,1) circle (0.1);
    \draw[thick] (0, 0) -- (2, 2);
    \draw[thick] (0, 2) -- (2, 0);
    
    \path
        (1,  2.4) node {$s_1^1$}
        (1,  1.6) node {$s_1^2$}
        (1,  0.4) node {$s_3^1$}
        (1,  -0.4) node {$s_3^2$}
        (-0.4,  1) node {$s_4^1$}
        (0.4,  1) node {$s_4^2$}
        (1.6,  1) node {$s_2^1$}
        (2.4,  1) node {$s_2^2$}
    ;
\end{tikzpicture}}
\end{subfloat}
\begin{subfloat}[]{
\begin{tikzpicture}[scale=0.5, every node/.style={scale=0.6}]
    \draw[thick] (0,0) rectangle (2,2);
    \draw[thick] (2,0) rectangle (4,2);
    \draw[thick] (-2,0) rectangle (0,2);
    \draw[thick] (0,2) rectangle (2,4);
    \draw[thick] (2,2) rectangle (4,4);
    \draw[thick] (0,-2) rectangle (2,0);
    
    \draw[latex-latex] (1, 0) -- (1, 2);
    \draw[latex-latex] (0, 1) -- (2, 1);
    \draw[darkred, -latex] (0.5, 0.5) -- (0.5, 2);
    \draw[darkred, -latex] (0.5, 0.5) -- (2, 0.5);
    \draw[gray, dotted, -latex] (0.5, 0.5) -- (0, 0.5);
    \draw[gray, dotted, -latex] (0.5, 0.5) -- (0.5, 0);
    
    \draw[gray, dotted, -latex] (0.5, 0) -- (0.5, -1);
    \draw[-latex] (1, 0) -- (1, -1);
    
    \draw[gray, dotted, -latex] (0, 0.5) -- (-1, 0.5);
    \draw[-latex] (0, 1) -- (-1, 1);
    
    \draw[latex-latex] (3, 2) -- (3, 4);
    \draw[latex-latex] (2, 3) -- (4, 3);
    \draw[darkgreen, -latex] (3.5, 3.5) -- (3.5, 2);
    \draw[darkgreen, -latex] (3.5, 3.5) -- (2, 3.5);
    \draw[gray, dotted, -latex] (3.5, 3.5) -- (4, 3.5);
    \draw[gray, dotted, -latex] (3.5, 3.5) -- (3.5, 4);
    
    \draw[darkgreen, -latex] (3.5, 2) -- (3.5, 1);
    \draw[-latex] (3, 2) -- (3, 1);
    \draw[darkred, -latex] (2, 0.5) -- (3, 0.5);
    \draw[-latex] (2, 1) -- (3, 1);
    
    \draw[darkgreen, -latex] (2, 3.5) -- (1, 3.5);
    \draw[-latex] (2, 3) -- (1, 3);
    \draw[darkred, -latex] (0.5, 2) -- (0.5, 3);
    \draw[-latex] (1, 2) -- (1, 3);

\end{tikzpicture}}
\end{subfloat}
\begin{subfloat}[] {
\begin{tikzpicture}

\definecolor{darkred}{RGB}{224,0,0}

    \fill[darkred] (1, 1) -- (0.5, 1.5) -- (1, 2);
    \fill[Red] (1, 1) -- (1.5, 1.5) -- (1, 2);
    \draw[thick] (1, 1) -- (1, 2);
    \fill[ForestGreen] (1, 1) -- (1.5, 1.5) -- (2, 1);
    \fill[LimeGreen] (1, 1) -- (1.5, 0.5) -- (2, 1);
    \draw[thick] (1, 1) -- (2, 1);
    \fill[RedViolet, nearly opaque] (1.5, 1.5) -- (2, 2) -- (2, 1);
    \fill[Purple, nearly opaque] (2, 1) -- (2.5, 1.5) -- (2, 2);
    \draw[thick] (2, 1) -- (2, 2);
    \fill[CornflowerBlue] (2, 2) -- (2.5, 2.5) -- (3, 2);
    \fill[SkyBlue] (2, 2) -- (2.5, 1.5) -- (3, 2);
    \draw[thick] (2, 2) -- (3, 2);
    \fill[Orange] (2, 2) -- (2.5, 2.5) -- (2, 3);
    \fill[Melon] (2,2) -- (1.5, 2.5) -- (2, 3);
    \draw[thick] (2, 2) -- (2, 3);
    \fill[Yellow] (1, 2) -- (1.5, 1.5) -- (2, 2);
    \fill[Goldenrod] (1, 2) -- (1.5, 2.5) -- (2, 2);
    \draw[thick] (1, 2) -- (2, 2);

\path
    (1.5,  1.2) node {2}
    (1.2,  1.5) node {1}
    (1.8,  1.5) node {5}
    (1.5,  1.8) node {6}
    (0.8,  1.5) node {12}
    (1.5,  0.8) node {11}
    (2.2,  1.5) node {8}
    (2.5,  1.8) node {9}
    (2.5,  2.2) node {4}
    (2.2,  2.5) node {3}
    (1.8,  2.5) node {10}
    (1.5,  2.2) node {7}
;

    \fill[Black] (0, 1.5) circle (0.05cm);
    \fill[Black] (0.5, 1) circle (0.05cm);
    \fill[Black] (0.5, 2) circle (0.05cm);
    \fill[Black] (1, 0.5) circle (0.05cm);
    \fill[Black] (1, 1.5) circle (0.05cm);
    \fill[Black] (1, 2.5) circle (0.05cm);
    \fill[Black] (1.5, 0) circle (0.05cm);
    \fill[Black] (1.5, 1) circle (0.05cm);
    \fill[Black] (1.5, 2) circle (0.05cm);
    \fill[Black] (1.5, 3) circle (0.05cm);
    \fill[Black] (2, 0.5) circle (0.05cm);
    \fill[Black] (2, 1.5) circle (0.05cm);
    \fill[Black] (2, 2.5) circle (0.05cm);
    \fill[Black] (2.5, 1) circle (0.05cm);
    \fill[Black] (2.5, 2) circle (0.05cm);
    \fill[Black] (2.5, 3) circle (0.05cm);
    \fill[Black] (3, 1.5) circle (0.05cm);
    \fill[Black] (3, 2.5) circle (0.05cm);

    \draw[thick] (1, 0) rectangle (2, 1);
    \draw[thick] (0, 1) rectangle (1, 2);
    \draw[thick] (1, 1) rectangle (2, 2);
    \draw[thick] (2, 1) rectangle (3, 2);
    \draw[thick] (1, 2) rectangle (2, 3);
    \draw[thick] (2, 2) rectangle (3, 3);
    \draw[thick] (1, 1) -- (3, 3);
    \draw[thick] (0, 1) -- (2, 3);
    \draw[thick] (1, 0) -- (3, 2);
    \draw[thick] (1, 2) -- (2, 1);
    \draw[thick] (0, 2) -- (2, 0);
    \draw[thick] (1, 3) -- (3, 1);
    \draw[thick] (2, 3) -- (3, 2);

\end{tikzpicture}}
\end{subfloat}

\caption{Lattice sites are at the tiles' edges. Each edge can only match with its complementary, so each site $i$ is described by a pairing of sides $(s_i^1, s_i^2)$. Following the convention, the plaquette shown in (a) reads as the following $4$-tuple of pairs: $P = ((s_1^1, s_1^2), (s_2^1, s_2^2), (s_3^1, s_3^2), (s_4^1, s_4^2))$. As an example, we depict in (c) how the arrangement of tiles in (b) would translate to this setting.  For clarity, the parity layer is not drawn in (b). The central plaquette in this case is $P=((7,6), (5,8), (2, 1), (12, 1))$, and the tile it fully determines, $T = (6, 5, 2, 1)$.}
\label{fig:coloring_example}

\end{figure}

We will denote this unique pairing as a function $match(s_1,s_2)=1$ if the pair is a valid match (those found in Table \ref{tab:edge_matching}), and $0$ if not. We use the notation $(s_1, s_2)$ as in the usual tile problems, and our convention will also be that the first element of the pair will be the topmost one (if in a vertical arrangement) or the leftmost one (if in horizontal). Two sides match if and only if their labels add 13, independently of their position. That is, $(12, 1)$ is a match, but $(1, 12)$ too. For a visual reference, see Figure \ref{fig:coloring_example}.

As these pairings are unique, one could naturally think that describing the edges (states of the Hilbert space) as a pair is unnecessarily complicated. Nevertheless, for deciding if the arising tile exists, it is important to see which part of the matching is on each side. For example, in Figure \ref{fig:coloring_example}, $(12, 1)$ appears, but $(1, 12)$ is also a valid match with the same sides involved. However, even though $(6, 5, 2, 1)$ is a tile that exists, $(6, 5, 2, 12)$ is not.

Then, the local Hilbert space has dimension $d=12$ and is described as
\begin{equation}\label{eq:tiling_hilbert_space}
    \mathcal{H}=\mathbb{C}^d=\operatorname{span}\{(s_1, s_2): s_1, s_2 \in \mathcal{S}, \;\; match(s_1,s_2)=1\}
\end{equation}
Tiles are consequently described by a $4$-tuple, conventionally starting from the topmost edge in a clockwise fashion. However, not all of these $4$-tuples represent elements from the set of valid Robinson tiles
\begin{equation}\label{eq:robinson_tiles}
    R = \{T = (s_1, s_2, s_3, s_4): s_i \in \mathcal{S}, \;T \text{  is a modified Robinson tile}\},
\end{equation}
so we penalize those combinations that do not yield a Robinson tile as described in the beginning of this section. We do this by the following plaquette interaction:
\begin{equation}\label{eq:tiling_penalty}
    h_c \ket{P} = h_c\ket{((s_1^1, s_1^2), (s_2^1, s_2^2), (s_3^1, s_3^2), (s_4^1, s_4^2))} =
        \begin{cases}
        0 & \text{if $(s_1^2, s_2^1, s_3^1, s_4^2) \in R$} \\
        1 & \text{otherwise}
        \end{cases}
\end{equation}

\begin{theorem}\label{th:tiling_layer}
Given a lattice $\Lambda(L\times H)$, where the sites are located in the middle point of the edges of the unit cells, the rotationally invariant Hamiltonian $H_c$ given by local interaction $h_c$ (Equation \ref{eq:tiling_penalty}) describes the tiling problem associated to $\mathcal{R}$, the set of modified Robinson tiles described by its edges (Equation \ref{eq:robinson_tiles}). That is, there exist zero-energy configurations that encode a valid tiling of the $L\times H$ section of the plane.
\end{theorem}
\begin{proof}
The Hamiltonian $H_c$ consists only of positive energy penalties. Therefore, the minimal energy occurs when the number of penalties is also minimized. By Equation \ref{eq:tiling_penalty}, a positive energy contribution can only emerge from invalid tiles. Additionally, as the modified Robinson tiles are designed to enforce a correct tiling of the whole plane (without fault lines), then a zero-penalty (with no invalid tiles) configuration can be achieved, translating into a zero-energy configuration.

An energy penalty is given when a certain tile configuration does not exist. As all rotations and reflections of the tiles are considered as part of the tile set, rotating a valid/invalid tile will still yield a valid/invalid tile, respectively. Therefore, $h_c\ket{P} = h_c(U_{\pi/2}\ket{P})$, and thus $H_c$ is rotationally invariant.
\end{proof}

\subsection{Red segments}\label{subsec:red_segments}
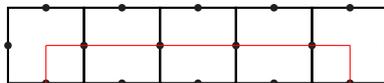
\begin{figure}[htbp]
\centering
\begin{tikzpicture}
    \draw[thick] (0, 0) rectangle (1, 1);
    \draw[thick] (1, 0) rectangle (2, 1);
    \draw[thick] (2, 0) rectangle (3, 1);
    \draw[thick] (3, 0) rectangle (4, 1);
    \draw[thick] (4, 0) rectangle (5, 1);

    \fill[Black] (0, 0.5) circle (0.05cm);
    \fill[Black] (0.5, 0) circle (0.05cm);
    \fill[Black] (0.5, 1) circle (0.05cm);
    \fill[Black] (1, 0.5) circle (0.05cm);
    \fill[Black] (1.5, 0) circle (0.05cm);
    \fill[Black] (1.5, 1) circle (0.05cm);
    \fill[Black] (2, 0.5) circle (0.05cm);
    \fill[Black] (2.5, 0) circle (0.05cm);
    \fill[Black] (2.5, 1) circle (0.05cm);
    \fill[Black] (3, 0.5) circle (0.05cm);
    \fill[Black] (3.5, 0) circle (0.05cm);
    \fill[Black] (3.5, 1) circle (0.05cm);
    \fill[Black] (4, 0.5) circle (0.05cm);
    \fill[Black] (4.5, 0) circle (0.05cm);
    \fill[Black] (4.5, 1) circle (0.05cm);
    \fill[Black] (5, 0.5) circle (0.05cm);

    \draw[red] (0.5, 0) -- (0.5, 0.5);
    \draw[red] (0.5, 0.5) -- (4.5, 0.5);
    \draw[red] (4.5, 0.5) -- (4.5, 0);
\end{tikzpicture}
\caption{A $4$-segment needs of $3$ red segment tiles and $2$ corner tiles, using a total of $6$ lattice sites.}
\label{fig:chain_tile_example}
\end{figure}

If we determine that the square tile size is the unit, all squares will have side length of $2^n$. In particular, $n$ odd for green squares and $n$ even for red ones, so, following Section 5 of \cite{CPW22}, we refer any red edge of a square as a $4^n$ segment. To form a $4^n$-segment, $4^n - 1$ consecutive red segment tiles and $2$ red corner tiles are needed, giving a chain of an even number of particles, as needed in Section \ref{sec:onedencoding} (see Figure \ref{fig:chain_tile_example} for an example).

Accordingly, when we talk about a $L \times H$ rectangle, we mean a section of the plane with a length of $L$ tiles, and a height of $H$ tiles. In order to give some energy results, we first need to lower and upper bound the number of (complete) red segments appearing in a $L \times H$ rectangle.

\begin{lemma}\label{lemma:segment_bound}
The number $s$ of red segments of size $4^n$ contained in a $L \times H$ rectangle (width $L$ and height $H$) tiled using modified Robinson tiles is lower and upper bounded by
\begin{equation}\label{eq:segment_bounds}
\begin{split}
    4\left\lfloor \dfrac{H}{2^{2n+1}}\right\rfloor \left\lfloor \dfrac{L}{2^{2n+1}} \right\rfloor - 2\left(\left\lfloor \dfrac{H}{2^{2n+1}}\right\rfloor + \left\lfloor \dfrac{L}{2^{2n+1}}\right\rfloor\right) \leq s \\
    \\
    s \leq 4\left\lfloor \dfrac{H}{2^{2n+1}}\right\rfloor \left\lfloor \dfrac{L}{2^{2n+1}} \right\rfloor + 2\left(\left\lfloor \dfrac{H}{2^{2n+1}}\right\rfloor + \left\lfloor \dfrac{L}{2^{2n+1}}\right\rfloor\right)
\end{split}
\end{equation}
\end{lemma}
\begin{proof}
By Lemma 48 in \cite{CPW22}, we know that the number of top red segments $s_T$ of size $4^n$ in a modified Robinson tiling of an $L \times H$ rectangle is contained in the interval $\lfloor H/2^{2n+1}\rfloor \bigl(\lfloor L/2^{2n+1}\rfloor -1\bigr), \bigl(\lfloor H/2^{2n+1}\rfloor +1 \bigr)\lfloor L/2^{2n+1}\rfloor]$ for all $n$. This interval also applies to bottom red segments $s_B$, as they can be seen as the top ones of a rotation of the original lattice. On the other hand, right and left segments $s_R$ and $s_L$ can be seen as the top ones in a rotation whose height and width are interchanged. Adding all four inequalities gives the stated result.
\end{proof}

The following Lemma is the key rigidity result needed later. It shows that a defect in the tiling (i.e., the presence of a non-valid tile), does alter the pattern of $4^n$ segments, but not outside a ball of size $4^{n+1}$ centered on the defect. 

\begin{lemma}\label{lemma:segment_rigidity}
In any tiling of an $L\times H$ rectangle (width $L$, height $H$) with $d$ defects using modified Robinson tiles, the total number of red segments of size $4^n$ is at least
\begin{equation}\label{eq:min_segments_defects}
    4 \left\lfloor \dfrac{H}{2^{2n+1}}\right\rfloor \left\lfloor \dfrac{L}{2^{2n+1}} \right\rfloor - 2\left(\left\lfloor \dfrac{H}{2^{2n+1}}\right\rfloor + \left\lfloor \dfrac{L}{2^{2n+1}}\right\rfloor\right) - 8d.
\end{equation}
\end{lemma}
\begin{proof}
By Lemma 49 in \cite{CPW22}, we know that the result for top segments is at least $\left\lfloor \dfrac{H}{2^{2n+1}}\right\rfloor \left(\left\lfloor \dfrac{L}{2^{2n+1}}\right\rfloor - 1 \right) - 2d$. Their defects are, however, pairs of non-matching adjacent pair of tiles. They achieve the previous bound by dividing the rectangle in blocks of size $2^{n+1} \times 2^{n+1}$, and analyzing the remaining shape after removing any blocks that contain a defect. Therefore, having a defect in our setting (a non-valid tile) would result in the same block removal. Consequently, their bound applies to our setting too, and as before, we can apply it to bottom red segments, and also to left and right red segments (interchanging height and width). And adding all these four inequalities gives the stated bound.
\end{proof}

\section{Fusing computation and tiling}\label{sec:fusion}
The Hamiltonian described in Section \ref{sec:onedencoding} relies on the use of penalty terms to describe a desired evolution, giving a $1$-dimensional computational Hamiltonian, which we called $H_q$. The Hamiltonian describing the tiling layer will be $H_c$, which also uses penalties in order to encode the valid tiling in a ground state.

We will now describe another Hamiltonian, $H_f$ which can be understood as the merger between $H_c$ and $H_q$. Given a certain tiling configuration $\ket{\varphi_C}$, we want the computational 1D Hamiltonian to only appear at specific regions of the tiling (on the edges of the red squares), so it adds extra penalties to undesirable overlappings of $H_c$ and $H_q$. This can be formalized as follows:

\begin{lemma}\label{lemma:ground_state}
Given a certain classical configuration of the tiling layer $\ket{\varphi_C}$ and a computational Hamiltonian $H_q$, there exists a rotationally invariant Hamiltonian $H_f$ whose ground state configurations restricted to the subspace $S=\ket{\varphi_C}\otimes\mathbb{C}^Q$ are configurations where every red segment $L$ present on the tiling layer is also a bracketed subspace $S_{br}$, and such that $H_q(L)|_{S_{br}}$ presents a ground state configuration (a correct evolution of the computation).
\end{lemma}
\begin{proof}

The first step is to locate the beginnings and ends of the red segments: the red corners. Each valid tile can be then described as an ordered group $(a, b, c, d)$ of four of Table \ref{tab:edge_matching} numbers. With this depiction of the tiles, we can easily detect all four possible red corners (bottom left, bottom right, top left and top right), as we just need to search for the plaquettes $(1,6,5,2)$, $(2,1,6,5)$, $(5,2,1,6)$ or $(6,5,2,1)$.

In order to do so, we denote the set of all computational states from the $1$-dimensional construction as $\mathcal{R}$ and define the following tile, that detects the boundaries of the construction, where $*$ denotes any possible computational state:
\begin{equation}\label{eq:corner_detection_tile}
    \tile{$\smallsymb{0.6}{\symbar}$}{*}{*}{$\smallsymb{0.6}{\symbar}$} = \sum_{r_1, r_2 \in \mathcal{R}} \tile{\smallsymb{0.6}{\symbar}}{$r_1$}{$r_2$}{\smallsymb{0.6}{\symbar}}
\end{equation}
We will use it to describe the following $4$-interactions. Note that $h_{n_2}$, $h_{n_3}$ and $h_{n_4}$ are the rotations of $h_{n_1}$ by $\frac{\pi}{2}$, $\pi$ and $\frac{3}{2}\pi$, respectively.
\begin{equation}\label{eq:corner_coupling_interaction}
\begin{split}
    h_{n_1}=\Bigg(\mathbbm{1} - \tile{1}{6}{5}{2}\Bigg) \otimes \tile{$\smallsymb{0.6}{\symbar}$}{*}{*}{$\smallsymb{0.6}{\symbar}$} + \Bigg(\mathbbm{1} - \tile{$\smallsymb{0.6}{\symbar}$}{*}{*}{$\smallsymb{0.6}{\symbar}$}\Bigg) \otimes \tile{1}{6}{5}{2} \\
    h_{n_2}=\Bigg(\mathbbm{1} - \tile{2}{1}{6}{5}\Bigg) \otimes \tile{$\smallsymb{0.6}{\symbar}$}{$\smallsymb{0.6}{\symbar}$}{*}{*} + \Bigg(\mathbbm{1} - \tile{$\smallsymb{0.6}{\symbar}$}{$\smallsymb{0.6}{\symbar}$}{*}{*}\Bigg) \otimes \tile{2}{1}{6}{5} \\
    h_{n_3}=\Bigg(\mathbbm{1} - \tile{5}{2}{1}{6}\Bigg) \otimes \tile{*}{$\smallsymb{0.6}{\symbar}$}{$\smallsymb{0.6}{\symbar}$}{*} + \Bigg(\mathbbm{1} - \tile{*}{$\smallsymb{0.6}{\symbar}$}{$\smallsymb{0.6}{\symbar}$}{*}\Bigg) \otimes \tile{5}{2}{1}{6} \\
    h_{n_4}=\Bigg(\mathbbm{1} - \tile{6}{5}{2}{1}\Bigg) \otimes \tile{*}{*}{$\smallsymb{0.6}{\symbar}$}{$\smallsymb{0.6}{\symbar}$} + \Bigg(\mathbbm{1} - \tile{*}{*}{$\smallsymb{0.6}{\symbar}$}{$\smallsymb{0.6}{\symbar}$}\Bigg) \otimes \tile{6}{5}{2}{1}
\end{split}
\end{equation}
These four interactions are in charge of guaranteeing that when a corner is detected in the classical tiling layer, the end of chain markers $\symbar$ must appear, and that next to them, the quantum or computational layer, must be not blank. Conversely, if end-of-chain markers appear in a region without a red corner, it will be penalized too.

Secondly, we would also want to ensure that when a red segment is encountered, a computation must be taking place. A red segment can be detected simply with a $(x, b, y, a)$ tile, where $a$ and $b$ are the labels of the edges with red side arrows that come from red arm tiles, and $x, y$ are other colors that complete the tuple to a valid tile. That is, $(x, b, y, a) \in \mathcal{A} := \{(x, b, y, a) \in\mathcal{R} \mid (a,b) = (6,7),(7,6),(5,8),(8,5)\}$.

To extend the computational $2$-body term to a plaquette, we introduce an additional state in the computational layer, that will represent a blank state (a non-computational state), and we use the notation - for it. Additionally, we introduce a scaling of $1/4$ (the reason for this will become apparent in Section \ref{sec:undecidability}), so the terms
\begin{equation}\label{eq:segments_coupling_interaction}
\begin{split}
    h_{s_h} = \dfrac{1}{4}\Bigg(\mathbbm{1} - \sum_{\mathcal{A}} \tile{x}{b}{y}{a}\Bigg) \otimes \tile{-}{*}{-}{*} + \dfrac{1}{4}\Bigg(\1 - \tile{-}{*}{-}{*}\Bigg) \otimes \sum_{\mathcal{A}}\tile{x}{b}{y}{a} \\
    h_{s_v} = \dfrac{1}{4}\Bigg(\mathbbm{1} - \sum_{\mathcal{A}}\tile{a}{x}{b}{y}\Bigg) \otimes \tile{*}{-}{*}{-} + \dfrac{1}{4}\Bigg(\1 - \tile{*}{-}{*}{-}\Bigg) \otimes \sum_{\mathcal{A}}\tile{a}{x}{b}{y}
\end{split}
\end{equation}
penalize a configuration where a vertical or horizontal red segment appears in the classical tiling, but has no computational state in the quantum layer (and vice versa).  Again, $h_{s_v}$ is the $\frac{\pi}{2}$ rotation of the $h_{s_h}$ term, and $*$ represents any computational state, acting as an identity over the set of computational states.

So far, we have associated the computational Hamiltonian $H_q$ to the red segments in the tiling. However, in section \ref{sec:onedencoding}, we modified the original $H_q$ for allowing a reflection symmetry, and therefore, having two different ground states (one corresponding to each direction of computation). Therefore, when merged to the tiling, if we have $k$ different red segments, we end up having $2^k$ different ground states per choice of tiling. Nevertheless, this is still not the final Hamiltonian to be considered. In Section \ref{sec:undecidability}, we will see how, by fusing with a trivial Hamiltonian, a unique $0$-energy ground state will appear in the gapped case, while lifting all this exponential degeneracy to a positive energy.

Even so, in order to avoid this exponential dependency, we can fix an orientation for the computation, depending on the segment it runs on, while also maintaining rotational invariance. This allows the final Hamiltonian to have an single ground state over tiling configuration $\ket{\varphi_C}$, and not an exponential number of them. For this purpose, we see that certain boundary terms (summarized in Table \ref{tab:ends}) allow us to detect if we are following a canonical or a reverse orientation. Using this idea, we add the following family of illegal terms: the terms below (and all their rotations) enforce a desired orientation for the computation depending on the detected red corner.

\begin{equation}\label{eq:computation_orientation_tiles}
    h_{c_1}=\scalebox{1.5}{\tile{$\smallsymb{0.6}{\symbar}$}{$\smallsymb{0.6}{\br}$}{*}{$\smallsymb{0.6}{\symbar}$}} \;\;\;\; h_{c_2}=\scalebox{1.5}{\tile{$\smallsymb{0.6}{\symbar}$}{$\smallsymb{0.6}{\al}$}{*}{$\smallsymb{0.6}{\symbar}$}} \;\;\;\; h_{c_3}=\scalebox{1.5}{\tile{$\smallsymb{0.6}{\symbar}$}{*}{$\smallsymb{0.6}{\ar}$}{$\smallsymb{0.6}{\symbar}$}} \;\;\;\; h_{c_4}=\scalebox{1.5}{\tile{$\smallsymb{0.6}{\symbar}$}{*}{$\smallsymb{0.6}{\bl}$}{$\smallsymb{0.6}{\symbar}$}}
\end{equation}
States $\ar$ or $\bl$ next to $\symbar$ mean that we are at the end where the computation started, whereas $\al$ and $\br$ indicates its returning point (see Table \ref{tab:ends}). Using these additional illegal plaquettes, we ensure that the final computation in the ground state can only evolve in a clockwise fashion (left to right in the top segment, top to bottom in the right one, right to left in the bottom, and bottom to top in the left), as any other combination will give an additional energy penalty.

Therefore, if $h_c$ is the classical tiling local interaction term and $h_q$, the computational one, the Hamiltonian $H_f$ described locally by
\begin{equation}\label{eq:hf_local_interaction}
    h_f = h_c + h_q + h_{s_h} + h_{s_v} + \sum_{i=1}^4 h_{n_i} + \sum_{i=1}^4 \sum_{r=0}^3 U_{\pi/2}^r  h_{c_i} (U_{\pi/2}^r)^\dag
\end{equation}

comprises all of the following energy penalties:

\begin{itemize}
    \item Those associated with the classical tiling ($h_c$ term).
    \item Those associated with the quantum computation ($h_q$ term).
    \item If end markers $\symbar$ and red corners are not both present at a site ($h_{n_i}$ terms).
    \item If a red segment and a computational state are not both present at a site ($h_{s_h}$ and $h_{s_v}$ terms).
    \item If the orientation of the computation is \textit{not} clockwise ($h_{c_i}$ terms and their rotations).
\end{itemize}

With this description, all the local terms that form $H_f$ have no energy bonuses, so the only way of minimizing the energy is reducing the number of penalties. There are no incompatibilities between them, so a state where all the restrictions are met is attainable. Therefore, the only energy present comes from both $H_c$ and $H_q$. Additionally, all terms are invariant under rotations, so $H_f$ is as well.
\end{proof}
\section{Undecidability results}\label{sec:undecidability}
We start by bounding the ground state energy of Hamiltonian $H_f$.

\begin{lemma}\label{lemma:gs_energy_bound}
Let $H_c$ be the Tiling Hamiltonian as described in Section \ref{sec:tiling}, with the distinguished state $\symbar$, and $H_q$ a computational Hamiltonian as in Section \ref{sec:onedencoding}. Then, there is a Hamiltonian $H_f$ on a lattice of width $L$ and height $H$ such that the ground state energy $\lambda_0(H_f^{\Lambda(L\times H)})$ is lower bounded by
\begin{equation}\label{eq:hf_energy_lower_bound}
    \lambda_0(H_f^{\Lambda(L\times H)}) \geq \sum_{n=1}^{\lfloor \log_4(\min\{L,H\}) \rfloor} \left(4 \left\lfloor \dfrac{H}{2^{2n+1}}\right\rfloor \left\lfloor \dfrac{L}{2^{2n+1}} \right\rfloor - 2\left(\left\lfloor \dfrac{H}{2^{2n+1}}\right\rfloor + \left\lfloor \dfrac{L}{2^{2n+1}}\right\rfloor\right)\right)\lambda_0(4^n)
\end{equation}
and upper bounded by 
\begin{equation}\label{eq:hf_energy_upper_bound}
    \lambda_0(H_f^{\Lambda(L\times H)}) \leq \sum_{n=1}^{\lfloor \log_4(\max\{L,H\}) \rfloor} \left(4 \left\lfloor \dfrac{H}{2^{2n+1}}\right\rfloor \left\lfloor \dfrac{L}{2^{2n+1}} \right\rfloor + 2\left(\left\lfloor \dfrac{H}{2^{2n+1}}\right\rfloor + \left\lfloor \dfrac{L}{2^{2n+1}}\right\rfloor\right)\right)
\lambda_0(4^n)
\end{equation}
\end{lemma}
\begin{proof}
Construct the Hamiltonian as in Lemma \ref{lemma:ground_state}. This implies that the lowest energy configuration for a given tiling (classical layer) is attained by putting blank states everywhere, except between the segments marked with a $\symbar$, where $h_q$ acts over the chain instead. In the modified Robinson tiling, this corresponds exactly to all the red edges, of size $4^n$. Therefore, we can restrict our analysis to classical configurations of Robinson tilings (not necessarily valid) with an eigenstate of $H_q$ along each complete red edge present.

If there are no defects present in the tiling, the minimum energy is given only by the computational part in each of the red edges $s \in S$: $\sum_{s \in E} \lambda_0(s)$. Lemma \ref{lemma:segment_bound} states the number of minimum and maximum segments we can have of size $4^n$, so we can see that $\sum_{s \in S} \lambda_0(s)$ is contained in the stated interval for every $L \times H$ lattice with no defects.

On the other hand, each defect present in the classical tile configuration contributes with energy at least $1$ from the $h_c$ term, and causes a decrease in the number of red segments. Lemma \ref{lemma:segment_rigidity} gives us a lower bound for the number of red segments in this case. If we write $\lambda_0(4^n)$ for the energy that the computational term gives in a red edge of size $4^n$, we have that the energy of an eigenstate with $d$ defects is at least

\[
d + \sum_{n=1}^{\lfloor \log_4(\min\{L,H\}) \rfloor} \left( 4 \left\lfloor \dfrac{H}{2^{2n+1}}\right\rfloor \left\lfloor \dfrac{L}{2^{2n+1}} \right\rfloor - 2\left(\left\lfloor \dfrac{H}{2^{2n+1}}\right\rfloor + \left\lfloor \dfrac{L}{2^{2n+1}}\right\rfloor\right) - 8d \right)\lambda_0(4^n) = \]

\[d + \sum_{n=1}^{\lfloor \log_4(\min\{L,H\}) \rfloor} \left(4 \left\lfloor \dfrac{H}{2^{2n+1}}\right\rfloor \left\lfloor \dfrac{L}{2^{2n+1}} \right\rfloor - 2\left(\left\lfloor \dfrac{H}{2^{2n+1}}\right\rfloor + \left\lfloor \dfrac{L}{2^{2n+1}}\right\rfloor\right)\right)
\lambda_0(4^n) - 8d\sum_{n=1}^{\lfloor \log_4(\min\{L,H\}) \rfloor} \lambda_0(4^n) \geq
\]

\[
d + \sum_{n=1}^{\lfloor \log_4(\min\{L,H\}) \rfloor} \left(4 \left\lfloor \dfrac{H}{2^{2n+1}}\right\rfloor \left\lfloor \dfrac{L}{2^{2n+1}} \right\rfloor - 2\left(\left\lfloor \dfrac{H}{2^{2n+1}}\right\rfloor + \left\lfloor \dfrac{L}{2^{2n+1}}\right\rfloor\right)\right)
\lambda_0(4^n) - 8d\dfrac{1}{8} =
\]

\[
\sum_{n=1}^{\lfloor \log_4(\min\{L,H\}) \rfloor} \left(4 \left\lfloor \dfrac{H}{2^{2n+1}}\right\rfloor \left\lfloor \dfrac{L}{2^{2n+1}} \right\rfloor - 2\left(\left\lfloor \dfrac{H}{2^{2n+1}}\right\rfloor + \left\lfloor \dfrac{L}{2^{2n+1}}\right\rfloor\right)\right)
\lambda_0(4^n),
\]
where we have used that $\sum_{n=1}^\infty\lambda_0(4^n) < 1/8$. This is true by Theorem 32 in \cite{CPW22}: in their construction, the bound for this sum in their computational Hamiltonian is $1/2$. However, when constructing $h_f$ in Section \ref{sec:fusion}, we have scaled the energy of computations over red segments by $1/4$, while embedding the 2-body computations in plaquette form. Therefore, bound from equation \ref{eq:hf_energy_lower_bound} still follows.

Note that the summatory limits can be refined, as we take the minimum segment size present in both vertical and horizontal orientations for the lower bound, and the maximum size present in any of both orientations for the upper bound. However, it is sufficient for our purposes.
\end{proof}

We use this lemma to construct a Hamiltonian $H_u$ with an undecidable ground state energy.

\begin{proposition}\label{prop:undecidable_energy_bound}
Choose any $\beta \in \mathbb{Q}^+$, as small as needed. There exists a local Hamiltonian $H_u$ and positive uncomputable functions $\delta_1(n), \delta_2(n)$ such that either:
\begin{itemize}
    \item $\lambda_0(H_u^{\Lambda(L)}) \leq -\dfrac{3\beta L}{4}$ for all $L$, or
    \item $\lambda_0(H_u^{\Lambda(L)}) \geq \beta(L^2\delta_2(n)-L\delta_1(n))$ for all $L \geq L_0(n)$ (with $L_0(n)$ uncomputable),
\end{itemize}
but determining which is undecidable. 
\end{proposition}
\begin{proof}
Let $h_q(n)$ be the local interactions of a computational Hamiltonian, obtained particularly by applying Theorem \ref{th:qtm_hamiltonian} with $K=3$ to the dovetailing of the QPE-machine $P_n$ described in Section \ref{subsec:qpe_machine} and a UTM. The input parameter $n$ is what determines the specific $P_n$ used, and thus creating a dependency of the computational interactions on the input. Let $h_f(n)$ be the local interactions obtained by constructing a Hamiltonian as described in Section \ref{sec:fusion}, merging $h_q(n)$ with a $h_c$ from a tiling layer. Finally, define the local interaction of $H_u$ as $h_u(n):=\beta h_f(n)$.

Notice that the UTM will start only if the QTM from Theorem 10 in \cite{CPW22} has enough tape and time to finish, therefore correctly initializing $n$. This happens when the segment length, $r$ is at least $|n| + 6$.

First, we consider the case in which the UTM does not halt. In that case, the only segments with positive energy will be those not long enough for achieving proper initialization. Using Lemma \ref{lemma:gs_energy_bound}, we have that:
\[\lambda_0(H_u^{\Lambda(L)}) \leq \beta \sum_{\substack{1\leq r \leq |n| + 6 \\ r=4^m, m \in \mathbb{N}}} \left(4\left(\dfrac{L}{2r}\right)^2+4\left(\dfrac{L}{2r}\right)\right)\lambda_0(r) = \beta L^2 \sum \dfrac{\lambda_0(r)}{r^2}+\beta L\sum \dfrac{2\lambda_0(r)}{r},\]
and if we name $\alpha_2(n)=\sum \dfrac{\lambda_0(r)}{r^2}$ and $\alpha_1(n)=\sum \dfrac{2\lambda_0(r)}{r}$, the expression reads as $\beta(L^2\alpha_2(n)+L\alpha_1(n))$.

We now perform an energy shift in the system: we give an additional $-\beta/2$ energy to each particle. As each plaquette has $4$ particles, our system of $L \times L$ plaquettes can be seen as $L$ rows of $L+1$ particles and $L$ columns of $L+1$ particles, all different (see figure \ref{fig:plaquette_system}). Therefore, we have $2L(L+1)$ particles, giving a total energy of $-\beta L(L+1)$. We also have $L^2$ plaquette interactions, and we give an energy of $-\beta(\alpha_2(n)-1)$ to each, adding another $-L^2\beta(\alpha_2(n)-1)$. Then, the previous $\beta(L^2\alpha_2(n)+L\alpha_1(n))$ transforms to:
\[\beta(L^2\alpha_2(n)+L\alpha_1(n)) -\beta L(L+1) -L^2\beta(\alpha_2(n)-1) = \beta L^2 \alpha_2(n) + \beta L \alpha_1(n) - \beta L - \beta L^2\alpha_2(n) = \]
\[\beta L\alpha_1(n) - \beta L = \beta L (\alpha_1(n) - 1) < -\dfrac{3}{4}\beta L,\]

where the last inequality follows from the fact that $\sum_{n=1}^\infty\lambda_0(4^n) < 1/8$, applied to $\alpha_1(n)$.

On the other hand, we have the case in which the UTM halts, and consider $r_1(n):=\min\{r=4^m \text{ such that}$ 
$\text{segment $r$ is large enough for UTM to halt on input $n$}\}$. Therefore, as entering the Halting state gives a $c^{-L}>0$ energy penalty from the $1$-dimensional computational Hamiltonian (see end of Section \ref{sec:onedencoding}, after Theorem \ref{th:qtm_hamiltonian}), we have that $\lambda_0(H_u^{\Lambda(L)}) > 0$ for all $r \geq r_1(n)$. After the same energy shift as before, the ground state energy can be lower bounded (again, by lemma \ref{lemma:gs_energy_bound}) by:

\[\lambda_0(H_u^{\Lambda(L)}) \geq -\beta L^2\alpha_2(n)-\beta L+\beta\sum_{\substack{1\leq r \leq |n| + 6 \\ r=4^m, m \in \mathbb{N}}} \left(\dfrac{L^2}{r^2}-\dfrac{2L}{r}\right)\lambda_0(r)+\beta\sum_{\substack{r \geq r_1(n) \\ r=4^m, m \in \mathbb{N}}}\left(\dfrac{L^2}{r^2}-\dfrac{2L}{r}\right)\lambda_0(r) = \]
\[-\beta L^2 \alpha_2(n) -\beta L + \beta L^2\alpha_2(n)-\beta L\alpha_1(n) + \beta L^2 \sum_{\substack{r \geq r_1(n) \\ r=4^m, m \in \mathbb{N}}}\dfrac{\lambda_0(r)}{r^2} -\beta L \sum_{\substack{r \geq r_1(n) \\ r=4^m, m \in \mathbb{N}}} \dfrac{2\lambda_0(r)}{r} = \]
\[-\beta L -\beta L\alpha_1(n) + \beta L^2 \sum_{\substack{r \geq r_1(n) \\ r=4^m, m \in \mathbb{N}}}\dfrac{\lambda_0(r)}{r^2} -\beta L \sum_{\substack{r \geq r_1(n) \\ r=4^m, m \in \mathbb{N}}} \dfrac{2\lambda_0(r)}{r} = \]
\[\beta L^2 \sum_{\substack{r \geq r_1(n) \\ r=4^m, m \in \mathbb{N}}}\dfrac{\lambda_0(r)}{r^2} - \beta L \left( 1 + \sum_{\substack{1\leq r \leq |n| + 6 \text{ or } r \geq r_1(n)  \\ r=4^m, m \in \mathbb{N}}} \dfrac{2\lambda_0(r)}{r} \right) := \beta (L^2 \delta_2(n) - L\delta_1(n))\]

where we have defined
\[\delta_2(n) = \sum_{\substack{r \geq r_1(n) \\ r=4^m, m \in \mathbb{N}}}\dfrac{\lambda_0(r)}{r^2} \text{\;\;and\;\;} \delta_1(n) = 1 + \sum_{\substack{1\leq r \leq |n| + 6 \text{ or } r \geq r_1(n)  \\ r=4^m, m \in \mathbb{N}}} \dfrac{2\lambda_0(r)}{r}\]

Finally, the proposition follows from the undecidability of the Halting problem.
\end{proof}

Now, by just looking at the definition of energy density, we can have the following result:

\begin{corollary}
Determining whether $E_\rho = 0$ or $E_\rho > 0$ is undecidable.
\end{corollary}
\begin{proof}
Apply the definition of energy density (Definition \ref{def:gs_energy_definitions}), $\lim_{L \rightarrow \infty} \lambda_0(H_u^{\Lambda(L)})/(2L(L+1))$, to the results of Proposition \ref{prop:undecidable_energy_bound}. In the non-halting case, $E_\rho = 0$. However, in the halting case, as $\lambda_0(n) > 0$ for all $r \geq r_1(n)$, then we also have $\delta_2(n) > 0$. And therefore, $E_\rho = \beta\delta_2(n) > 0$.
\end{proof}

\begin{figure}[hbtp]
  \centering
  \includegraphics[width=0.9\columnwidth]{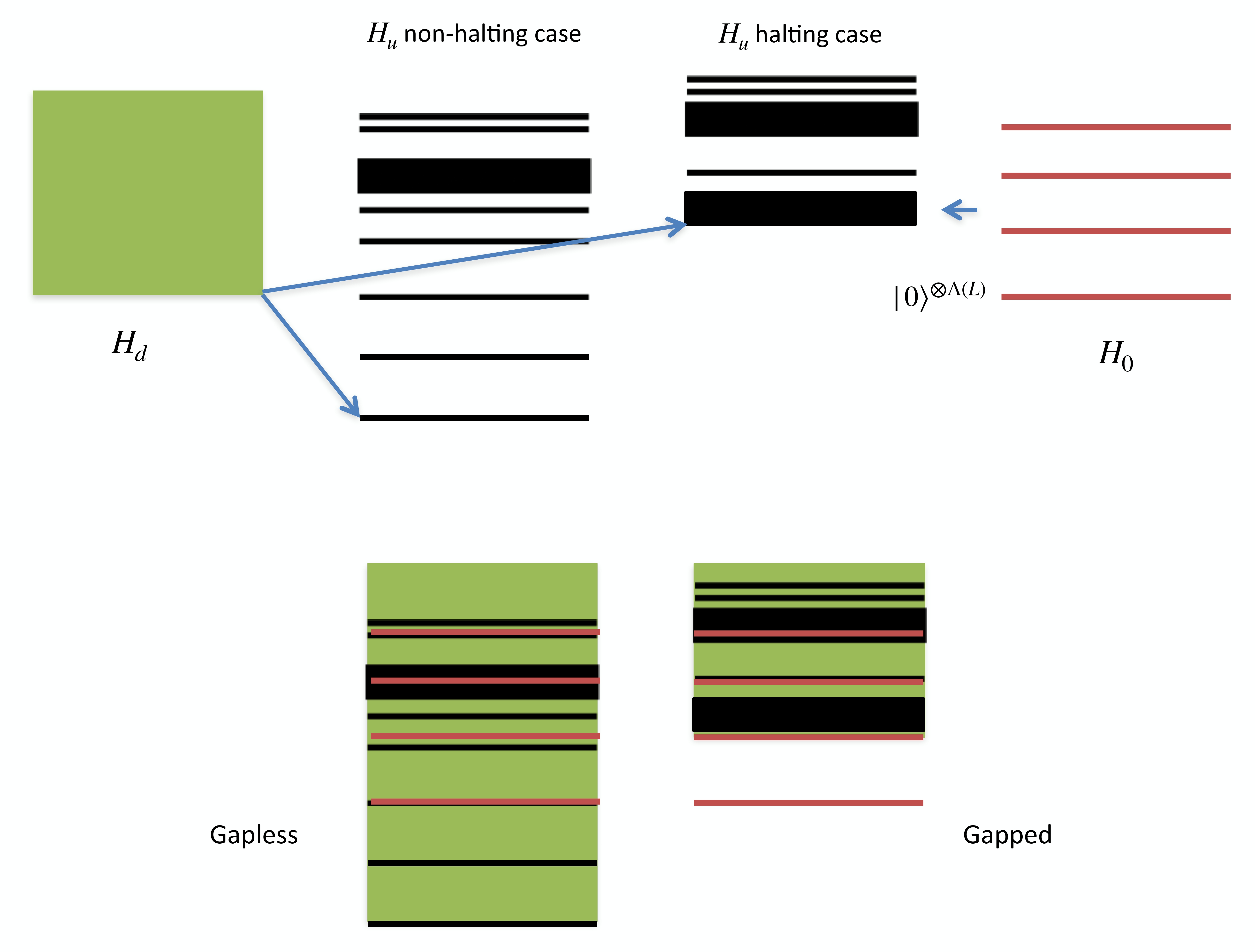}
  \caption{Image taken from \cite{CPW22} (Section 6). Starting with the Hamiltonian $H_u$ constructed in Proposition \ref{prop:undecidable_energy_bound}, we consider two additional Hamiltonians, $H_d$ and $H_0$, with dense and trivial spectrum respectively. These are combined into a final Hamiltonian $H$ such that the different spectra get combined as indicated by the arrows in the figure. This results in an overall Hamiltonian $H$ with gapped or gapless behavior, as shown in the bottom figure, depending on whether the Turing Machine encoded in Hamiltonian $H_u$ halts or not.}
  \label{fig:spectral_merging}
\end{figure}

The last step is then to prove the main result: the undecidability of the spectral gap problem. The idea behind the proof is illustrated in Figure \ref{fig:spectral_merging}, as in Section 6 of \cite{CPW22}, but we just need to use Hamiltonians $H_0$ and $H_d$ that are also rotationally symmetric. 

\begin{theorem}\label{theorem:undecidability_spectral_gap}
For any given universal Turing Machine UTM, we can construct a family of 2-dimensional Hamiltonians $H(n)$ with 4-body plaquette interactions with rotational symmetry such that:
    \begin{itemize}
        \item If UTM halts on input $n$, then the family $\{H^{\Lambda(L)}(n)\}$ is gapped as in Definition \ref{def:gs_energy_definitions}.
        \item If UTM does not halt on input $n$, then the family $\{H^{\Lambda(L)}(n)\}$ is gapless as in Definition \ref{def:gs_energy_definitions}.
    \end{itemize}
\end{theorem}
\begin{proof}
Let $h_u^{(i,j,k,l)}(n)$ the plaquette interactions obtained in Proposition \ref{prop:undecidable_energy_bound} and $h_u^{(i)}(n)$ the one-body ones. Take a gapless Hamiltonian $H_d$, with ground state energy $0$, and rotationally invariant (as in Definition \ref{def:rotational_invariance}). In particular, consider the spin-1/2 ferromagnetic Heisenberg model described in a plaquette form, as shown in Example \ref{ex:gapless_hamiltonian}, with local Hilbert space $\mathcal{H}_d$.

We now assign a Hilbert space $\mathcal{H}^{(i)} := \ket{0}\oplus \mathcal{H}_u \otimes \mathcal{H}_d$ to each site $i \in \Lambda$. Define the plaquette and one-body interactions of Hamiltonian $H(n)$ as
\begin{subequations}\label{eq:promise_H}
\begin{align}
h^{(i,j,k,l)}(n):=&\sum_{(a,b,c,d)} \ketbra{0}{0}^{(a)} \otimes \Pi^{(b,c,d)}_{ud} + \Pi^{(a)}_{ud} \otimes \ketbra{0}{0}^{(b,c,d)} \label{eq:hzero} \\
                &+ \mathbbm{1}_u^{(i,j,k,l)} \otimes h_d^{(i,j,k,l)}\label{eq:hd} \\
                &+ h_u^{(i,j,k,l)}(n) \otimes \mathbbm{1}_d^{(i,j,k,l)} \label{eq:hu}, \\
h^{(i)}(n):=&-\beta(1+\alpha_2(n))\Pi_{ud}^{(i)},\label{eq:one_site}
\end{align}
\end{subequations}
where $\beta$ and $\alpha_2(n)$ are those of Proposition \ref{prop:undecidable_energy_bound}, $(a,b,c,d)$ is taken over all four cyclic permutations of $(i,j,k,l)$, and $\Pi_{ud}$ is the projection of $\mathcal{H}$ onto its subspace $\mathcal{H}_u \otimes \mathcal{H}_d$. Decompose the global Hamiltonian in the square lattice $\Lambda(L)$ as $H^{\Lambda(L)}:=\tilde{H}_0+\tilde{H}_d+\tilde{H}_u$, taken over (\ref{eq:hzero}), (\ref{eq:hd}) and (\ref{eq:hu}) + (\ref{eq:one_site}), respectively. The three terms commute with each other and
\[ \spec \tilde{H}_d =\spec H_d\; , \;\; \spec \tilde{H}_u=\{0\} \cup \spec H_u\; , \;\; \spec \tilde{H}_0\subset \Z_{\ge 0}\]
Let us analyze the spectrum of $H^{\Lambda(L)}$ depending on the behavior of the UTM:
\begin{itemize}
    \item In the halting case, take $L \geq L(n)$ as the minimal $L$ from Proposition \ref{prop:undecidable_energy_bound} such that $L^2\delta_2(n)-L\delta_1(n) \geq 1/\beta$. In that case, $\tilde{H}_d, \tilde{H}_u \geq 0$ and hence $H^{\Lambda(L)}\geq \tilde{H}_0$. Since $\ket{0}^{\otimes \Lambda(L)}$ is the unique ground state of $\tilde{H}_0$, with energy $0$, and is also a 0-energy state for $H^{\Lambda(L)}$, we have that the spectral gap of $H^{\Lambda(L)}$ is at least as large as the spectral gap of $\tilde{H}_0$, which is $1$.
    \item In the non-halting case, we observe the following, given by the structure of the Hamiltonians:
    \[\spec \tilde{H}_u + \spec \tilde{H}_d\subset \spec H^{\Lambda(L)}\subset \spec \tilde{H}_u + \spec \tilde{H}_d + \spec \tilde{H}_0\]
    As $\spec \tilde{H}_0$ is contained in the set of non-negative integers, $\spec \tilde{H}_d\subset [0,+\infty)$ and converges in the thermodynamic limit to $[0,+\infty)$, and $\lambda_0(\tilde{H}_u)\le 0$ for all $L$, the Theorem follows.
\end{itemize}
\end{proof}
\section{Discussion}\label{sec:discussion}
We presented a family of 2D 4-body Hamiltonians, which respects the translation and rotational symmetry of the square lattice, and whose spectral gap depends on the halting property of a Turing machine, making it impossible for an algorithm to predict whether the spectral gap is positive or vanishes in the thermodynamic limit. The local Hamiltonians we construct, in the gapped instances, have a unique ground state, which is then necessarily invariant under the three discrete symmetries considered.

A small remark is that, as explained in \cite{CPW22} (Section 2), one can fuse a tiling problem and a gapless frustration-free Hamiltonian, and, due to the result of undecidability of the tiling problem itself (\cite{Berger1966}), the problem of the spectral gap for classical Hamiltonians with unconstrained local dimension is undecidable as well. Our construction uses a tiling and a gapless frustration-free Hamiltonian, both of them rotationally invariant, so the same conclusion can be reached for classical rotationally invariant Hamiltonians.

We believe that, by incorporating lattice symmetry constraints,  our construction brings the undecidability of the spectral gap result a bit closer to more physically realistic models, although it still suffers from a very large local dimension, as well as requiring 4-body instead of 2-body interactions as in the previous results. We hope that this is a step forward in understanding whether symmetries play a role in undecidability results. While our result is in a sense negative, given that we show that rotational symmetry is not an obstacle for undecidability, we think it is an important question to understand how much can one restrict the problem until it becomes decidable.

There are many future research directions and open problems that arise from our work. The first and most immediate is what happens in the case of 1D systems. Here the only natural symmetry of the lattice, beyond translations, is the reflection symmetry. In this work we use the results of Section 6 in \cite{GI10} to obtain a family of 1D interactions with reflection symmetry which encode the behavior of a QTM. In order to obtain a undecidability result in the 1D case, one could hope to adapt the construction of \cite{Bausch_2020}, in order to amplify the exponentially small gap of the history state Hamiltonian to a constant. We leave this for a future work.

A second open problem is to consider the reflection symmetry on the 2D lattice. Our construction is in a sense chiral, as our Hamiltonian description relies on the fact that the tiles have a sense of ``inner orientation''. While we have no fundamental reason to believe that reflection symmetry would be incompatible with undecidability of the spectral gap, our construction depends heavily on the lack of this particular symmetry. In both the 1D and 2D cases with reflection symmetry, a non trivial problem is to guarantee a unique gapped ground state while at the same time respecting the symmetry constraint.

The third open problem is whether the restriction on 4-body interactions is really required. This is again, on the one hand, an artifact of the proof, due to the necessity of encoding an invariant tiling problem into an invariant local Hamiltonian. If one could perform this step with a smaller interaction length, then most probably the rest of the proof could be adapted.

Finally, there is the question of what happens if we consider other symmetries that do not arise from the lattice, e.g., invariance under a global symmetry of the action of a Lie group, such as $SU(2)$-invariant models in the theory of quantum magnetism. Imposing a non-trivial symmetry of this kind breaks all of the constructions of history states and tiling Hamiltonians, so a very different approach would be needed to encode undecidable problems in this class of models. Similarly to the case of reflection symmetry, one would need to find a mechanism to guarantee a unique (and therefore invariant under the symmetry) ground state for the gapped instances, since otherwise the breaking of the continuous symmetry would immediately imply a gapless system \cite{Landau1981, Koma1994}.

\section*{Acknowledgements}
The authors acknowledge financial support from grants PID2020-113523GB-I00, PID2023-146758NB-I00 and CEX2023-001347-S, funded by MICIU/AEI/10.13039/501100011033. L.C. acknowledges support from grant PRE2021-098747 funded by MICIU/AEI/10.13039/501100011033 and ``ESF+'', and grant TEC-2024/COM-84-QUITEMAD-CM, funded by Comunidad de Madrid.
A.\,L.~acknowledges support from  grant RYC2019-026475-I funded by MICIU/AEI/10.13039/501100011033 and ``ESF Investing in your future'', and from the Italian Ministry of University and Research (MUR), through ``Programma per Giovani Ricercatori Rita Levi Montalcini'', as well as the grant ''Dipartimento di Eccellenza 2023-2027'' of Dipartimento di Matematica, Politecnico di Milano. This work has been financially supported by the Ministry for Digital Transformation and the Civil Service of the Spanish Government through the QUANTUM ENIA project call – Quantum Spain project, and by the European Union through the Recovery, Transformation and Resilience Plan – NextGenerationEU within the framework of the Digital Spain 2026 Agenda.

\clearpage

\printbibliography

\newpage
\begin{appendices}
\clearpage

\clearpage
\section{Tables}\label{appendix:tables}
The $1$-dimensional chain is divided into $7$ tracks.
\begin{center}
  \begin{tabular}{|l|clc|r|}
    \hline
    $\symbar$ & $\cdots$ & Track 0: Reflection track & $\cdots$ & $\symbar$ \\
    \hline
    $\symbar$ & $\cdots$ & Track 1: Clock oscillator & $\cdots$ & $\symbar$ \\
    \hline
    $\symbar$ & $\cdots$ & Track 2: Counter TM head and state & $\cdots$ & $\symbar$\\
    \hline
    $\symbar$ & $\cdots$ & Track 3: Counter TM tape & $\cdots$ & $\symbar$\\
    \hline
    $\symbar$ & $\cdots$ & Track 4: QTM head and state & $\cdots$ & $\symbar$\\
    \hline
    $\symbar$ & $\cdots$ & Track 5: QTM tape & $\cdots$ & $\symbar$\\
    \hline
    $\symbar$ & $\cdots$ & Track 6: Time-wasting tape & $\cdots$ & $\symbar$\\
    \hline
  \end{tabular}
\end{center}
And their local Hilbert spaces are:
\begin{equation}\label{eq:tracks_hilbert_spaces}
\begin{split}
    \mathcal{H}_{0} := &\linspan\bigl\{\pa,\pb,\ar,\al,\br,\bl\bigr\} \oplus \ket{\symbar}, \\
    \mathcal{H}_{1} := &\linspan\bigl\{\ket{s}\bigr\} \oplus \ket{\symbar}\\
    &s \in \{\arrRzero,\arrLzero,\arrLzeroi,\arrRone,\arrLone,
    \blankL,\blankR\} \text{ for } i\in\{1,\dots,K\},\\
    \mathcal{H}_{2} := &\linspan\bigl\{\ket{p}\bigr\} \oplus \ket{\symbar}\\
    &p \in P'\cup\{\blankL,\blankR\}
    \text{ where } P:=P_L\cup P_N\cup P_R \text{ and } P':=P\cup P'_R,\\
    \mathcal{H}_{3}:= &\linspan\bigl\{\ket{\tau}\bigr\} \oplus\ket{\symbar}\\
    &\tau \in \Xi
    \text{ where } \Xi := \{\vdash,\#, 0,\ldots, \zeta-1\}
    \text{ with } \zeta = \abs{\Sigma\times Q},\\
    \mathcal{H}_{4}:= &\linspan\bigl\{\ket{q}\bigr\} \oplus\ket{\symbar}\\
    &q \in Q'\cup P\cup P'_L\cup \{r_x\}\cup \{\blankL,\blankR\}
    \text{ where } Q:=Q_L\cup Q_N\cup Q_R, Q':=Q\cup Q'_L\\
    &\text{ and } x \in Q,\\
    \mathcal{H}_{5}:= &\linspan\bigl\{\ket{\sigma}\bigr\} \oplus\ket{\symbar}\\
    &\sigma \in \Sigma,\\
    \mathcal{H}_{6}:= &\linspan\bigl\{\ket{\gamma}\bigr\} \oplus\ket{\symbar}\\
    &\gamma \in \Xi \cup \{\vdash_q\} \text{ where } q\in Q'.
\end{split}
\end{equation}
$K$ is a constant that is fixed later. $\Sigma$ is the tape alphabet of the given QTM $M$. $\Xi$ is the alphabet of the counter TM that will drive the clock. $P_L,P_N,P_R$ are the sets of internal states of the counter TM that can be entered by the TM head moving left, not moving, or moving right (respectively). The states $p'\in P_R'$ duplicate the states $p\in P_R$, and $p'\in P'_L$ duplicate those in $P_L$. Similarly for the internal states $Q_L,Q_N,Q_R$ of the QTM, with $q'\in Q'_L$ duplicating the states $q\in Q_L$.

The $\blankL$ and $\blankR$ Track 2 and 4 symbols are used for cells that do not currently hold the head.\footnote{There are two blank symbols because different symbols are needed to the left and right of the head, in order to enforce the constraint that there is only one head on the track.} The general marker state $\ket{\symbar}$ that extends to all tracks, as in Theorem \ref{th:qtm_hamiltonian} is just the state $\ket{\symbar} = \bigotimes_{i=0}^6\ket{\symbar}_{\text{track}\; i}$. This set of states defines a standard basis for the single-site Hilbert space $\mathcal{H}$. The product states over this single-site basis then give a basis over Hilbert space $\mathcal{H}^{\otimes L}$ of the chain.

\clearpage

\vspace*{-3.5cm}
\setlength{\tabcolsep}{2pt} 
\renewcommand{\arraystretch}{0.5} 
\vspace*{0cm}
\setlength\LTleft{0cm}
\hfuzz=0pt
\newcolumntype{P}[1]{>{\centering\arraybackslash}p{#1}}
\begin{longtable}{|P{1cm}|P{4.5cm}P{4.5cm}P{4.5cm}|}
\caption{Clock rules for the canonical orientation. Rules marked with a $*$ are modified in Table \ref{tab:can_quantum}.}\label{tab:can_rules} \\
\hline & & & \\
    & 1. $\fourcells{\ar}{\pa}{\arrRzero}{\blankR} \longrightarrow \fourcells{\pa}{\br}{\blankL}{\arrRzero}$
        & 4. $\fourcells{\pa}{\al}{\blankL}{\arrLzeroi} \longrightarrow \fourcells{\bl}{\pa}{\arrLzeroipo}{\blankR}$
        & 8. $\fourcells{\pa}{\al}{\blankL}{\arrLzero} \longrightarrow \fourcells{\al}{\pb}{\arrLzero}{\blankR}$ \\ & & & \\

    Tracks 0, 1 & 2. $\fourcells{\br}{\pb}{\arrRzero}{\blankR} \longrightarrow \fourcells{\pb}{\ar}{\blankL}{\arrRzero}$
                & 5. $\fourcells{\pb}{\bl}{\blankL}{\arrLzeroi} \longrightarrow \fourcells{\al}{\pb}{\arrLzeroipo}{\blankR}$
                & 9. $\fourcells{\pb}{\bl}{\blankL}{\arrLzero} \longrightarrow \fourcells{\bl}{\pa}{\arrLzero}{\blankR}$ \\ & & & \\

    & 3. $\threecellsRsym{\br}{\arrRzero} \longrightarrow \threecellsRsym{\al}{\arrLzeroz}$
        & 6. $\fourcells{\pa}{\al}{\blankL}{\arrLzeroK} \longrightarrow \fourcells{\bl}{\pa}{\arrLzero}{\blankR}$
        & 10. $\fourcells{\pa}{\al}{\blankL}{\arrLone} \overset{*}{\longrightarrow} \fourcells{\bl}{\pa}{\arrLone}{\blankR}$ \\ & & & \\

    & & 7. $\fourcells{\pb}{\bl}{\blankL}{\arrLzeroK} \longrightarrow \fourcells{\al}{\pb}{\arrLzero}{\blankR}$
        & 11. $\fourcells{\pb}{\bl}{\blankL}{\arrLone} \overset{*}{\longrightarrow} \fourcells{\al}{\pb}{\arrLone}{\blankR}$ \\ & & & \\
        
\hline & & & \\
    & 12. $\fourcellsLsym{\bl}{\arrLzero}{p_\alpha} \longrightarrow \fourcellsLsym{\ar}{\arrRone}{p_\alpha}$
        & 15. $\sixcellsvert{\ar}{\pa}{\arrRone}{\blankR}{\blankL}{\blankL} \longrightarrow \sixcellsvert{\pa}{\br}{\blankL}{\arrRone}{\blankL}{\blankL}$
        & 17. $\sixcellsvert{\ar}{\pa}{\arrRone}{\blankR}{\blankR}{\blankR} \longrightarrow \sixcellsvert{\pa}{\br}{\blankL}{\arrRone}{\blankR}{\blankR}$ \\ & & & \\

    Tracks 0, 1, 2 & 13. $\fourcellsLsym{\bl}{\arrLone}{\neg p_\alpha} \overset{*}{\longrightarrow} \fourcellsLsym{\ar}{\arrRone}{\neg p_\alpha}$
                & 16. $\sixcellsvert{\br}{\pb}{\arrRone}{\blankR}{\blankL}{\blankL} \longrightarrow \sixcellsvert{\pb}{\ar}{\blankL}{\arrRone}{\blankL}{\blankL}$
                & 18. $\sixcellsvert{\br}{\pb}{\arrRone}{\blankR}{\blankR}{\blankR} \longrightarrow \sixcellsvert{\pb}{\ar}{\blankL}{\arrRone}{\blankR}{\blankR}$ \\ & & & \\

    & 14. $\fourcellsRsym{\br}{\arrRone}{\blankR} \overset{*}{\longrightarrow} \fourcellsRsym{\al}{\arrLone}{\blankR}$ & & \\ & & & \\

\hline & & & \\
    & 19. $\eightcellsvert{\ar}{\pa}{\arrRone}{\blankR}{\blankL}{p}{\wc}{\sigma} \longrightarrow \eightcellsvert{\pa}{\br}{\blankL}{\arrRone}{p_L}{\blankR}{\wc}{\tau}$
        & 23. $\eightcellsvert{\ar}{\pa}{\arrRone}{\blankR}{p}{\blankR}{\sigma}{\wc} \longrightarrow \eightcellsvert{\pa}{\br}{\blankL}{\arrRone}{\blankL}{p'_R}{\tau}{\wc}$
        & 27. $\fivecellsRsym{\br}{\arrRone}{p}{\sigma} \overset{*}{\longrightarrow} \fivecellsRsym{\al}{\arrLone}{p_N}{\tau}$ \\ & & & \\

    Tracks 0, 1, 2, 3 & 20. $\eightcellsvert{\br}{\pb}{\arrRone}{\blankR}{\blankL}{p}{\wc}{\sigma} \longrightarrow \eightcellsvert{\pb}{\ar}{\blankL}{\arrRone}{p_L}{\blankR}{\wc}{\tau}$
        & 24. $\eightcellsvert{\br}{\pb}{\arrRone}{\blankR}{p}{\blankR}{\sigma}{\wc} \longrightarrow \eightcellsvert{\pb}{\ar}{\blankL}{\arrRone}{\blankL}{p'_R}{\tau}{\wc}$
        & 28. $\fivecellsRsym{\br}{\arrRone}{p'_R}{\wc} \overset{*} {\longrightarrow} \fivecellsRsym{\al}{\arrLone}{p_R}{\wc}$ \\ & & & \\

    & 21. $\eightcellsvert{\ar}{\pa}{\arrRone}{\blankR}{p}{\blankR}{\sigma}{\wc} \longrightarrow \eightcellsvert{\pa}{\br}{\blankL}{\arrRone}{p_N}{\blankR}{\tau}{\wc}$
        & 25. $\eightcellsvert{\ar}{\pa}{\arrRone}{\blankR}{p'_R}{\blankR}{\wc}{\wc} \longrightarrow \eightcellsvert{\pa}{\br}{\blankL}{\arrRone}{p_R}{\blankR}{\wc}{\wc}$
        & \\ & & & \\

    & 22. $\eightcellsvert{\br}{\pb}{\arrRone}{\blankR}{p}{\blankR}{\sigma}{\wc} \longrightarrow \eightcellsvert{\pb}{\ar}{\blankL}{\arrRone}{p_N}{\blankR}{\tau}{\wc}$
        & 26. $\eightcellsvert{\br}{\pb}{\arrRone}{\blankR}{p'_R}{\blankR}{\wc}{\wc} \longrightarrow \eightcellsvert{\pb}{\ar}{\blankL}{\arrRone}{p_R}{\blankR}{\wc}{\wc}$
        & \\ & & & \\
\hline
\end{longtable}
\clearpage
\setlength{\tabcolsep}{2pt} 
\renewcommand{\arraystretch}{0.9} 
\vspace*{-3cm}
\newcolumntype{P}[1]{>{\centering\arraybackslash}p{#1}}
\begin{longtable}{|P{6cm}|P{8cm}|}
\caption{Canonical clock illegal pairs. $x,y$ denotes any canonically oriented pair of track 0.} \label{tab:can_illegal} \\
\hline & \\
    Tracks 0, 1, 2 & 1. $\fourcellsnotLsym{\ar}{\arrRzero}{\neg\blankR}$ $\;\;$ 2. $\fourcellsLsym{\ar}{\arrRzero}{\neg p_\alpha}$ $\;\;$ 3. $\fourcellsLsym{\bl}{\arrLone}{p_\alpha}$
        $\;\;$ 4. $\threecellsvert{\wc}{\neg\arrRone}{p'_R}$ \\ & \\
\hline & \\
    Tracks 0, 1, 3 & 5. $\fourcellsnotLsym{\ar}{\arrRzero}{\neg \#}$ \\ & \\
\hline & \\
    Tracks 0, 2 & 6. $\threecellsnotLsym{\ar}{p_\alpha}$ $\;\;$ 7. $\threecellsnotLsym{\bl}{p_\alpha}$ \\ & \\
\hline & \\
    Tracks 0, 2, 3 & 8. $\sixcellsvert{x}{y}{p_\alpha}{\wc}{\wc}{\neg\#}$ $\;\;\;$ 9. $\sixcellsvert{x}{y}{\blankL}{p}{\#}{\#}$ \\ & \\
\hline & \\
    Tracks 0, 3 & 10. $\fourcells{x}{y}{0}{\#}$ $\;\;\;$ 11. $\threecellsRsym{x}{0}$ $\;\;\;$ \\ & \\
\hline & \\
    Tracks 0, 2, 3 for undefined transitions in \textsc{Base-$\zeta$ Counter} (\cite{CPW22}, (69)) & 12. $\threecellsvert{\wc}{p}{\tau}$ $\;\;\;$ 13. $\sixcellsvert{x}{y}{\blankL}{p_R}{\tau}{\wc}$ $\;\;\;$ 14. $\sixcellsvert{x}{y}{p_L}{\blankR}{\wc}{\tau}$ $\;\;\;$ 15. $\sixcellsvert{x}{y}{p_N}{\blankR}{\tau}{\wc}$ \\ & \\
\hline & \\
    Tracks 1, 2, 3 if $\delta(p,\sigma)=(p_L,\tau,L)$ & 16. $\threecellsvert{\arrRone}{p}{\sigma}$ \\ & \\
\hline & \\
    Tracks 0, 1, 4 & 17. $\fourcellsnotLsym{\ar}{\arrRzero}{\neg\blankR}$ $\;\;\;$ 18. $\fourcellsLsym{\ar}{\arrRzero}{\neg q_0}$ \\ & \\
\hline & \\
    Tracks 1, 5 & 19. $\twocellsvert{\arrLzero}{\neg 1}$ $\;\;\;$ 20. $\twocellsvert{\arrLzeroi}{\neg \#}$ \\ & \\
\hline & \\
    Tracks 0, 1, 6 & 21. $\fourcellsnotLsym{\ar}{\arrRzero}{\neg\#}$ $\;\;\;$ 22. $\fourcellsLsym{\ar}{\arrRzero}{\neg\vdash}$ \\ & \\
\hline
\end{longtable}
\clearpage
\setlength{\tabcolsep}{2pt} 
\renewcommand{\arraystretch}{0.75} 
\vspace*{-2cm}
\setlength\LTleft{-2cm}
\hfuzz=0pt
\newcolumntype{P}[1]{>{\centering\arraybackslash}p{#1}}
\begin{longtable}{|P{1.5cm}|P{18cm}|}
\caption{Quantum rules for the canonical orientation.}\label{tab:can_quantum} \\
\hline & \\
    Tracks 0, 1, 4 & 1. $\sixcellsvert{\pa}{\al}{\blankL}{\arrLone}{\blankL}{\blankL} \longrightarrow \sixcellsvert{\bl}{\pa}{\arrLone}{\blankR}{\blankL}{\blankL}$ \;\;
                   3. $\sixcellsvert{\pa}{\al}{\blankL}{\arrLone}{\blankR}{\blankR} \longrightarrow \sixcellsvert{\bl}{\pa}{\arrLone}{\blankR}{\blankR}{\blankR}$ \;\;
                   5. $\sixcellsvert{\pa}{\al}{\blankL}{\arrLone}{\blankL}{r_q} \longrightarrow \sixcellsvert{\bl}{\pa}{\arrLone}{\blankR}{r_q}{\blankR}$ \;\;
                   7. $\sixcellsvert{\pa}{\al}{\blankL}{\arrLone}{\blankL}{p} \longrightarrow \sixcellsvert{\bl}{\pa}{\arrLone}{\blankR}{\blankL}{p}$ \\
                   & \\
                   & 2. $\sixcellsvert{\pb}{\bl}{\blankL}{\arrLone}{\blankL}{\blankL} \longrightarrow \sixcellsvert{\al}{\pb}{\arrLone}{\blankR}{\blankL}{\blankL}$ \;\;
                   4. $\sixcellsvert{\pb}{\bl}{\blankL}{\arrLone}{\blankR}{\blankR}\longrightarrow \sixcellsvert{\al}{\pb}{\arrLone}{\blankR}{\blankR}{\blankR}$ \;\;
                   6. $\sixcellsvert{\pb}{\bl}{\blankL}{\arrLone}{\blankL}{r_q}\longrightarrow \sixcellsvert{\al}{\pb}{\arrLone}{\blankR}{r_q}{\blankR}$ \;\;
                   8. $\sixcellsvert{\pb}{\bl}{\blankL}{\arrLone}{\blankL}{p} \longrightarrow \sixcellsvert{\al}{\pb}{\arrLone}{\blankR}{\blankL}{p}$ \\
                   & \\ 
\hline & \\
                      & 9. $\Ket{\,\eightcellsvert{\pa}{\al}{\blankL}{\arrLone}{p}{\blankR}{\sigma}{\wc}\,} \longrightarrow \displaystyle\sum_{\tau,q_R}\delta(p,\sigma,\tau,q_R,R) \Ket{\,\eightcellsvert{\bl}{\pa}{\arrLone}{\blankR}{\blankL}{q_R}{\tau}{\wc}\,} 
                      + \displaystyle\sum_{\tau,q_N}\delta(p,\sigma,\tau,q_N,N)\Ket{\,\eightcellsvert{\bl}{\pa}{\arrLone}{\blankR}{q_N}{\blankR}{\tau}{\wc}\,}
                      + \displaystyle\sum_{\tau,q_L}\delta(p,\sigma,\tau,q_L,L)\Ket{\,\eightcellsvert{\bl}{\pa}{\arrLone}{\blankR}{q'_L}{\blankR}{\tau}{\wc}\,}$ \\ & \\
    Tracks 0, 1, 4, 5 & 10. $\Ket{\,\eightcellsvert{\pb}{\bl}{\blankL}{\arrLone}{p}{\blankR}{\sigma}{\wc}\,} \longrightarrow 
                      \displaystyle\sum_{\tau,q_R}\delta(p,\sigma,\tau,q_R,R) \Ket{\,\eightcellsvert{\al}{\pb}{\arrLone}{\blankR}{\blankL}{q_R}{\tau}{\wc}\,} 
                      + \displaystyle\sum_{\tau,q_N}\delta(p,\sigma,\tau,q_N,N)\Ket{\,\eightcellsvert{\al}{\pb}{\arrLone}{\blankR}{q_N}{\blankR}{\tau}{\wc}\,}
                      + \displaystyle\sum_{\tau,q_L}\delta(p,\sigma,\tau,q_L,L)\Ket{\,\eightcellsvert{\al}{\pb}{\arrLone}{\blankR}{q'_L}{\blankR}{\tau}{\wc}\,}$ \\ & \\
                      & 11. $\Ket{\,\eightcellsvert{\pa}{\al}{\blankL}{\arrLone}{\blankL}{q'_L}{\wc}{\wc}\,} \longrightarrow \Ket{\,\eightcellsvert{\bl}{\pa}{\arrLone}{\blankR}{q_L}{\blankR}{\wc}{\wc}\,}$ \;\; 12. $\Ket{\,\eightcellsvert{\pb}{\bl}{\blankL}{\arrLone}{\blankL}{q'_L}{\wc}{\wc}\,}
                      \longrightarrow \Ket{\,\eightcellsvert{\al}{\pb}{\arrLone}{\blankR}{q_L}{\blankR}{\wc}{\wc}\,}$ \\ & \\
\hline & \\
                      & 13. $\eightcellsvert{\pa}{\al}{\blankL}{\arrLone}{q_f}{\blankR}{\vdash}{\wc} \longrightarrow \eightcellsvert{\bl}{\pa}{\arrLone}{\blankR}{p_\alpha}{\blankR}{\;\vdash_{q_f}}{\wc}$ \;\;
                      15. $\eightcellsvert{\pa}{\al}{\blankL}{\arrLone}{p}{\blankR}{\sigma}{\wc} \longrightarrow \eightcellsvert{\bl}{\pa}{\arrLone}{\blankR}{\blankL}{p_R}{\tau}{\wc}$ \;\;
                      17. $\eightcellsvert{\pa}{\al}{\blankL}{\arrLone}{p}{\blankR}{\sigma}{\wc} \longrightarrow \eightcellsvert{\bl}{\pa}{\arrLone}{\blankR}{p_N}{\blankR}{\tau}{\wc}$\;\;
                      19. $\eightcellsvert{\pa}{\al}{\blankL}{\arrLone}{p}{\blankR}{\wc}{\sigma} \longrightarrow \eightcellsvert{\bl}{\pa}{\arrLone}{\blankR}{p'_L}{\blankR}{\wc}{\tau}$ \;\;
                      \\ & \\
    Tracks 0, 1, 4, 6 & 14. $\eightcellsvert{\pb}{\bl}{\blankL}{\arrLone}{q_f}{\blankR}{\vdash}{\wc} \longrightarrow \eightcellsvert{\al}{\pb}{\arrLone}{\blankR}                          {p_\alpha} 
                      {\blankR}{\;\vdash_{q_f}}{\wc}$ \;\;
                      16. $\eightcellsvert{\pb}{\bl}{\blankL}{\arrLone}{p}{\blankR}{\sigma}{\wc} \longrightarrow \eightcellsvert{\al}{\pb}{\arrLone}{\blankR}{\blankL}{p_R}{\tau}{\wc}$ \;\;
                      18. $\eightcellsvert{\pb}{\bl}{\blankL}{\arrLone}{p}{\blankR}{\sigma}{\wc} \longrightarrow \eightcellsvert{\al}{\pb}{\arrLone}{\blankR}{p_N}{\blankR}{\tau}{\wc}$ \;\;
                      20. $\eightcellsvert{\pb}{\bl}{\blankL}{\arrLone}{p}{\blankR}{\wc}{\sigma} \longrightarrow \eightcellsvert{\al}{\pb}{\arrLone}{\blankR}{p'_L}{\blankR}{\wc}{\tau}$ \;\;
                      \\ & \\
                      & 21. $\eightcellsvert{\pa}{\al}{\blankL}{\arrLone}{\blankL}{p'_L}{\wc}{\wc} \longrightarrow \eightcellsvert{\bl}{\pa}{\arrLone}{\blankR}{p_L}{\blankR}{\wc}{\wc}$ \;\; 22. $\eightcellsvert{\pb}{\bl}{\blankL}{\arrLone}{\blankL}{p'_L}{\wc}{\wc} \longrightarrow \eightcellsvert{\al}{\pb}{\arrLone}{\blankR}{p_L}{\blankR}{\wc}{\wc}$
                      \\ & \\
\hline & \\
    Tracks 0, 1, 2, 4 & 23. $\fivecellsRsym{\br}{\arrRone}{\blankR}{q} \longrightarrow \fivecellsRsym{\al}{\arrLone}{\blankR}{r_q}$ \\ & \\
\hline & \\
    Tracks 0, 1, 2, 4, 6 & 24. $\sixcellsLsym{\bl}{\arrLone}{\neg p_\alpha}{q'_L}{\vdash} \longrightarrow \sixcellsLsym{\ar}{\arrRone}{\neg p_\alpha}{p_\alpha}                                     {\;\vdash_{q'_L}}$ \;\;
                           25. $\sixcellsLsym{\bl}{\arrLone}{\neg p_\alpha}{r_q}{\vdash} \longrightarrow \sixcellsLsym{\ar}{\arrRone}{\neg p_\alpha}{p_\alpha}  
                               {\;\vdash_q}$ \\ & \\
\hline & \\
    Tracks 0, 1, 2, 3, 4 & 26. $\sixcellsRsym{\br}{\arrRone}{p}{\sigma}{q} \longrightarrow \sixcellsRsym{\al}{\arrLone}{p_N}{\tau}{r_q}$ \;\;
                           27. $\sixcellsRsym{\br}{\arrRone}{p'_R}{\cdot}{q} \longrightarrow \sixcellsRsym{\al}{\arrLone}{p_R}{\cdot}{r_q}$ \\ & \\
\hline
\end{longtable}
\setlength{\tabcolsep}{2pt} 
\renewcommand{\arraystretch}{0.9} 
\vspace*{2cm}
\newcolumntype{P}[1]{>{\centering\arraybackslash}p{#1}}
\begin{longtable}{|P{6cm}|P{8cm}|}
\caption{Illegal pairs for the quantum tracks in canonical orientation.} \label{tab:can_quantum_illegal} \\
\hline & \\
Tracks 0, 1, 4 & 1. $\fourcellsnotLsym{\ar}{\arrRzero}{\neg\blankR}$ \;\; 2. $\fourcellsLsym{\ar}{\arrRzero}{\neg q_0}$ \\ & \\ \hline & \\
Tracks 1, 5 & 3. $\twocellsvert{\arrLzero}{\neg 1}$ \;\; 4. $\twocellsvert{\arrLzeroi}{\neg\#}$ \\ & \\ \hline & \\
Tracks 0, 1, 6 & 5. $\fourcellsnotLsym{\ar}{\arrRzero}{\neg\#}$ \;\; 6. $\fourcellsLsym{\ar}{\arrRzero}{\neg\vdash}$ \\ & \\
\hline
\end{longtable}

\clearpage
\setlength{\tabcolsep}{2pt} 
\renewcommand{\arraystretch}{0.7} 
\vspace*{-3cm}
\setlength\LTleft{0cm}
\hfuzz=0pt
\newcolumntype{P}[1]{>{\centering\arraybackslash}p{#1}}
\begin{longtable}{|P{1cm}|P{4.5cm}P{4.5cm}P{4.5cm}|}
\caption{Clock rules for the reverse orientation. Rules marked with a $*$ are modified in Table \ref{tab:rev_quantum}.}\label{tab:rev_rules} \\
\hline & & & \\
    & 1. $\fourcells{\pa}{\ar}{\blankR}{\arrRzero} \longrightarrow \fourcells{\br}{\pa}{\arrRzero}{\blankL}$
        & 4. $\fourcells{\al}{\pa}{\arrLzeroi}{\blankL} \longrightarrow \fourcells{\pa}{\bl}{\blankR}{\arrLzeroipo}$
        & 8. $\fourcells{\al}{\pa}{\arrLzero}{\blankL} \longrightarrow \fourcells{\pb}{\al}{\blankR}{\arrLzero}$ \\ & & & \\

    Tracks 0, 1 & 2. $\fourcells{\pb}{\br}{\blankR}{\arrRzero} \longrightarrow \fourcells{\ar}{\pb}{\arrRzero}{\blankL}$
                & 5. $\fourcells{\bl}{\pb}{\arrLzeroi}{\blankL} \longrightarrow \fourcells{\pb}{\al}{\blankR}{\arrLzeroipo}$
                & 9. $\fourcells{\bl}{\pb}{\arrLzero}{\blankL} \longrightarrow \fourcells{\pa}{\bl}{\blankR}{\arrLzero}$ \\ & & & \\

    & 3. $\threecellsLsym{\br}{\arrRzero} \longrightarrow \threecellsLsym{\al}{\arrLzeroz}$
        & 6. $\fourcells{\al}{\pa}{\arrLzeroK}{\blankL} \longrightarrow \fourcells{\pa}{\bl}{\blankR}{\arrLzero}$
        & 10. $\fourcells{\al}{\pa}{\arrLone}{\blankL} \overset{*}{\longrightarrow} \fourcells{\pa}{\bl}{\blankR}{\arrLone}$ \\ & & & \\

    & & 7. $\fourcells{\bl}{\pb}{\arrLzeroK}{\blankL} \longrightarrow \fourcells{\pb}{\al}{\blankR}{\arrLzero}$
        & 11. $\fourcells{\bl}{\pb}{\arrLone}{\blankL} \overset{*}{\longrightarrow} \fourcells{\pb}{\al}{\blankR}{\arrLone}$ \\ & & & \\
        
\hline & & & \\
    & 12. $\fourcellsRsym{\bl}{\arrLzero}{p_\alpha} \longrightarrow \fourcellsRsym{\ar}{\arrRone}{p_\alpha}$
        & 15. $\sixcellsvert{\pa}{\ar}{\blankR}{\arrRone}{\blankL}{\blankL} \longrightarrow \sixcellsvert{\br}{\pa}{\arrRone}{\blankL}{\blankL}{\blankL}$
        & 17. $\sixcellsvert{\pa}{\ar}{\blankR}{\arrRone}{\blankR}{\blankR} \longrightarrow \sixcellsvert{\br}{\pa}{\arrRone}{\blankL}{\blankR}{\blankR}$ \\ & & & \\

    Tracks 0, 1, 2 & 13. $\fourcellsRsym{\bl}{\arrLone}{\neg p_\alpha} \overset{*}{\longrightarrow} \fourcellsRsym{\ar}{\arrRone}{\neg p_\alpha}$
                & 16. $\sixcellsvert{\pb}{\br}{\blankR}{\arrRone}{\blankL}{\blankL} \longrightarrow \sixcellsvert{\ar}{\pb}{\arrRone}{\blankL}{\blankL}{\blankL}$
                & 18. $\sixcellsvert{\pb}{\br}{\blankR}{\arrRone}{\blankR}{\blankR} \longrightarrow \sixcellsvert{\ar}{\pb}{\arrRone}{\blankL}{\blankR}{\blankR}$ \\ & & & \\

    & 14. $\fourcellsLsym{\br}{\arrRone}{\blankR} \overset{*}{\longrightarrow} \fourcellsLsym{\al}{\arrLone}{\blankR}$ & & \\ & & & \\

\hline & & & \\
    & 19. $\eightcellsvert{\pa}{\ar}{\blankR}{\arrRone}{p}{\blankL}{\sigma}{\wc} \longrightarrow \eightcellsvert{\br}{\pa}{\arrRone}{\blankL}{\blankR}{p_L}{\tau}{\wc}$
        & 23. $\eightcellsvert{\pa}{\ar}{\blankR}{\arrRone}{\blankR}{p}{\wc}{\sigma} \longrightarrow \eightcellsvert{\br}{\pa}{\arrRone}{\blankL}{p'_R}{\blankL}{\wc}{\tau}$
        & 27. $\fivecellsLsym{\br}{\arrRone}{p}{\sigma} \overset{*}{\longrightarrow} \fivecellsLsym{\al}{\arrLone}{p_N}{\tau}$ \\ & & & \\

    Tracks 0, 1, 2, 3 & 20. $\eightcellsvert{\pb}{\br}{\blankR}{\arrRone}{p}{\blankL}{\sigma}{\wc} \longrightarrow \eightcellsvert{\ar}{\pb}{\arrRone}{\blankL}{\blankR}{p_L}{\tau}{\wc}$
        & 24. $\eightcellsvert{\pb}{\br}{\blankR}{\arrRone}{\blankR}{p}{\wc}{\sigma} \longrightarrow \eightcellsvert{\ar}{\pb}{\arrRone}{\blankL}{p'_R}{\blankL}{\wc}{\tau}$
        & 28. $\fivecellsLsym{\br}{\arrRone}{p'_R}{\wc} \overset{*} {\longrightarrow} \fivecellsLsym{\al}{\arrLone}{p_R}{\wc}$ \\ & & & \\

    & 21. $\eightcellsvert{\pa}{\ar}{\blankR}{\arrRone}{\blankR}{p}{\wc}{\sigma} \longrightarrow \eightcellsvert{\br}{\pa}{\arrRone}{\blankL}{\blankR}{p_N}{\wc}{\tau}$
        & 25. $\eightcellsvert{\pa}{\ar}{\blankR}{\arrRone}{\blankR}{p'_R}{\wc}{\wc} \longrightarrow \eightcellsvert{\br}{\pa}{\arrRone}{\blankL}{\blankR}{p_R}{\wc}{\wc}$
        & \\ & & & \\

    & 22. $\eightcellsvert{\pb}{\br}{\blankR}{\arrRone}{\blankR}{p}{\wc}{\sigma} \longrightarrow \eightcellsvert{\ar}{\pb}{\arrRone}{\blankL}{\blankR}{p_N}{\wc}{\tau}$
        & 26. $\eightcellsvert{\pb}{\br}{\blankR}{\arrRone}{\blankR}{p'_R}{\wc}{\wc} \longrightarrow \eightcellsvert{\ar}{\pb}{\arrRone}{\blankL}{\blankR}{p_R}{\wc}{\wc}$
        & \\ & & & \\
\hline
\end{longtable}
\clearpage
\setlength{\tabcolsep}{2pt} 
\renewcommand{\arraystretch}{0.9} 
\vspace*{-3cm}
\newcolumntype{P}[1]{>{\centering\arraybackslash}p{#1}}
\begin{longtable}{|P{6cm}|P{8cm}|}
\caption{Reverse clock illegal pairs. $y,x$ denotes any reversibly oriented pair of track 0.} \label{tab:rev_illegal} \\
\hline & \\
    Tracks 0, 1, 2 & 1. $\fourcellsnotRsym{\ar}{\arrRzero}{\neg\blankR}$ $\;\;$ 2. $\fourcellsRsym{\ar}{\arrRzero}{\neg p_\alpha}$ $\;\;$ 3. $\fourcellsRsym{\bl}{\arrLone}{p_\alpha}$
        $\;\;$ 4. $\threecellsvert{\wc}{\neg\arrRone}{p'_R}$ \\ & \\
\hline & \\
    Tracks 0, 1, 3 & 5. $\fourcellsnotRsym{\ar}{\arrRzero}{\neg \#}$ \\ & \\
\hline & \\
    Tracks 0, 2 & 6. $\threecellsnotRsym{\ar}{p_\alpha}$ $\;\;$ 7. $\threecellsnotRsym{\bl}{p_\alpha}$ \\ & \\
\hline & \\
    Tracks 0, 2, 3 & 8. $\sixcellsvert{y}{x}{\wc}{p_\alpha}{\neg\#}{\wc}$ $\;\;\;$ 9. $\sixcellsvert{y}{x}{p}{\blankL}{\#}{\#}$ \\ & \\
\hline & \\
    Tracks 0, 3 & 10. $\fourcells{y}{x}{\#}{0}$ $\;\;\;$ 11. $\threecellsLsym{y}{0}$ \\ & \\
\hline & \\
    Tracks 0, 2, 3 for undefined transitions in \textsc{Base-$\zeta$ Counter} (\cite{CPW22}, 69) & 12. $\threecellsvert{\wc}{p}{\tau}$ $\;\;\;$ 13. $\sixcellsvert{y}{x}{p_R}{\blankL}{\wc}{\tau}$ $\;\;\;$ 14. $\sixcellsvert{y}{x}{\blankR}{p_L}{\tau}{\wc}$ $\;\;\;$ 15. $\sixcellsvert{y}{x}{\blankR}{p_N}{\wc}{\tau}$ \\ & \\
\hline & \\
    Tracks 1, 2, 3 if $\delta(p,\sigma)=(p_L,\tau,L)$ & 16. $\threecellsvert{\arrRone}{p}{\sigma}$ \\ & \\
\hline & \\
    Tracks 0, 1, 4 & 17. $\fourcellsnotRsym{\ar}{\arrRzero}{\neg\blankR}$ $\;\;\;$ 18. $\fourcellsRsym{\ar}{\arrRzero}{\neg q_0}$ \\ & \\
\hline & \\
    Tracks 1, 5 & 19. $\twocellsvert{\arrLzero}{\neg 1}$ $\;\;\;$ 20. $\twocellsvert{\arrLzeroi}{\neg \#}$ \\ & \\
\hline & \\
    Tracks 0, 1, 6 & 21. $\fourcellsnotRsym{\ar}{\arrRzero}{\neg\#}$ $\;\;\;$ 22. $\fourcellsRsym{\ar}{\arrRzero}{\neg\vdash}$ \\ & \\
\hline
\end{longtable}
\vspace*{-3cm}
\clearpage
\setlength{\tabcolsep}{2pt} 
\renewcommand{\arraystretch}{0.75} 
\vspace*{-2cm}
\setlength\LTleft{-2cm}
\hfuzz=0pt
\newcolumntype{P}[1]{>{\centering\arraybackslash}p{#1}}
\begin{longtable}{|P{1.5cm}|P{18cm}|}
\caption{Quantum rules for the reverse orientation.}\label{tab:rev_quantum} \\
\hline & \\
    Tracks 0, 1, 4 & 1. $\sixcellsvert{\al}{\pa}{\arrLone}{\blankL}{\blankL}{\blankL} \longrightarrow \sixcellsvert{\pa}{\bl}{\blankR}{\arrLone}{\blankL}{\blankL}$ \;\;
                   3. $\sixcellsvert{\al}{\pa}{\arrLone}{\blankL}{\blankR}{\blankR} \longrightarrow \sixcellsvert{\pa}{\bl}{\blankR}{\arrLone}{\blankR}{\blankR}$ \;\;
                   5. $\sixcellsvert{\al}{\pa}{\arrLone}{\blankL}{r_q}{\blankL} \longrightarrow \sixcellsvert{\pa}{\bl}{\blankR}{\arrLone}{\blankR}{r_q}$ \;\;
                   7. $\sixcellsvert{\al}{\pa}{\arrLone}{\blankL}{p}{\blankL} \longrightarrow \sixcellsvert{\pa}{\bl}{\blankR}{\arrLone}{p}{\blankL}$ \\
                   & \\
                   & 2. $\sixcellsvert{\bl}{\pb}{\arrLone}{\blankL}{\blankL}{\blankL} \longrightarrow \sixcellsvert{\pb}{\al}{\blankR}{\arrLone}{\blankL}{\blankL}$ \;\;
                   4. $\sixcellsvert{\bl}{\pb}{\arrLone}{\blankL}{\blankR}{\blankR}\longrightarrow \sixcellsvert{\pb}{\al}{\blankR}{\arrLone}{\blankR}{\blankR}$ \;\;
                   6. $\sixcellsvert{\bl}{\pb}{\arrLone}{\blankL}{r_q}{\blankL} \longrightarrow \sixcellsvert{\pb}{\al}{\blankR}{\arrLone}{\blankR}{r_q}$ \;\;
                   8. $\sixcellsvert{\bl}{\pb}{\arrLone}{\blankL}{p}{\blankL} \longrightarrow \sixcellsvert{\pb}{\al}{\blankR}{\arrLone}{p}{\blankL}$ \\
                   & \\ 
\hline & \\
                      & 9. $\Ket{\,\eightcellsvert{\al}{\pa}{\arrLone}{\blankL}{\blankR}{p}{\wc}{\sigma}\,} \longrightarrow \displaystyle\sum_{\tau,q_R}\delta(p,\sigma,\tau,q_R,R) \Ket{\,\eightcellsvert{\pa}{\bl}{\blankR}{\arrLone}{q_R}{\blankL}{\wc}{\tau}\,} 
                      + \displaystyle\sum_{\tau,q_N}\delta(p,\sigma,\tau,q_N,N)\Ket{\,\eightcellsvert{\pa}{\bl}{\blankR}{\arrLone}{\blankR}{q_N}{\wc}{\tau}\,}
                      + \displaystyle\sum_{\tau,q_L}\delta(p,\sigma,\tau,q_L,L)\Ket{\,\eightcellsvert{\pa}{\bl}{\blankR}{\arrLone}{\blankR}{q'_L}{\wc}{\tau}\,}$ \\ & \\
    Tracks 0, 1, 4, 5 & 10. $\Ket{\,\eightcellsvert{\bl}{\pb}{\arrLone}{\blankL}{\blankR}{p}{\wc}{\sigma}\,} \longrightarrow 
                      \displaystyle\sum_{\tau,q_R}\delta(p,\sigma,\tau,q_R,R) \Ket{\,\eightcellsvert{\pb}{\al}{\blankR}{\arrLone}{q_R}{\blankL}{\wc}{\tau}\,} 
                      + \displaystyle\sum_{\tau,q_N}\delta(p,\sigma,\tau,q_N,N)\Ket{\,\eightcellsvert{\pb}{\al}{\blankR}{\arrLone}{\blankR}{q_N}{\wc}{\tau}\,}
                      + \displaystyle\sum_{\tau,q_L}\delta(p,\sigma,\tau,q_L,L)\Ket{\,\eightcellsvert{\pb}{\al}{\blankR}{\arrLone}{\blankR}{q'_L}{\wc}{\tau}\,}$ \\ & \\
                      & 11. $\Ket{\,\eightcellsvert{\al}{\pa}{\arrLone}{\blankL}{q'_L}{\blankL}{\wc}{\wc}\,} \longrightarrow \Ket{\,\eightcellsvert{\pa}{\bl}{\blankR}{\arrLone}{\blankR}{q_L}{\wc}{\wc}\,}$ \;\; 12. $\Ket{\,\eightcellsvert{\pb}{\bl}{\arrLone}{\blankL}{q'_L}{\blankL}{\wc}{\wc}\,}
                      \longrightarrow \Ket{\,\eightcellsvert{\pb}{\al}{\blankR}{\arrLone}{\blankR}{q_L}{\wc}{\wc}\,}$ \\ & \\
\hline & \\
                      & 13. $\eightcellsvert{\al}{\pa}{\arrLone}{\blankL}{\blankR}{q_f}{\wc}{\vdash} \longrightarrow \eightcellsvert{\pa}{\bl}{\blankR}{\arrLone}{\blankR}{p_\alpha}{\wc}{\;\vdash_{q_f}}$ \;\;
                      15. $\eightcellsvert{\al}{\pa}{\arrLone}{\blankL}{\blankR}{p}{\wc}{\sigma} \longrightarrow \eightcellsvert{\pa}{\bl}{\blankR}{\arrLone}{p_R}{\blankL}{\wc}{\tau}$ \;\;
                      17. $\eightcellsvert{\al}{\pa}{\arrLone}{\blankL}{\blankR}{p}{\wc}{\sigma} \longrightarrow \eightcellsvert{\pa}{\bl}{\blankR}{\arrLone}{\blankR}{p_N}{\wc}{\tau}$\;\;
                      19. $\eightcellsvert{\al}{\pa}{\arrLone}{\blankL}{\blankR}{p}{\sigma}{\wc} \longrightarrow \eightcellsvert{\pa}{\bl}{\blankR}{\arrLone}{\blankR}{p'_L}{\tau}{\wc}$ \;\;
                      \\ & \\
    Tracks 0, 1, 4, 6 & 14. $\eightcellsvert{\bl}{\pb}{\arrLone}{\blankL}{\blankR}{q_f}{\wc}{\vdash} \longrightarrow \eightcellsvert{\pb}{\al}{\blankR}{\arrLone}                         {\blankR}{p_\alpha}{\wc}{\;\vdash_{q_f}}$ \;\;
                      16. $\eightcellsvert{\bl}{\pb}{\arrLone}{\blankL}{\blankR}{p}{\wc}{\sigma} \longrightarrow \eightcellsvert{\pb}{\al}{\blankR}{\arrLone}{p_R}{\blankL}{\wc}{\tau}$ \;\;
                      18. $\eightcellsvert{\bl}{\pb}{\arrLone}{\blankL}{\blankR}{p}{\wc}{\sigma} \longrightarrow \eightcellsvert{\pb}{\al}{\blankR}{\arrLone}{\blankR}{p_N}{\wc}{\tau}$ \;\;
                      20. $\eightcellsvert{\bl}{\pb}{\arrLone}{\blankL}{\blankR}{p}{\sigma}{\wc} \longrightarrow \eightcellsvert{\pb}{\al}{\blankR}{\arrLone}{\blankR}{p'_L}{\tau}{\wc}$ \;\;
                      \\ & \\
                      & 21. $\eightcellsvert{\al}{\pa}{\arrLone}{\blankL}{p'_L}{\blankL}{\wc}{\wc} \longrightarrow \eightcellsvert{\pa}{\bl}{\blankR}{\arrLone}{\blankR}{p_L}{\wc}{\wc}$ \;\; 22. $\eightcellsvert{\bl}{\pb}{\arrLone}{\blankL}{p'_L}{\blankL}{\wc}{\wc} \longrightarrow \eightcellsvert{\pb}{\al}{\blankR}{\arrLone}{\blankR}{p_L}{\wc}{\wc}$
                      \\ & \\
\hline & \\
    Tracks 0, 1, 2, 4 & 23. $\fivecellsLsym{\br}{\arrRone}{\blankR}{q} \longrightarrow \fivecellsLsym{\al}{\arrLone}{\blankR}{r_q}$ \\ & \\
\hline & \\
    Tracks 0, 1, 2, 4, 6 & 24. $\sixcellsRsym{\bl}{\arrLone}{\neg p_\alpha}{q'_L}{\vdash} \longrightarrow \sixcellsRsym{\ar}{\arrRone}{\neg p_\alpha}{p_\alpha}                                     {\;\vdash_{q'_L}}$ \;\;
                           25. $\sixcellsRsym{\bl}{\arrLone}{\neg p_\alpha}{r_q}{\vdash} \longrightarrow \sixcellsRsym{\ar}{\arrRone}{\neg p_\alpha}{p_\alpha}  
                               {\;\vdash_q}$ \\ & \\
\hline & \\
    Tracks 0, 1, 2, 3, 4 & 26. $\sixcellsLsym{\br}{\arrRone}{p}{\sigma}{q} \longrightarrow \sixcellsLsym{\al}{\arrLone}{p_N}{\tau}{r_q}$ \;\;
                           27. $\sixcellsLsym{\br}{\arrRone}{p'_R}{\cdot}{q} \longrightarrow \sixcellsLsym{\al}{\arrLone}{p_R}{\cdot}{r_q}$ \\ & \\
\hline
\end{longtable}
\setlength{\tabcolsep}{2pt} 
\renewcommand{\arraystretch}{0.9} 
\vspace*{2cm}
\newcolumntype{P}[1]{>{\centering\arraybackslash}p{#1}}
\begin{longtable}{|P{6cm}|P{8cm}|}
\caption{Illegal pairs for the quantum tracks in reverse orientation.} \label{tab:rev_quantum_illegal} \\
\hline & \\
Tracks 0, 1, 4 & 1. $\fourcellsnotRsym{\ar}{\arrRzero}{\neg\blankR}$ \;\; 2. $\fourcellsRsym{\ar}{\arrRzero}{\neg q_0}$ \\ & \\ \hline & \\
Tracks 1, 5 & 3. $\twocellsvert{\arrLzero}{\neg 1}$ \;\; 4. $\twocellsvert{\arrLzeroi}{\neg\#}$ \\ & \\ \hline & \\
Tracks 0, 1, 6 & 5. $\fourcellsnotRsym{\ar}{\arrRzero}{\neg\#}$ \;\; 6. $\fourcellsRsym{\ar}{\arrRzero}{\neg\vdash}$ \\ & \\
\hline
\end{longtable}

\end{appendices}

\end{document}